\documentclass[aps,prb,twocolumn,superscriptaddress,showpacs,english,nofootinbib]{revtex4-1}
\usepackage{caption}
\usepackage{float}
\usepackage{subcaption}
\usepackage[T1]{fontenc}
\usepackage[utf8]{inputenc}
\usepackage{babel}
\usepackage{amsmath}
\usepackage{amssymb}
\usepackage{wasysym}
\usepackage{graphicx}
\usepackage{multirow}
\usepackage{gensymb}
\usepackage[dvipsnames]{xcolor}
\usepackage{pifont}
\usepackage{microtype}
\usepackage{xcolor}
\usepackage{soul}
\usepackage[version=3]{mhchem} 
\usepackage{chemformula}[2014/04/08] 
\usepackage{booktabs}    
\usepackage{dcolumn}     
\usepackage{bm}  

\usepackage[linktocpage=true,
  colorlinks=true, 
  pdfborder={0 0 0},
  linkcolor=blue,
  citecolor=red,
  filecolor=yellow,
  urlcolor=blue,
  bookmarks,
  pdfauthor={},
]{hyperref}

\begin{document}


\title{ An assessment of the structural resolution of various fingerprints
commonly used in machine learning}
\author{Behnam Parsaeifard}
\affiliation{Department\ of\ Physics,\ University\ of\ Basel,\ Klingelbergstrasse\ 82,\ CH-4056\ Basel,\ Switzerland}
\affiliation{National Center for Computational Design and Discovery of Novel Materials (MARVEL), Switzerland}
\author{Deb Sankar De}
\affiliation{Department\ of\ Physics,\ University\ of\ Basel,\ Klingelbergstrasse\ 82,\ CH-4056\ Basel,\ Switzerland}
\affiliation{National Center for Computational Design and Discovery of Novel Materials (MARVEL), Switzerland}
\author{Anders S. Christensen}
\affiliation{Institute\ of\ Physical\ Chemistry,\ Department\ of\ Chemistry,\
University\ of\ Basel,\ Klingelbergstr.\ 80,\ CH-4056\ Basel,\ Switzerland}
\author{Felix A. Faber}
\affiliation{Institute\ of\ Physical\ Chemistry,\ Department\ of\ Chemistry,\
University\ of\ Basel,\ Klingelbergstr.\ 80,\ CH-4056\ Basel,\ Switzerland}
\author{Emir Kocer}
\affiliation{Universit\"at G\"ottingen, Institut f\"ur Physikalische Chemie, Theoretische Chemie, Tammannstr. 6, 37077 G\"{o}ttingen, Germany}

\author{Sandip De}
\affiliation{Present address BASF SE, 67056 Ludwigshafen am Rhein, Germany}
\affiliation{Laboratory of Computational Science and Modelling, Institute of Materials, Ecole Polytechnique F\'{e}d\'{e}rale de Lausanne, Lausanne, Switzerland}
\affiliation{National Center for Computational Design and Discovery of Novel Materials (MARVEL), Switzerland}

\author{J\"{o}rg Behler}
\affiliation{Universit\"at G\"ottingen, Institut f\"ur Physikalische Chemie, Theoretische Chemie, Tammannstr. 6, 37077 G\"{o}ttingen, Germany}
\author{O. Anatole von Lilienfeld}
\affiliation{Institute\ of\ Physical\ Chemistry,\ Department\ of\ Chemistry,\
University\ of\ Basel,\ Klingelbergstr.\ 80,\ CH-4056\ Basel,\ Switzerland}
\affiliation{National Center for Computational Design and Discovery of Novel Materials (MARVEL), Switzerland}
\author{Stefan Goedecker}
\affiliation{Department\ of\ Physics,\ University\ of\ Basel,\ Klingelbergstrasse\ 82,\ CH-4056\ Basel,\ Switzerland}
\affiliation{National Center for Computational Design and Discovery of Novel Materials (MARVEL), Switzerland}


\date{\today}

\begin{abstract}
Atomic environment fingerprints are widely used in computational materials science, from machine learning potentials to the quantification of 
similarities between atomic configurations. 
Many approaches to the construction of such 
fingerprints, also called 
structural
descriptors, have been proposed. 
In this work, we compare the performance of fingerprints based on the Overlap Matrix (OM), 
the Smooth Overlap of Atomic Positions (SOAP), Behler-Parrinello atom-centered symmetry functions (ACSF),  modified Behler-Parrinello symmetry functions (MBSF) used in the ANI-1ccx potential
and the Faber-Christensen-Huang-Lilienfeld (FCHL) fingerprint
under various aspects. We study their ability to resolve differences in local environments and in 
particular examine whether there are certain atomic movements that leave the fingerprints exactly or nearly 
invariant.  
For this purpose, 
we introduce a sensitivity matrix whose eigenvalues quantify the effect 
of atomic displacement modes on the fingerprint. 
Further, we check whether these displacements correlate with
the variation of localized physical quantities such as forces.
Finally, we 
extend our examination to the correlation between molecular fingerprints 
obtained from the atomic fingerprints and global quantities of entire molecules.

\end{abstract} 

\maketitle

\section{introduction}
Materials sciences and chemistry are becoming data driven sciences~\cite{morgan2004high,saal2013materials,curtarolo2012aflow,jain2013commentary,de2015charting,qu2015electrolyte,kirklin2015open,blum,rupp2012fast}. 
Both experimental and theoretical data often contain similar, or duplicate structures  
which differ only by the noise which is present in any experimental measurements as 
well as in theoretical structure predictions~\cite{lyakhov2010crystal,goedecker2004minima,amsler2010crystal,neumann2008major, oganov2009quantify}.
Such structures can be eliminated based on fingerprint 
distances.  If the structures differ by more than just noise, 
one frequently wants to quantify their dissimilarity. This is particularly important 
for applications
of supervised machine learning in materials science~\cite{P2559,P3033,P4938,P4643,behler2017},
where fingerprints form in most schemes the 
input for neural networks or other machine learning schemes, but also for eliminating redundant structures e.g. in the global exploration of potential-energy surfaces. Both, for the detection of duplicate structures as well as for machine learning 
various atomic environment descriptors 
have been proposed to date.
In the pioneering work of Behler and Parrinello~\cite{behler2007generalized,behler2011atom}
so-called symmetry functions 
have been introduced to explore the chemical environment of each atom 
and to form the input to 
atomic neural networks.
Two schemes 
related
to the original Behler-Parrinello 
atom-centered symmetry functions (ACSF)
will also be used here and 
denoted as 
MBSF~\cite{smith2017ani} and FCHL~\cite{faber2018alchemical}. 
The numerically more efficient discretized version of 
the FCHL fingerprint~\cite{christensen2020fchl} is used in our study.
Another fingerprint that is widely used in the 
context of machine learning is the Smooth Overlap of Atomic Positions (SOAP)
atomic environment descriptor~\cite{bartok2013representing}.
The last fingerprint that is included in our tests is the Overlap Matrix (OM) fingerprint~\cite{zhu2016fingerprint}  that 
has been used to find duplicate structures in minima hopping based structures 
predictions~\cite{MH} and to bias the potential energy landscape to find chemical reaction 
pathways~\cite{PhysRevLett.123.206102}, as well as in machine learning~\cite{joost,sadeghi}.
Many other types of fingerprints have been proposed in the literature to date~\cite{P3136,P5292,P5075,P5420,P4862,P4644,P5645,doi:10.1021,huang2017dna,mallat,huan}.
In the following all these descriptors will be called fingerprints, Cartesian coordinates of atoms in structures, augmented in the crystalline case with the vectors 
describing the unit cell, form an elementary representation of a configuration or atomic environment. However 
such Cartesian descriptors are problematic since they are not invariant under
translations, rotation and atomic index permutations. So, other descriptors are needed which must be invariant under translations, rotations, and other symmetry operations as well as permutation of identical atoms~\cite{behler2007generalized}. 
All the fingerprints considered in this work are invariant under these operations.
The fingerprint distance between two structures can for instance be calculated as the Euclidean norm 
of the difference between the two fingerprint vectors.
In this work, we compare the structural resolution of various fingerprints, i.e.  their ability to recognize and quantify 
differences in atomic environments based on such fingerprint distances.

\section{Description of fingerprints used}

  In this section we give a very brief summary of the fingerprints used in this study.  
  For a complete description of the fingerprints, the reader is referred to the original publications on OM~\cite{zhu2016fingerprint}, SOAP~\cite{bartok2013representing}, FCHL~\cite{christensen2019fchl}, ACSF~\cite{behler2011atom}, and MBSF~\cite{smith2017ani}.
  
  The OM method is inspired by the experimental approach to identify structures. 
  Experimental approaches typically use some spectrum such as a vibrational spectrum or an electronic 
  excitation spectrum to identify structures. Both are related to the eigenvalues of certain matrices.
  As was shown by Sadeghi et al.~\cite{sadeghi2013metrics} 
  eigenvalues of the Hessian matrix or of the Kohn Sham Hamiltonian matrix are excellent fingerprints 
  for molecular structures, but these matrices are quite expensive to calculate. Fortunately, it turns out 
  that the eigenvalues of a matrix that is extremely fast to calculate, namely the overlap matrix which contains the full structural information are of 
  comparable quality.
  To calculate the fingerprint of an atom $k$ in the OM scheme, a sphere of radius $R_c$ 
  is centered on it. We place a minimal basis set of four Gaussian type  orbitals (GTOs) $G_{\nu}(\mathbf{r}-\mathbf{R}_i)$ 
  (i.e. radial Gaussians times spherical harmonics) on 
  each atom $i$ in the sphere, namely one s-type GTO ($ \nu=1$), and 3 p-type GTOs ($ \nu=2,3,4$) {\color{black} shown by OM[sp]}. 
  The width of the radial Gaussian is given by the 
  covalent radius of the element. Then the overlap between all atoms in the sphere is calculated as $S^k_{i,\nu,j,\mu}=\int G_{\nu}(\mathbf{r}-\mathbf{R}_i)G_{\mu}(\mathbf{r}-\mathbf{R}_j)d\bm{r}$.

  The off-diagonal elements of the overlap matrix decay quite fast with respect to distance from the central atom. This decay is also exploited in the linear electronic structure calculation~\cite{rmp}.
  Such a fast decay has been shown in a similar context to be advantageous compared to a slower inverse power law decay~\cite{bing}. 
  Each element $S^k_{i,j}$  of this matrix is then multiplied by two amplitudes ~$f_c(|\mathbf{R}_k-\mathbf{R}_i|)$ and ~$f_c(|\mathbf{R}_k-\mathbf{R}_j|)$ where 
  $f_c(r)= \left( 1-\frac{1}{4} ( \frac{r}{w} )^2 \right)^2$ 
  is a cutoff function which smoothly tends to zero at $r=2w =R_c$.
  So the width $w$ which determines the cutoff radius is the only parameter in this scheme.
  
  The vector $\mathbf{F}^k$ containing all the eigenvalues of this matrix is then the fingerprint of atom $k$. 
  The fingerprint distance between two atoms $I$ and $J$ is defined to be the Euclidean distance between their fingerprint 
  vectors~\cite{sadeghi2013metrics}: $\Delta_{IJ}=|\mathbf{F}^I-\mathbf{F}^J|$.
  
  
  The above defined fingerprint distance has a discontinuity in the first derivative when 
  two eigenvalues cross. This is an extremely rare event~\cite{Neumann-Wigner}
  and does not cause problems in most applications.
  If a completely continuous distance is desired the following post-processing of the eigenvalues 
  can be used to generate a new set $\mathbf{\tilde{F}}$ of fingerprints that gives rise to completely continuous 
  fingerprint distances:
  \begin{equation}
   \tilde{F}_i = \frac{  \sum_l F_l \exp \left(-\frac{1}{2} \left( \frac{F_l-F_i}{a} \right)^2   \right)}
                     {\sum_l \exp \left(-\frac{1}{2} \left( \frac{F_l-F_i}{a} \right)^2   \right)} 
   \end{equation}
  

  In the SOAP (Smooth Overlap of Atomic Positions) scheme,
  a Gaussian of width $\sigma$ is centered on each atom within the cutoff distance around the central
  atom $k$ at position $\mathbf{r}$. The resulting density of atoms
  $\rho^k(\mathbf{r})=\sum_{i} \exp\left( -\frac{(\mathbf{r} -\mathbf{R}_{ki})^2}{2\sigma^2}\right) \times f_{cut}(|\mathbf{r}-\mathbf{R}_{ki}|)$, 
  multiplied with a cutoff function, which goes smoothly to zero at the cutoff radius over a characteristic width $r_\delta$,
  is then expanded in terms of orthogonal radial functions $g_{n}(r)$ and spherical harmonics $Y_{lm}(\theta,\phi)$ as $\rho^k(\mathbf{r})=\sum_{nlm}{c^k_{nlm}g_n(r)Y_{lm}(\theta,\phi)}$,
  where $c^k_{nlm}=<g_nY_{lm}|\rho^k>$.
  $p^k_{nn'l}=\sqrt{\frac{8\pi^2}{2l+1}}\sum_{m}{c^k_{nlm}(c^{k}_{n'lm})^{*}}$ is invariant under rotations and the vector $\mathbf{F}^k$ containing all $p^k_{nn'l}$'s with $n,n' \leq n_{max}$ and $l \leq l_{max}$ is the SOAP fingerprint vector of atom $k$.
  The fingerprint distance between atoms $I$ and $J$ can then either  be defined as  $\Delta_{IJ}=|\mathbf{F^I}-\mathbf{F^J}|$ or 
  $\Delta_{IJ}=(1-\mathbf{F^I}\cdot \mathbf{F^J})^{1/2}$.
  Since the second definition is used in the majority of machine learning applications and since we could not find any difference in preliminary tests, for SOAP we
  use the second definition of the fingerprint distance. 
  This definition requires the fingerprint vector to be normalized to 1 such that $\sum_{i} F_i^2=1$.

  
  This has the strange side effect that the $N$ fingerprints of a system of $N$ atoms 
  remain identical if $N$  additional atoms are placed on top of the 
  original $N$ atoms.
  Further, the fingerprint vectors are the same for 
  a dimer where the two atoms are at a very large and zero distance.
  
  The QUIPPY~\cite{quippy} software was used to generate the SOAP fingerprints, with the following parameters: 
  $n_{max}=l_{max}=12$ and $\sigma=0.5$, $r_\delta=4.0$ \r{A}.
  
  
  The atom-centered symmetry functions (ACSF) proposed by Behler and Parrinello in 2007 have been the first descriptors suitable as input for ML methods for the description of high-dimensional multi-atom systems~\cite{behler2007generalized,behler2011atom}. They form atomic fingerprint vectors consisting of sets of atom-centered many-body radial and angular functions, which describe the chemical environments of the atoms in the system.
 
 Radial functions are the sum of two-body terms and describe the radial environment of an atom $i$. 
  They have, for instance, the analytical form $G^2_i=\sum_{j}{e^{-\eta(R_{ij}-R_s)^2}f_c(R_{ij})}$. 
  
  The angular functions are sums of three-body terms and describe the angular environment of the atom. Two 
examples are defined below:

\begin{widetext}
  \begin{equation}
  G^4_i=2^{1-\zeta}\sum_{j,k\neq i}^{all}{(1+\lambda \cos(\theta_{ijk}))^{\zeta}e^{-\eta (R^2_{ij}+R^2_{ik}+R^2_{jk})}f_c(R_{ij})f_c(R_{ik})f_c(R_{jk})} 
  \end{equation}
  
\end{widetext}
\begin{widetext}
  \begin{equation}
      G^5_i =2^{1-\zeta}\sum_{j,k\neq i}^{all}{(1+\lambda \cos(\theta_{ijk}))^{\zeta}e^{-\eta (R^2_{ij}+R^2_{ik})}f_c(R_{ij})f_c(R_{ik})} \
  \end{equation}
\end{widetext}
  where $\theta_{ijk}$ is the angle between $\mathbf{R_{ij}}$ and $\bold{R_{ik}}$ and $f_c(r)$ is a smooth cutoff function~\cite{behler2011atom}. The vector $\mathbf{F}^i$ containing all the $G_i$'s for various values of $\eta$, $\lambda$, $R_s$, and $\zeta$ is the fingerprint vector of atom $i$. 
  In the present work, we used 10 radial symmetry functions of type $G_2$ and 48 angular symmetry functions of type $G_4$, which have been generated with the software RuNNer~\cite{behler2015constructing,behler2017}. We have used CUR to find the most relevant symmetry functions~\cite{P5398}, as we found that larger sets did not lead to significant improvements.

  
  Isayev et al. made two modifications to the original Behler-Parrinello angular symmetry functions to obtain modified Behler-Parrinello symmetry functions (MBSFs)~\cite{smith2017ani} while retaining the form of the radial functions. 
  These modifications are the addition of a reference angle $\theta_s$ to the term $\cos(\theta_{ijk})$ which allows an arbitrary number of shifts in the angular environment and $R_s$ to the exponential term in the angular symmetry functions. The $R_s$ addition allows the angular environment to be considered within radial shells based on the average of the distance from the neighboring atoms~\cite{smith2017ani} similar to the radial shift $R_s$ in the original Behler-Parrinello radial functions. So their modified angular symmetry function is 
\begin{widetext}
  \begin{equation}
    G^A_i=2^{1-\zeta}\sum_{j,k\neq i}^{all}{(1+\lambda  \cos(\theta_{ijk}-\theta_s))^{\zeta} e^{-\eta (\frac{R_{ij}+R_{ik}}{2}-R_s)^2}f_c(R_{ij})f_c(R_{ik})}
  \end{equation}
\end{widetext}
  In this approach, a single $\eta$ and multiple values of $R_s$ and $\theta_s$ are used to generate the fingerprint vector $\mathbf{F}^i$.
  We used 32 evenly spaced radial shifting parameters for the radial part, and a total of 8 radial and 8 angular shifting parameters
  for the angular part for the MBSF resulting in a total 96 symmetry functions. 
  The QML~\cite{christensen2017qml} software package was then used to generate the MBSF fingerprints.

The last fingerprint that we study is the discretized FCHL fingerprint introduced  by Faber et al.~\cite{christensen2019fchl}. 
FCHL encodes geometric elemental information into the fingerprint  
with up to three-body terms included.
%
The 2-body terms consist of sums of log-normal radial functions on the form
\begin{widetext}
\begin{equation}
    G^\text{2-body} =  \xi_2\left(r_{IJ}\right)f_\text{cut}\left(r_{IJ}\right) \tfrac{1}{R_s\sigma\left(r_{ij}\right)\sqrt{2\pi}} \exp\left(- \frac{\left( \ln R_s - \mu\left(r_{ij}\right) \right)^2}{2\sigma\left(r_{ij}\right)^2}\right)
\end{equation}
\end{widetext}
%
where $ f_\text{cut}\left(r_{IJ}\right) $ is a smooth cut-off function, $\xi_2\left(r_{IJ}\right)$ is a weight function on the form $\frac{1}{r_{ij} ^{N_2}}$ which serves to put a higher weight in the regression to effects from atoms at closer distances, $\mu\left(r_{ij}\right)=\ln{\left(\frac{r_{IJ}}{\sqrt{1+\frac{w}{r_{IJ}^2}}} \right)}$, and $\sigma\left(r_{ij}\right)^2=1+\frac{w}{r_{IJ}^2}$.
The three-body term in FCHL is the product of a radial part,
but uses a (truncated) Fourier expansion for the angular spectrum on the form:

\begin{widetext}
\begin{equation}
    G^\text{3-body} =  \xi_3 G^{3-body}_{Radial}G^{3-body}_{Angular} f_\text{cut}\left(r_{IJ}\right) f_\text{cut}\left(r_{JK}\right)f_\text{cut}\left(r_{KI}\right) 
\end{equation}
\end{widetext}
Where 
\begin{equation}
    G^{3-body}_{Radial}=\sqrt{\frac{\eta_3}{\pi}}\exp{\left( -\eta_3 \left( \frac{1}{2} \left( r_{IJ}+r_{IK}\right)-R_s \right)^2\right)}
\end{equation}
and $G^{3-body}_{Angular}$ contains the below sine and cosine terms with $n=1$:
\begin{align}
    G_n^{\mathrm{cos}}&=\exp\left(-\frac{\left(\zeta  n\right)^2}{2}\right) \left(\cos{\left(n \theta_{KIJ}\right)} - \cos\left(n\left(\theta_{KIJ}+\pi\right)\right)\right)\\
    G_n^{\mathrm{sin}}&= \exp\left(-\frac{\left(\zeta  n\right)^2}{2}\right)\left(\sin{\left(n \theta_{KIJ}\right)} - \sin\left(n\left(\theta_{KIJ}+\pi\right)\right)\right)
\end{align}
where $\theta_{KIJ} $ is the angle between the atoms I, J and K.
Furthermore, the three-body symmetry functions
are weighted with an Axilrod-Teller-Muto term~\cite{Muto1943,AxilrodTeller} defined as:
\begin{equation}
    \xi_3 = c_3 \frac{1 + 3\cos\left(\theta_{KIJ}\right)\cos\left(\theta_{IJK}\right)\cos\left(\theta_{JKI}\right) }{\left(r_{IK}r_{JK}r_{KI}\right)^{N_3}}
\end{equation}
This again serves to attribute a higher weight to atomic configuration that likely to more strongly interacting~\cite{faber2018alchemical,bing}. 
We used the default parameters described in~\cite{faber2018alchemical} and~\cite{christensen2020fchl} and the 
QML~\cite{christensen2017qml} software to generate the FCHL fingerprints.

For all fingerprints related to the Behler-Parrinello symmetry functions, i.e. for ACSF, MBSF  and FCHL we use the Euclidean norm of the difference of the fingerprint vectors 
as the fingerprint distance. 

  


For a fair comparison  we have chosen for all fingerprints the same cutoff radius, namely 6.0
\r{A}. This or very similar values were used in numerous studies~\cite{behler2015constructing,christensen2020fchl,dragoni2018achieving,smith2017ani}.
So all the methods see exactly the same environment and could therefore in principle encode the same information in their resulting fingerprint vectors. 
With this choice of parameters, the length of the fingerprints was 240 for OM, 1015 for SOAP, 58 for ACSF, 96 for MBSF and 64 for FCHL.



\begin{table*}[]
    \centering
    \begin{tabular}{|c|c|c|c|c|c|}
         \hline
         Fingerprint type & Name & number & value & unit & description  \\
         \hline
         \multirow{8} {2em}{MBSF} & $R_s (G^R)$ & 32 & \footnote{[0.8, 0.968, 1.135, 1.303, 1.471, 1.639, 1.806, 1.974, 2.142, 2.31, 2.477, 2.645, 2.813, 2.981, 3.148, 3.316, 3.484, 3.652, 3.819, 3.987, 4.155 4.323, 4.490, 4.658, 4.826, 4.994, 5.161, 5.329, 5.497, 5.665, 5.832, 6.0]} & \r{A} & Two-body radial bins \\
         & $R_s (G^A)$ & 8 & \footnote{ [0.8, 1.543, 2.286, 3.0286, 3.771, 4.514, 5.257, 6.0]} & \r{A} & Three-body radial bins \\
         & $\theta_s$ & 8 & \footnote{[0.0, 0.449, 0.898, 1.346, 1.795, 2.244, 2.693, 3.142]} &  &     Three-body angular bins \\
         & $r_\mathrm{cut}$  & & 6.0 &  \r{A}         & Radial cutoff (two-body) \\
         & $a_\mathrm{cut}$  & & 6.0 &  \r{A}     &     Radial cutoff (three-body) \\
         & $\eta (G^R)$  &  & 1.0  &  \r{A}$^{-2}$ &  Two-body width parameter \\
         & $\eta (G^A)$  &  & 1.0  & \r{A}$^{-2}$ &  Three-body width parameter \\
         & $\zeta$   &  & 1.0  &       &  Angular exponent \\
         \hline
         \multirow{11} {2em}{FCHL} & $n_{Rs2}$ & 24 & \footnote{ [0.25, 0.5, 0.75, 1.0, 1.25, 1.5, 1.75, 2.0, 2.25, 2.5, 2.75, 3.0, 3.25, 3.5, 3.75, 4.0, 4.25, 4.5, 4.75, 5.0, 5.25, 5.5, 5.75, 6.0]}& \AA & Two-body radial bins \\
         & $n_{Rs3}$    & 20 & \footnote{[0.3, 0.6, 0.9, 1.2, 1.5, 1.8, 2.1, 2.4, 2.7, 3.0, 3.3, 3.6, 3.9, 4.2, 4.5, 4.8, 5.1, 5.4, 5.7, 6.0]}& \AA & Sin three-body radial bins \\
         & $n_{Rs3}$    & 20 & \footnote{[0.3, 0.6, 0.9, 1.2, 1.5, 1.8, 2.1, 2.4, 2.7, 3.0, 3.3, 3.6, 3.9, 4.2, 4.5, 4.8, 5.1, 5.4, 5.7, 6.0]} &  \AA& Cos three-body radial bins \\
         & w    &  & 0.32 &  \AA$^2$    &  Two-body width parameter \\
         
         & $\eta_3$ & &2.7  &  \AA$^{-2}$ &  Three-body width parameter \\
         & $N_2$    &  &1.8 &    &     Two-body scaling exponent \\
         & $N_3$    &  &0.57 &   &      Three-body scaling exponent \\
        & $c_3$     &   &   13.4 &  \AA$^{N_3}$ & Three body-weight \\
        & $\zeta$   &   &  $\pi$ &   &  Angular exponent \\
        & $r_\mathrm{cut}$ & & 6.0 &  \AA   &       Radial cutoff (two-body) \\
       & $a_\mathrm{cut}$ & &6.0 &  \AA   &      Radial cutoff (three-body) \\
       \hline
       \multirow{5}{2em}{SOAP} & $\sigma$ & & 0.5 & \AA & atom sigma\\ 
       & $l_{max}$ & & 12 & &  \\
       & $n_{max}$ & & 12& & \\
       & $r_\delta$ & & 4.0 & \AA & Characteristic decay length\\
       & $R_c$ & & 6.0 & \AA & Cutoff radius \\
       \hline
       \multirow{4}{2em}{OM} & w & & 3.0 & \AA & Gaussian width \\
       & $R_c=2w$ & & 6.0 & \AA & Cutoff radius \\
       & s-type orbitals & 1& $s$& & \\
       & p-type orbitals & 3& $p_x$,$p_y$,$p_z$ & & \\
       \hline
       \multirow{5}{2em}{ACSF} & $\eta (G_2)$ & 10 & \footnote{[0.003, 0.018, 0.036, 0.054, 0.071, 0.089, 0.125, 0.161, 0.214, 0.285]}& \AA$^{-2}$ & Two-body width parameter\\
       & $\eta (G_4)$ & 6 & \footnote{[0.0, 0.004, 0.018, 0.071, 0.214, 0.285]}& \AA$^{-2}$ & Three-body width parameter \\
       & $\lambda$ & 2 & -1,1&  &\\
       & $\zeta$ & 4 & 1,2,4,16 & & Angular exponent\\
       & $R_c$ & & 6.0 & \AA & Cutoff radius \\
       \hline
    \end{tabular}
    \caption{The parameters used for each fingerprint.}
    \label{tab:param_tab}
\end{table*}

\section{Results}

In this section we will introduce some criteria to assess the performance of the various fingerprints.
First, we derive a formalism that allows to check the behavior of the different fingerprints 
under infinitesimal changes of the atomic coordinates. We show that there is a matrix, that 
we baptize sensitivity matrix, that describes this behavior. In particular, the displacement modes 
of this matrix that belong to zero eigenvalues give rise to constant fingerprints for movements 
along these modes and indicate therefore a failure of the fingerprint to detect geometry changes. 
Next we will compare for a test set the distances obtained by different fingerprints.
This test helps us to find cases where a certain fingerprint can not recognize differences between different chemical environments.
In addition we will correlate in both cases changes in fingerprint distances with changes of 
physical quantities such as forces, energies and densities of states.

\subsection{Behavior of fingerprints under infinitesimal displacements}\label{derivatives}
To study the evolution of fingerprint distances under small displacements, we 
consider the change of the squared fingerprint distance up to second order in a Taylor expansion 
around a reference configuration.
Denoting the fingerprint of the reference configuration by $\bold{F^0}$ and the 
fingerprint of a configuration displaced by $\Delta \bold{R}$ by $\bold{F(R)}$  we get 
\begin{equation}
(\bold{F(R)} - \bold{F^0})^2 = \sum_{\alpha,\beta}
\Delta {R}_{\alpha} \left( \sum_i  g_{i,\alpha}  g_{i,\beta} \right)  \Delta {R}_{\beta}
\end{equation}

where $ g_{i,\alpha}$ is the gradient of the $i$-th component of the fingerprint vector with respect to 
the three Cartesian components $\alpha$ (x, y, and z) of the position vector $\bold{R}$, i.e.
\begin{equation}
    g_{i,\alpha} = \left.  \frac{ \partial F_i}{\partial R_{\alpha} } \right|_{\bold{R} = \bold{R}_0} \
\end{equation}
In taking this derivative we have to consider only the atomic positions within the sphere around the central atom because by construction 
atoms outside the sphere have no influence on the fingerprint. 
We call this matrix $ \sum_i  g_{i,\alpha}  g_{i,\beta} $ sensitivity matrix.
It has the dimension $3N\times 3N$ where $N$ is the number of atoms within the cutoff sphere around the reference atom. In the following, we will examine its eigenvalues and eigenvectors. To allow a meaningful comparison of the fingerprints obtained by different methods 
we have scaled all the eigenvalues such that the largest eigenvalue is one.  
Since the fingerprint is invariant under a uniform translation and rotation of all the atoms in the sphere, the sensitivity matrix has always at least 6 zero eigenvalues.  More than 6 zero eigenvalues indicate 
that there are other displacement modes which will leave the fingerprint invariant. 
This is highly problematic since it indicates that one can generate different atomic environments which will not change the fingerprint. By calculating iteratively these zero eigenvalue displacement modes and then moving the system by an infinitesimal amount along those consecutive modes one can construct from a sequence of 
infinitesimal small displacements a finite displacement 
which will leave the fingerprint 
invariant~\cite{sadeghi2013metrics}. Equally problematic are eigenvalues that are very small.
In this case the fingerprint variation will not exactly be zero, but will be extremely small. 
\begin{figure}[!h]
     \centering
     \begin{subfigure}[b]{0.2\textwidth}
         \centering
         \includegraphics[width=\textwidth]{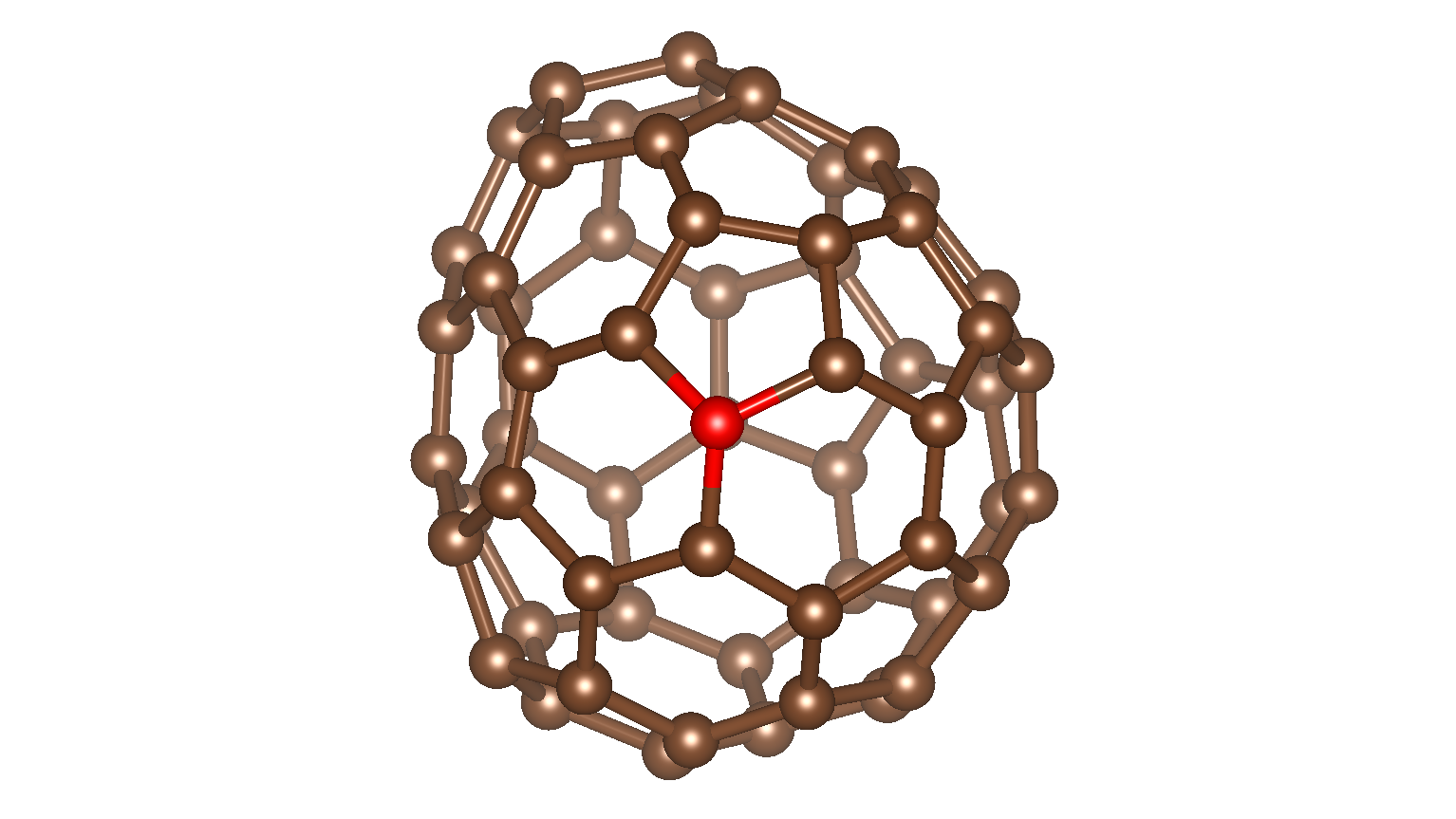}
         \caption{}
         \label{fig:conf}
     \end{subfigure}
     \begin{subfigure}[b]{0.2\textwidth}
         \centering
         \includegraphics[width=\textwidth]{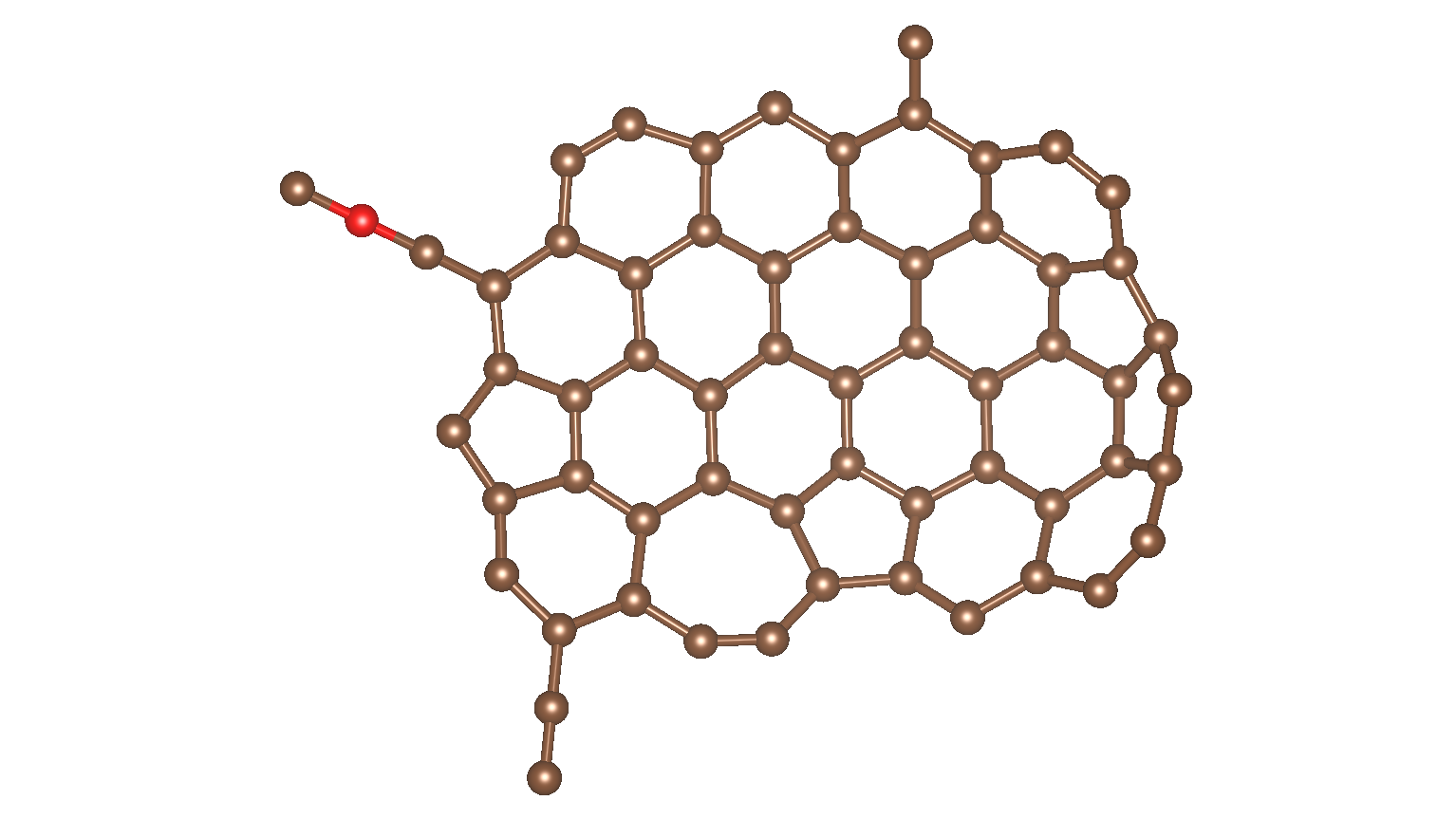}
         \caption{}
         \label{fig:conf22}
     \end{subfigure}
     \caption{Two environments which are used for studying the behavior of various fingerprints.
     The two atoms whose environment needs to be described are shown in red. Both structures are 
     meta-stable. }
     \label{fig:confandconf2}
\end{figure}
We now study the sensitivity matrix for the two configurations of 60 carbon atoms shown in Fig.~\ref{fig:confandconf2}. {\color{black} An analogous analysis will be presented in the supplementary information for two more structures. }  

In Fig.~\ref{fig:conf} the 
reference atom
forms three bonds with its three nearest neighbors and is surrounded by one pentagon and two hexagons, while in Fig.~\ref{fig:conf22} the 
atom of interest resides on a chain and has fewer neighbors compared to the 
atom in Fig.~\ref{fig:conf}. 
\begin{figure*}[tp!]
    \centering
     \begin{subfigure}[b]{0.48\textwidth}
         \centering
         \includegraphics[width=\textwidth]{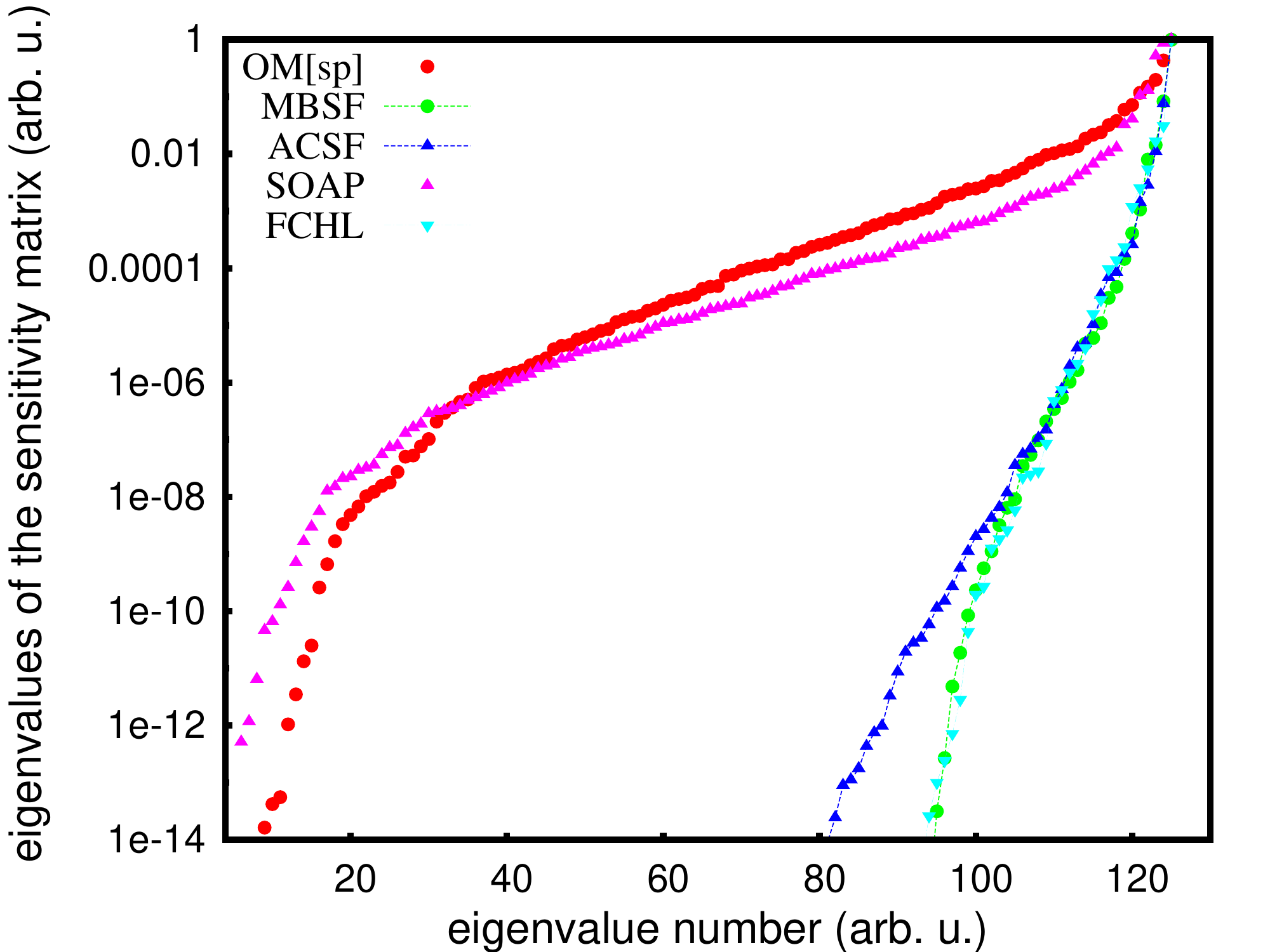}
         \caption{}
         \label{fig:confeig}
     \end{subfigure}
     \begin{subfigure}[b]{0.48\textwidth}
         \centering
         \includegraphics[width=\textwidth]{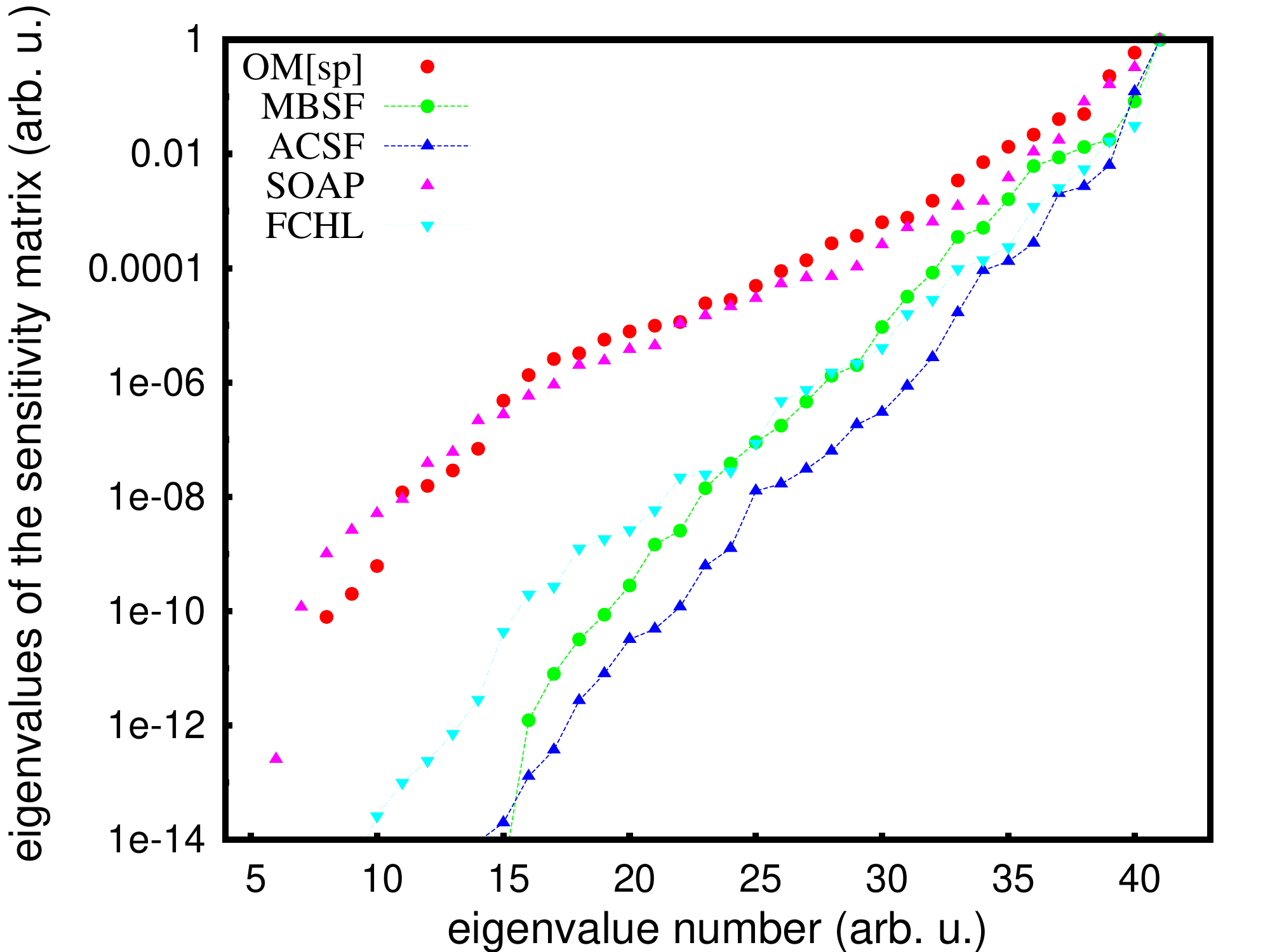}
         \caption{}
         \label{fig:conf2eig}
     \end{subfigure}
    \label{fig:eigs}
    \caption{The eigenvalues of the sensitivity matrix in \textbf{a}: for the 
    reference atom of \ref{fig:conf} and in \textbf{b}: for the 
    reference atom in \ref{fig:conf22}. The configuration in \ref{fig:conf} has 42 atoms in the sphere around the central atom giving rise to $3\times42-6=120$ non-zero eigenvalues whereas the configuration in \ref{fig:conf22} has 14 atoms in the sphere giving rise to $3\times14-6=36$ non-zero eigenvallues. All non-zero eigenvalues up to machine precision are shown.}
\end{figure*}

\begin{figure*}[p!]
     \centering
     \begin{subfigure}[b]{0.2\textwidth}
         \centering
         \includegraphics[width=\textwidth]{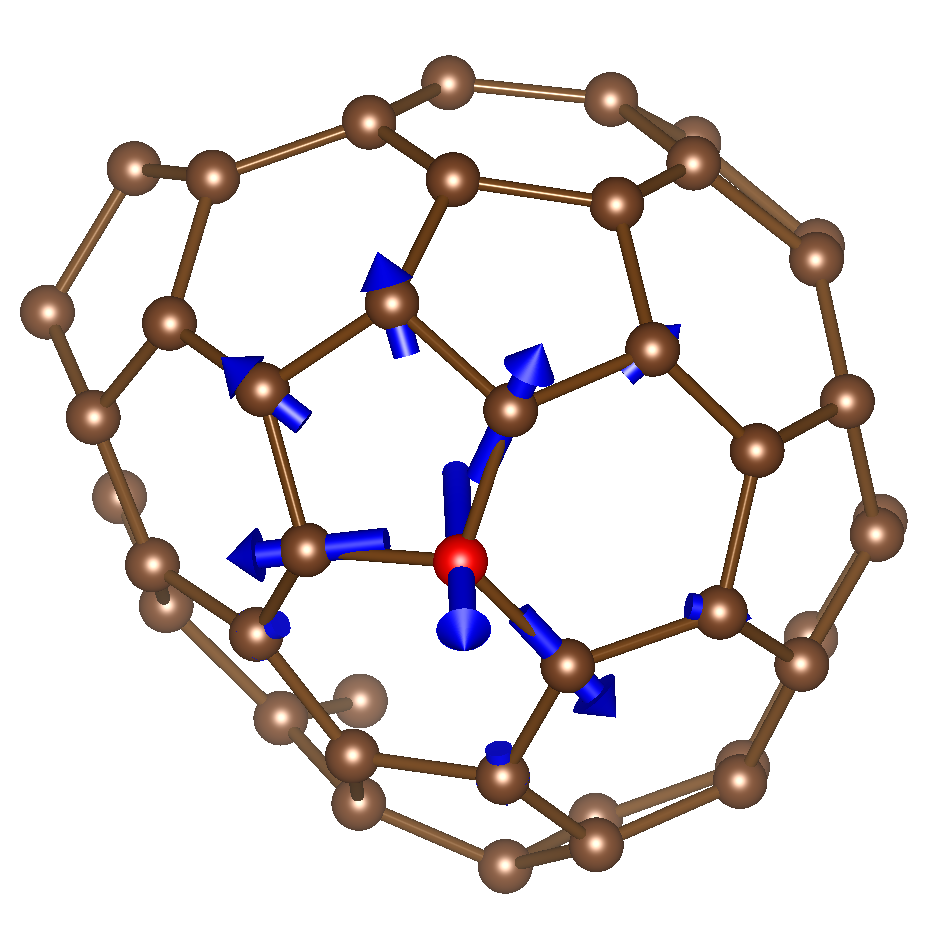}
         \caption{OM[sp], $\lambda=1.0$}
         \label{fig:eigenmodeconfom1}
     \end{subfigure}
     \begin{subfigure}[b]{0.2\textwidth}
         \centering
         \includegraphics[width=\textwidth]{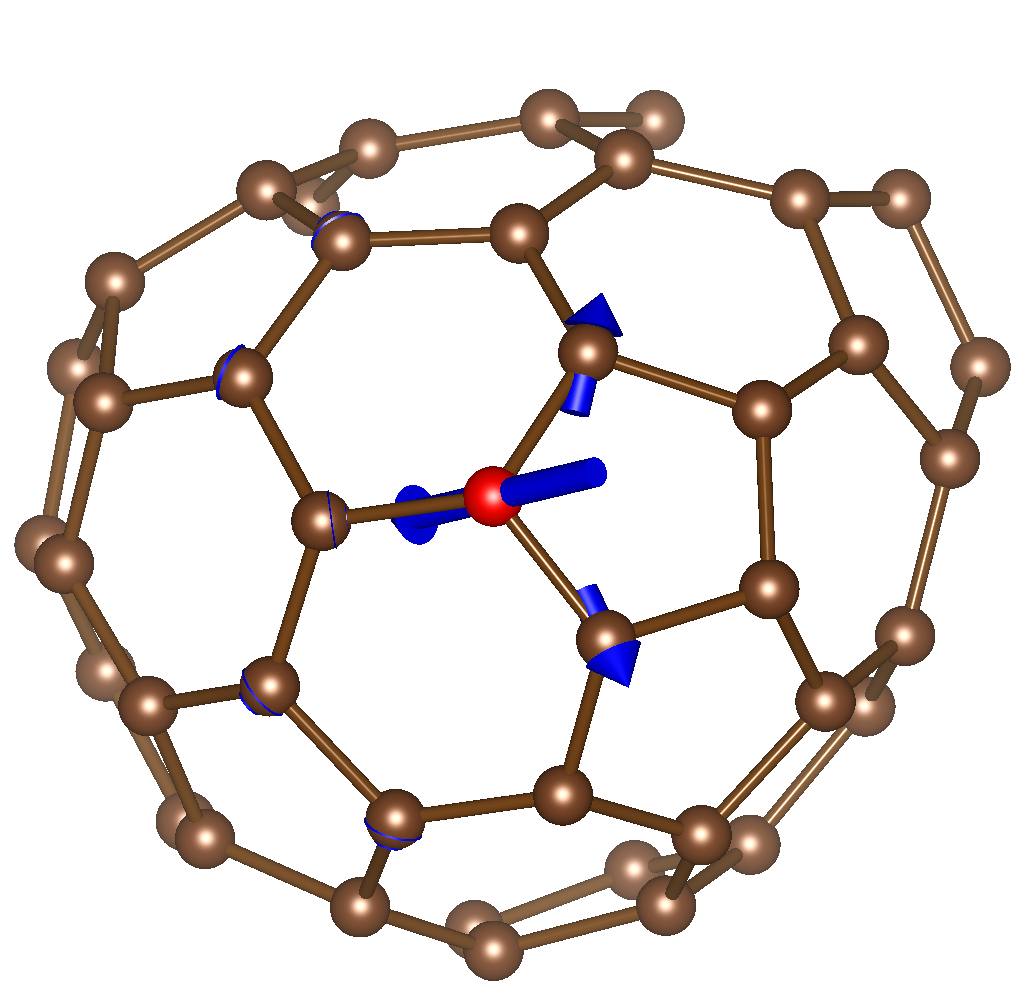}
         \caption{OM[sp], $\lambda=0.433$}
         \label{fig:eigenmodeconfom2}
     \end{subfigure}
     \begin{subfigure}[b]{0.2\textwidth}
         \centering
         \includegraphics[width=\textwidth]{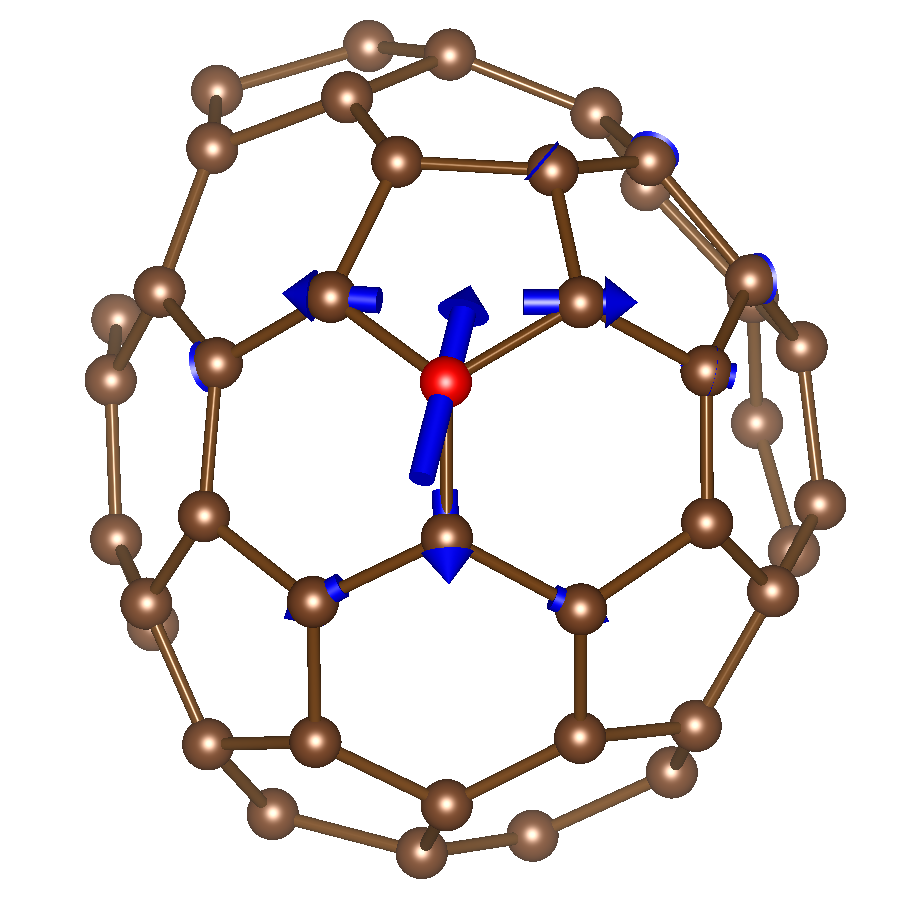}
         \caption{OM[sp], $\lambda=0.199$}
         \label{fig:eigenmodeconfom3}
     \end{subfigure}
     \begin{subfigure}[b]{0.2\textwidth}
         \centering
         \includegraphics[width=\textwidth]{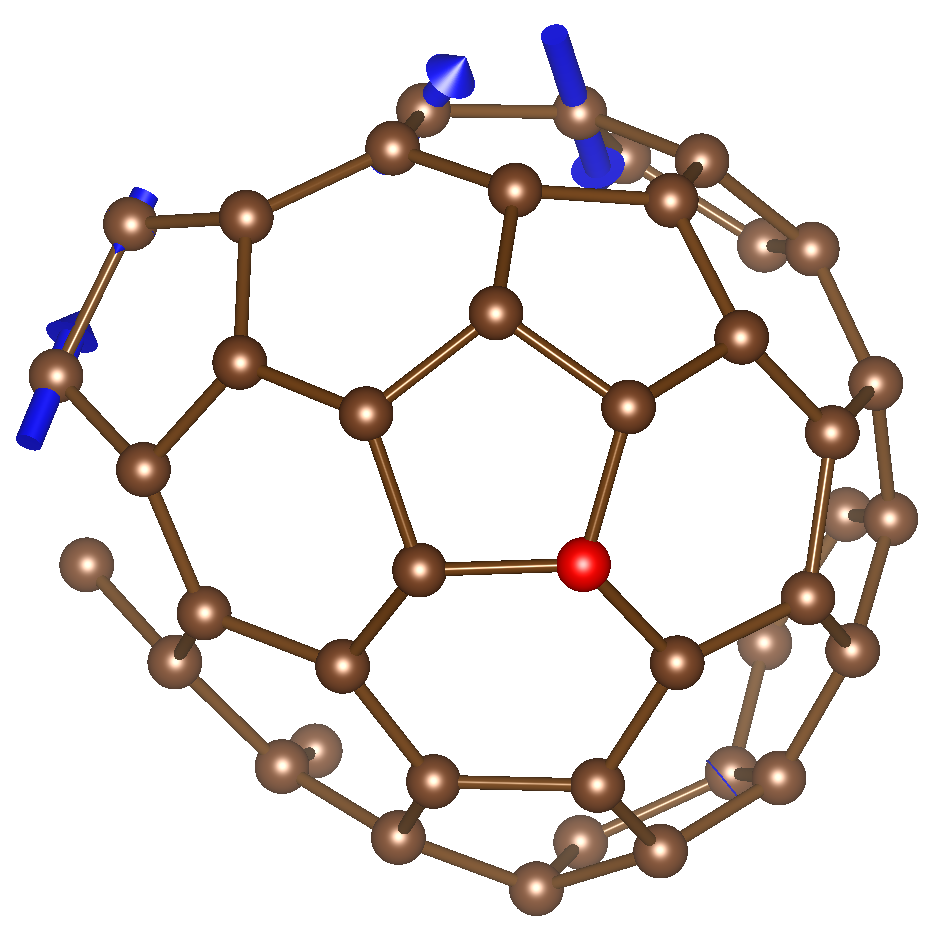}
         \caption{OM[sp], $\lambda \sim 1\times 10^{-6}$}
         \label{fig:eigenmodeconfom4}
     \end{subfigure}
     
     \begin{subfigure}[b]{0.2\textwidth}
         \centering
         \includegraphics[width=\textwidth]{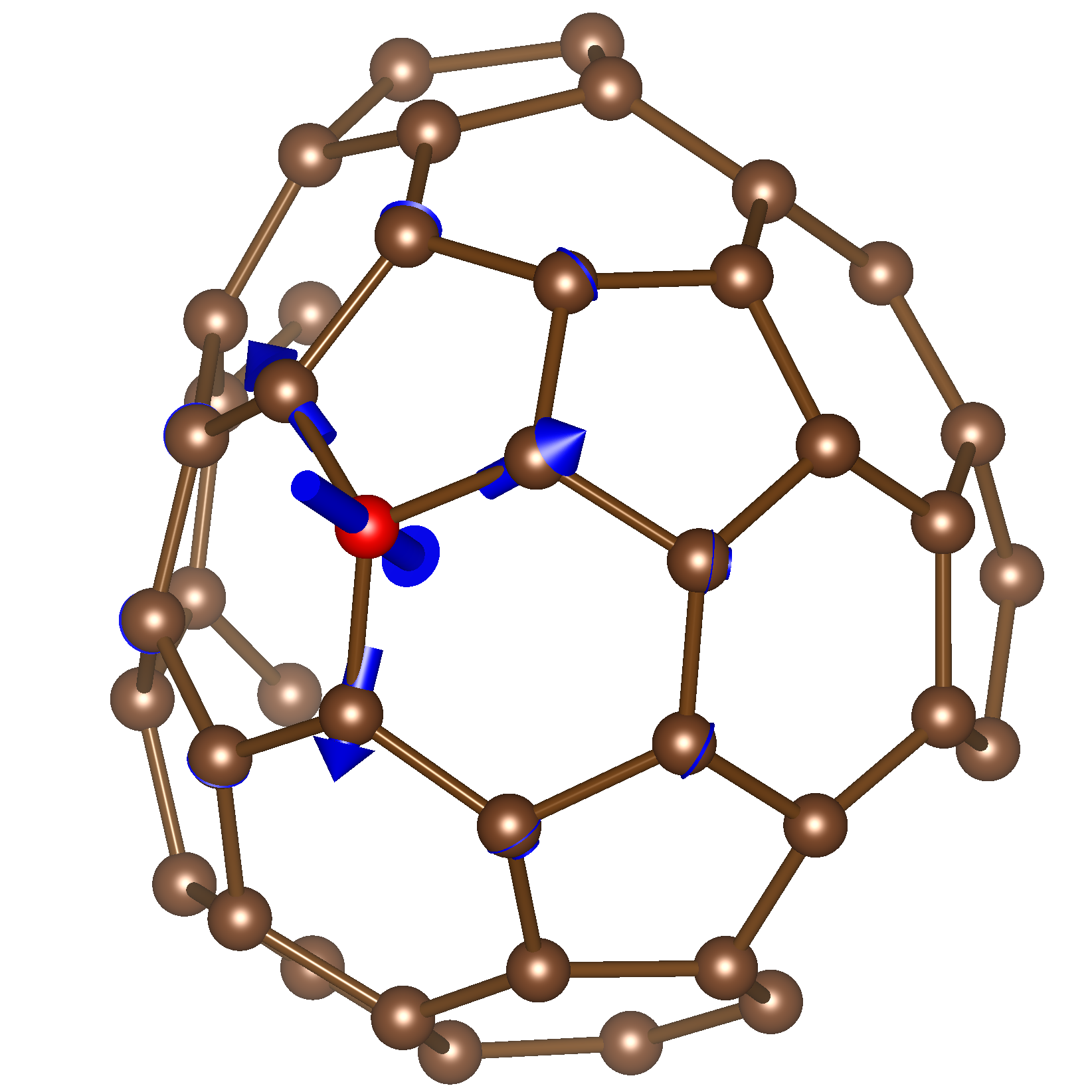}
         \caption{SOAP, $\lambda=1.0$}
         \label{fig:}
     \end{subfigure}
     \begin{subfigure}[b]{0.2\textwidth}
         \centering
         \includegraphics[width=\textwidth]{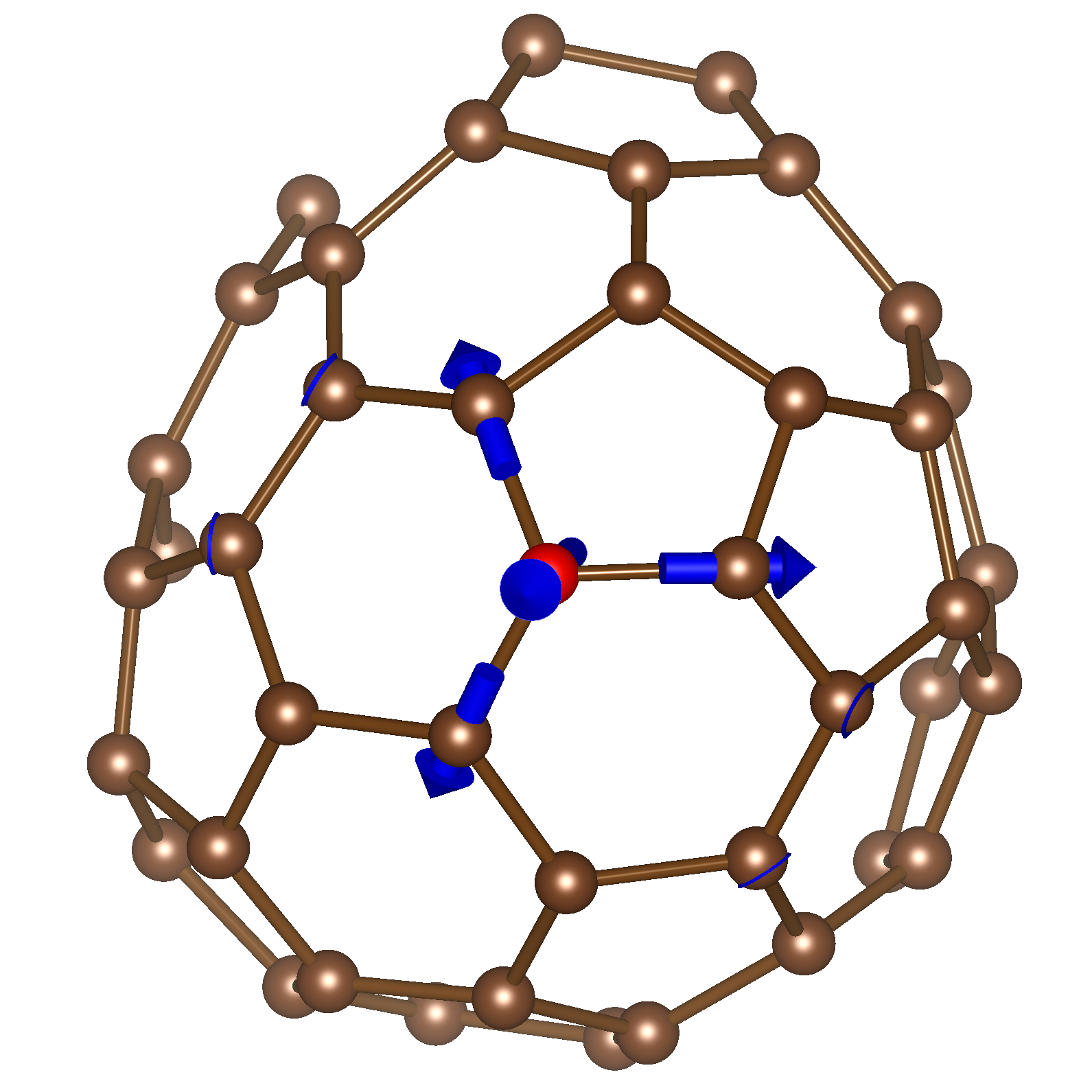}
         \caption{SOAP, $\lambda=0.867$}
         \label{fig:}
     \end{subfigure}
     \begin{subfigure}[b]{0.2\textwidth}
         \centering
         \includegraphics[width=\textwidth]{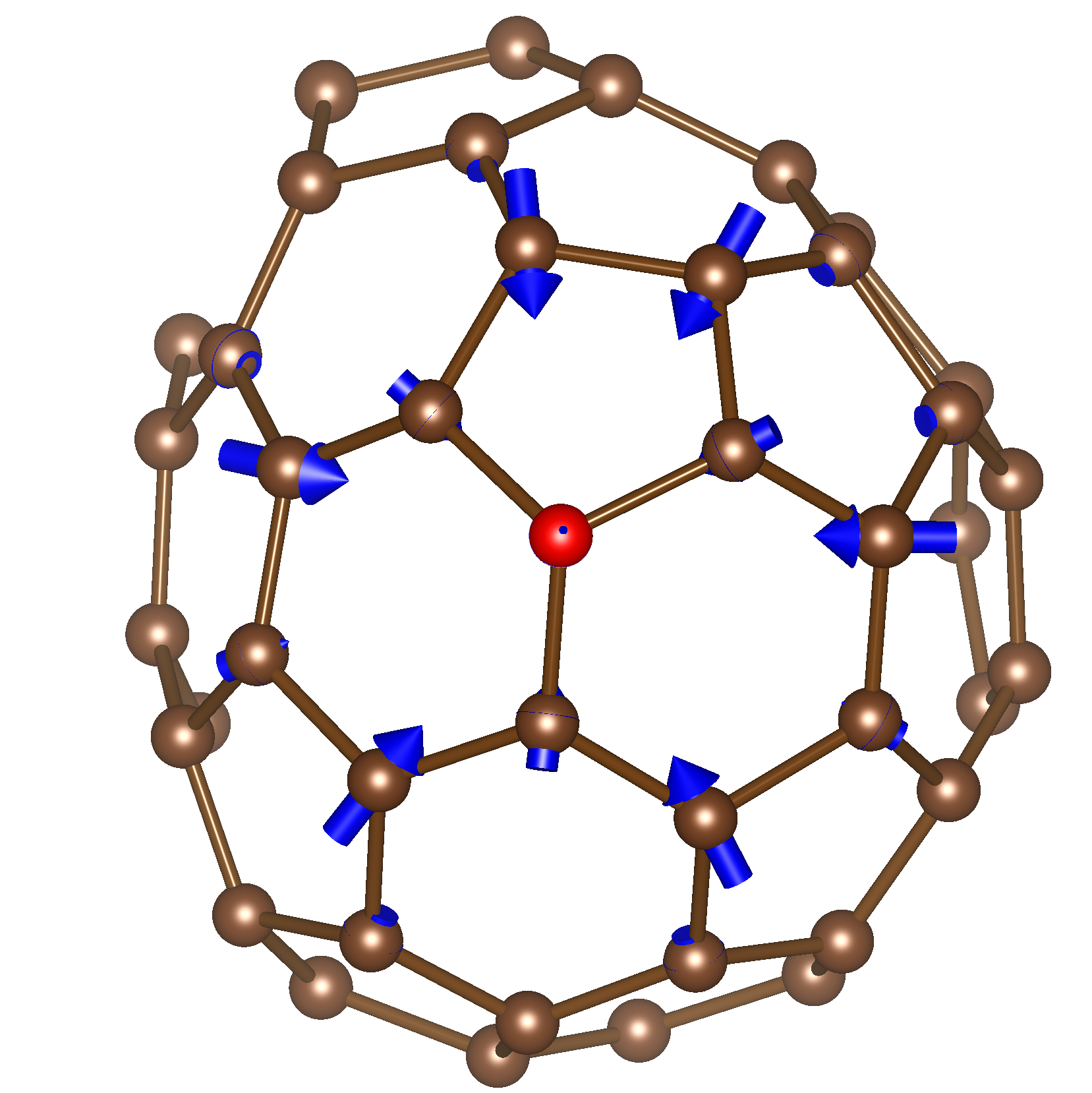}
         \caption{SOAP, $\lambda=0.520$}
         \label{fig:}
     \end{subfigure}
     \begin{subfigure}[b]{0.2\textwidth}
         \centering
         \includegraphics[width=\textwidth]{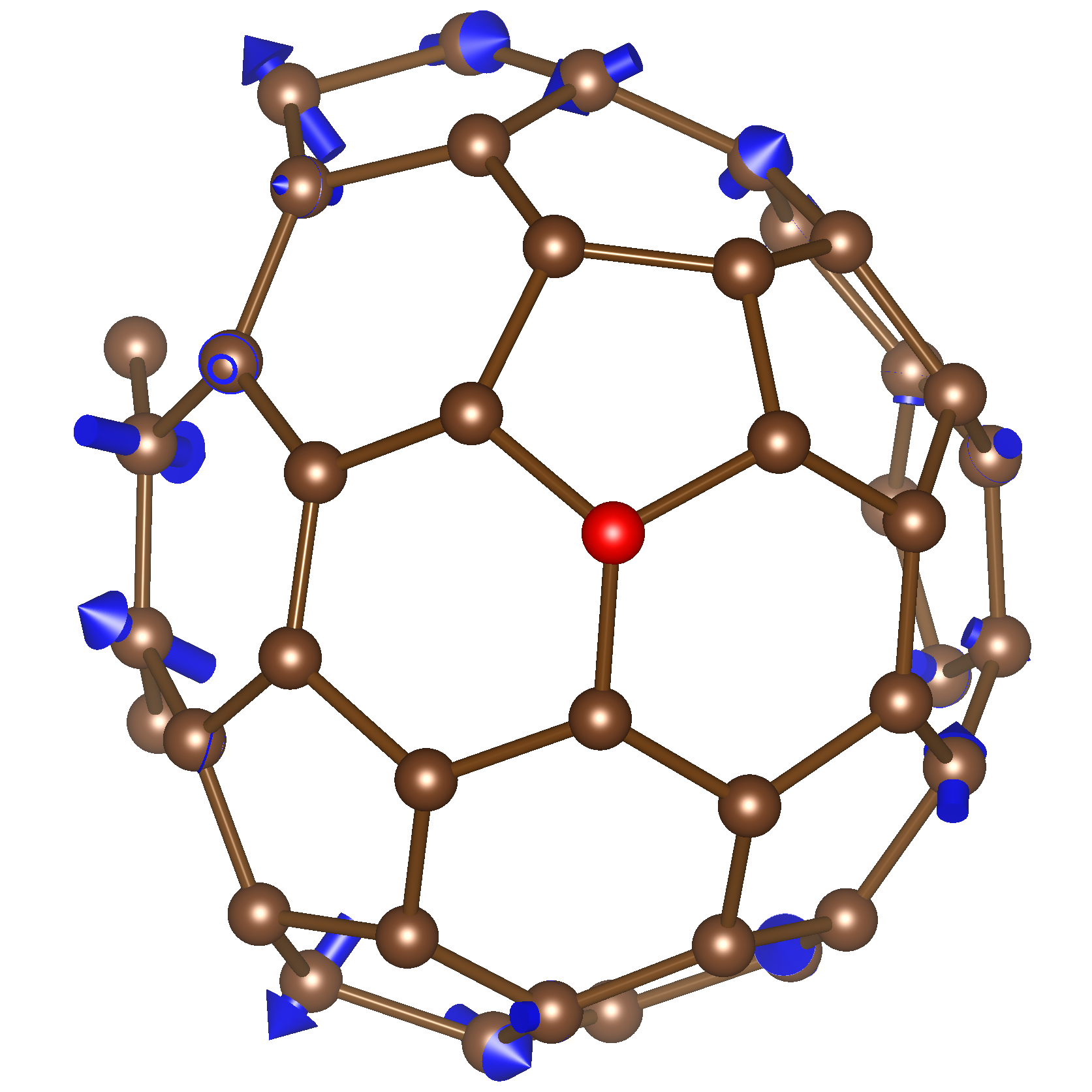}
         \caption{SOAP, $\lambda \sim 1 \times 10^{-6}$}
         \label{fig:}
     \end{subfigure}
     
     \begin{subfigure}[b]{0.2\textwidth}
         \centering
         \includegraphics[width=\textwidth]{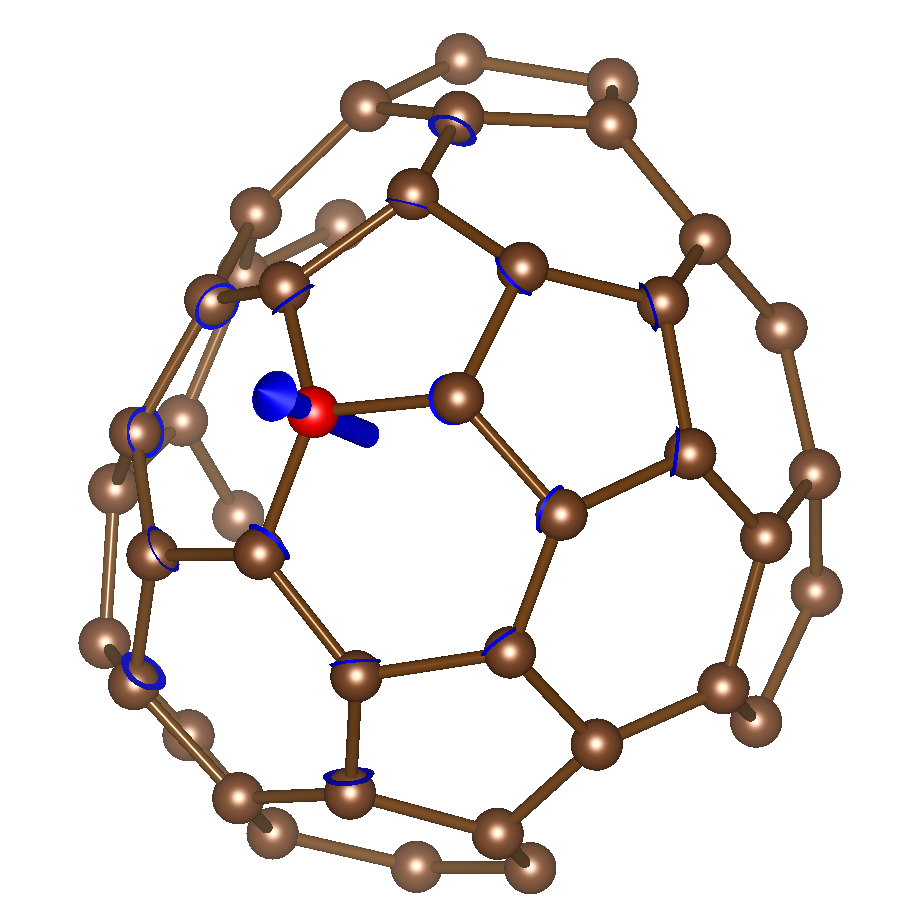}
         \caption{ACSF, $\lambda=1.0$}
         \label{fig:}
     \end{subfigure}
     \begin{subfigure}[b]{0.2\textwidth}
         \centering
         \includegraphics[width=\textwidth]{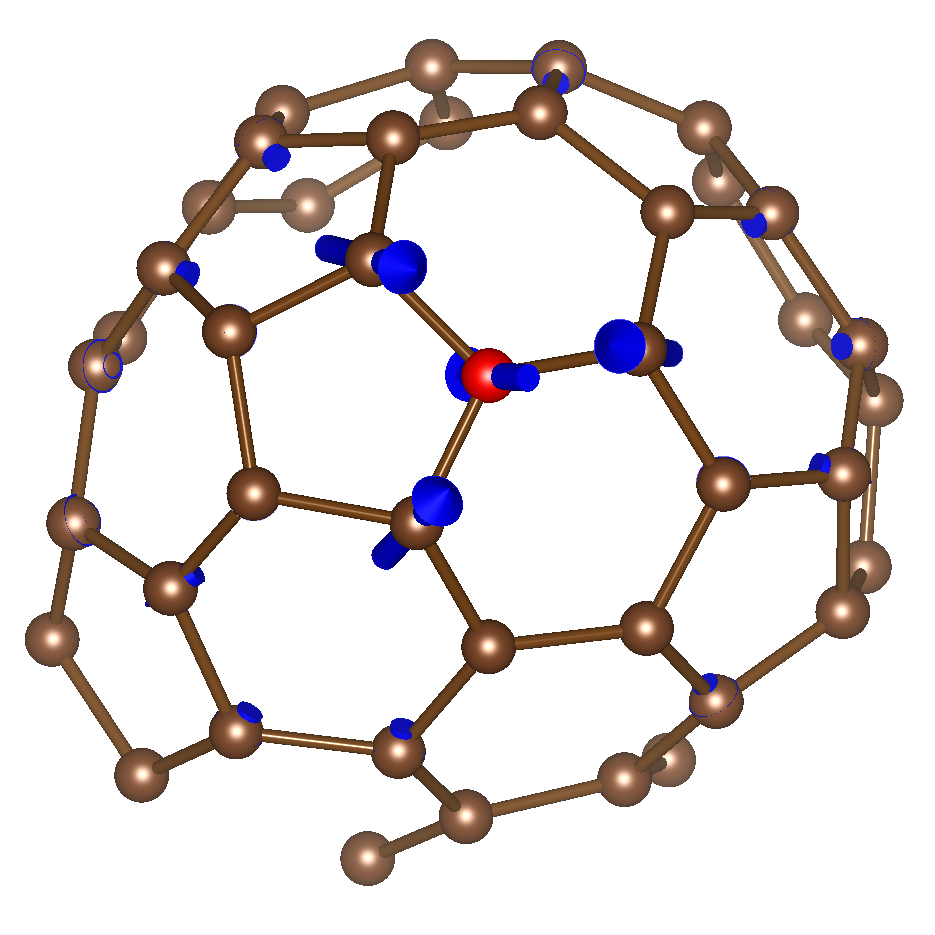}
         \caption{ACSF, $\lambda=0.076$}
         \label{fig:}
     \end{subfigure}
     \begin{subfigure}[b]{0.2\textwidth}
         \centering
         \includegraphics[width=\textwidth]{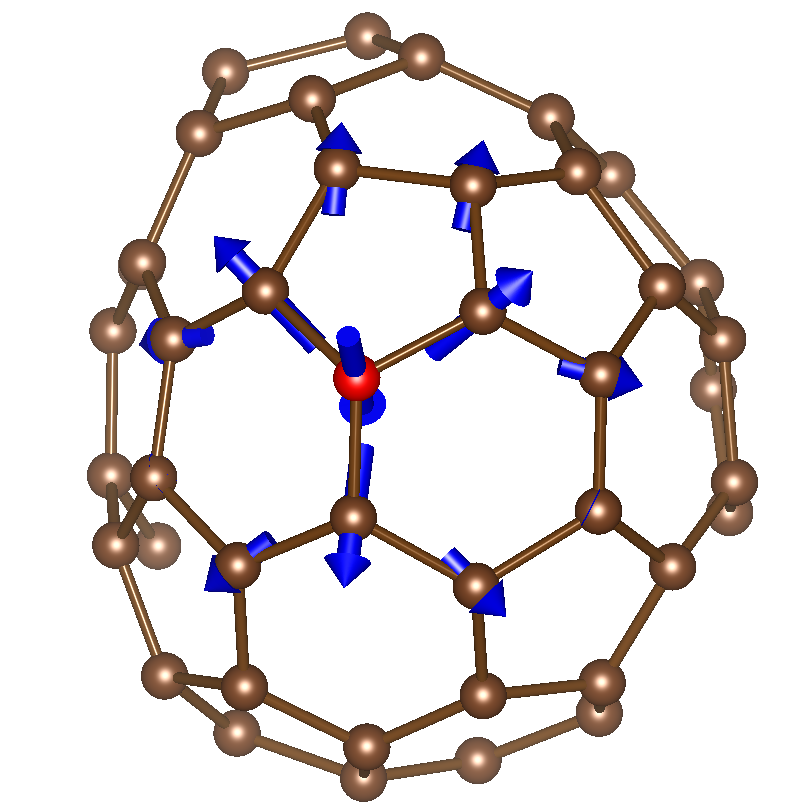}
         \caption{ACSF, $\lambda=0.011$}
         \label{fig:}
     \end{subfigure}
     \begin{subfigure}[b]{0.2\textwidth}
         \centering
         \includegraphics[width=\textwidth]{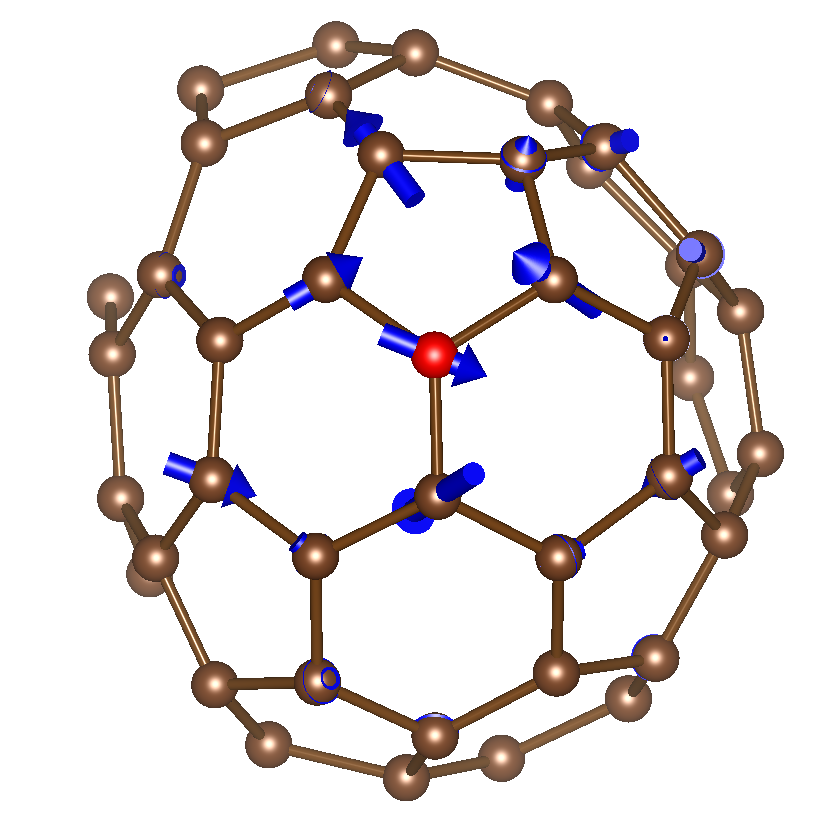}
         \caption{ACSF, $\lambda \sim 1 \times 10^{-6}$}
         \label{fig:}
     \end{subfigure}
     
     \begin{subfigure}[b]{0.2\textwidth}
         \centering
         \includegraphics[width=\textwidth]{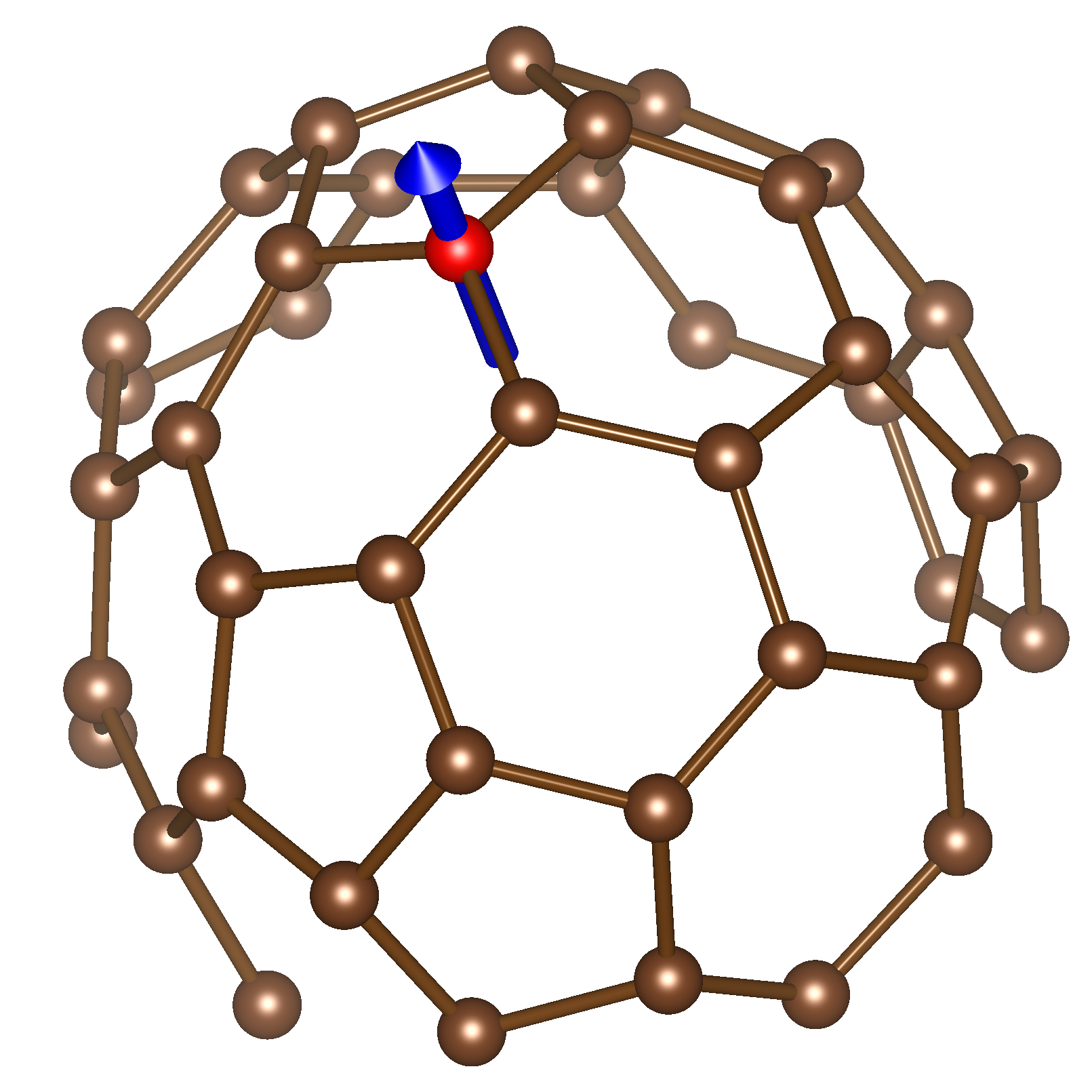}
         \caption{MBSF, $\lambda=1.0$ }
         \label{fig:}
     \end{subfigure}
     \begin{subfigure}[b]{0.2\textwidth}
         \centering
         \includegraphics[width=\textwidth]{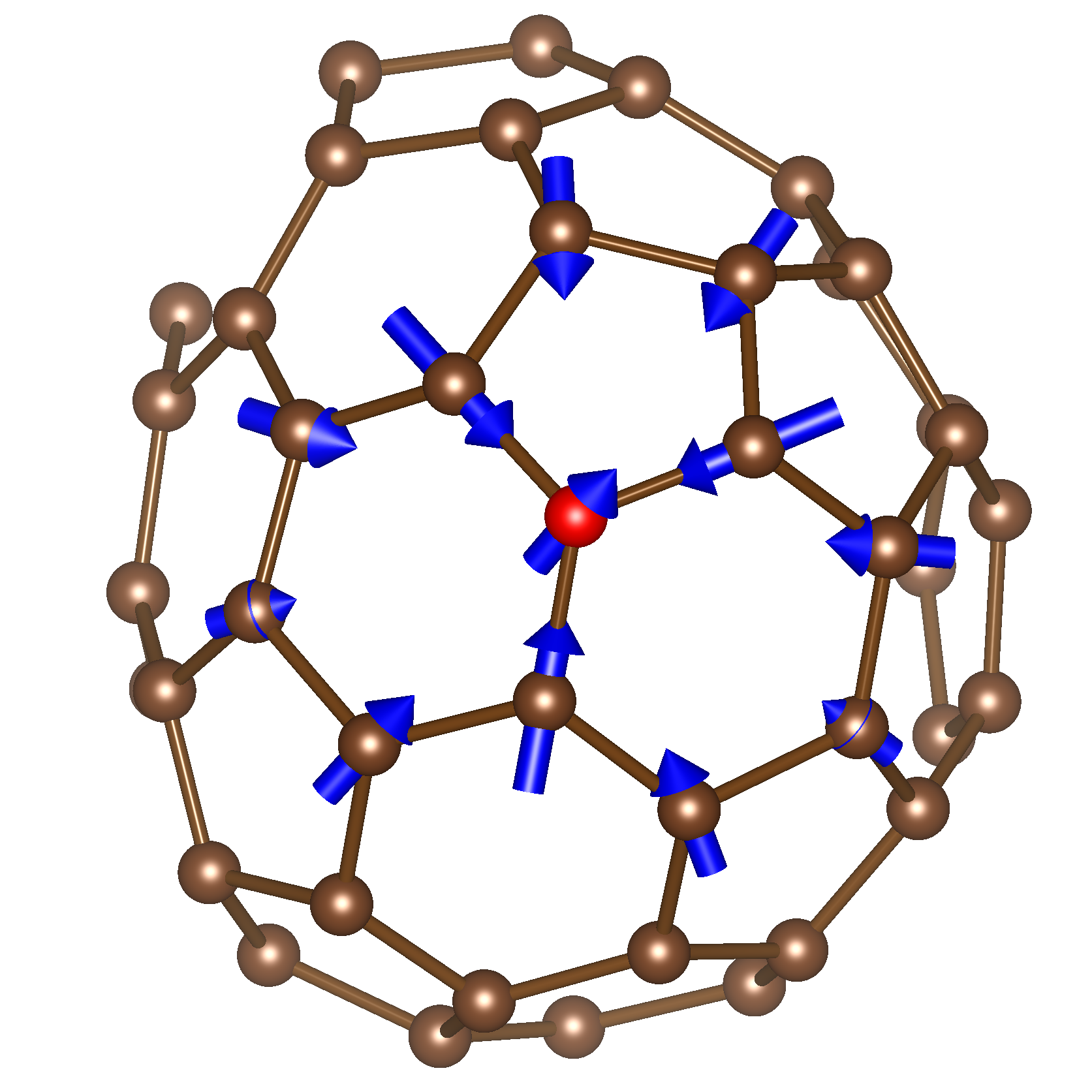}
         \caption{MBSF, $\lambda=0.084$ }
         \label{fig:}
     \end{subfigure}
     \begin{subfigure}[b]{0.2\textwidth}
         \centering
         \includegraphics[width=\textwidth]{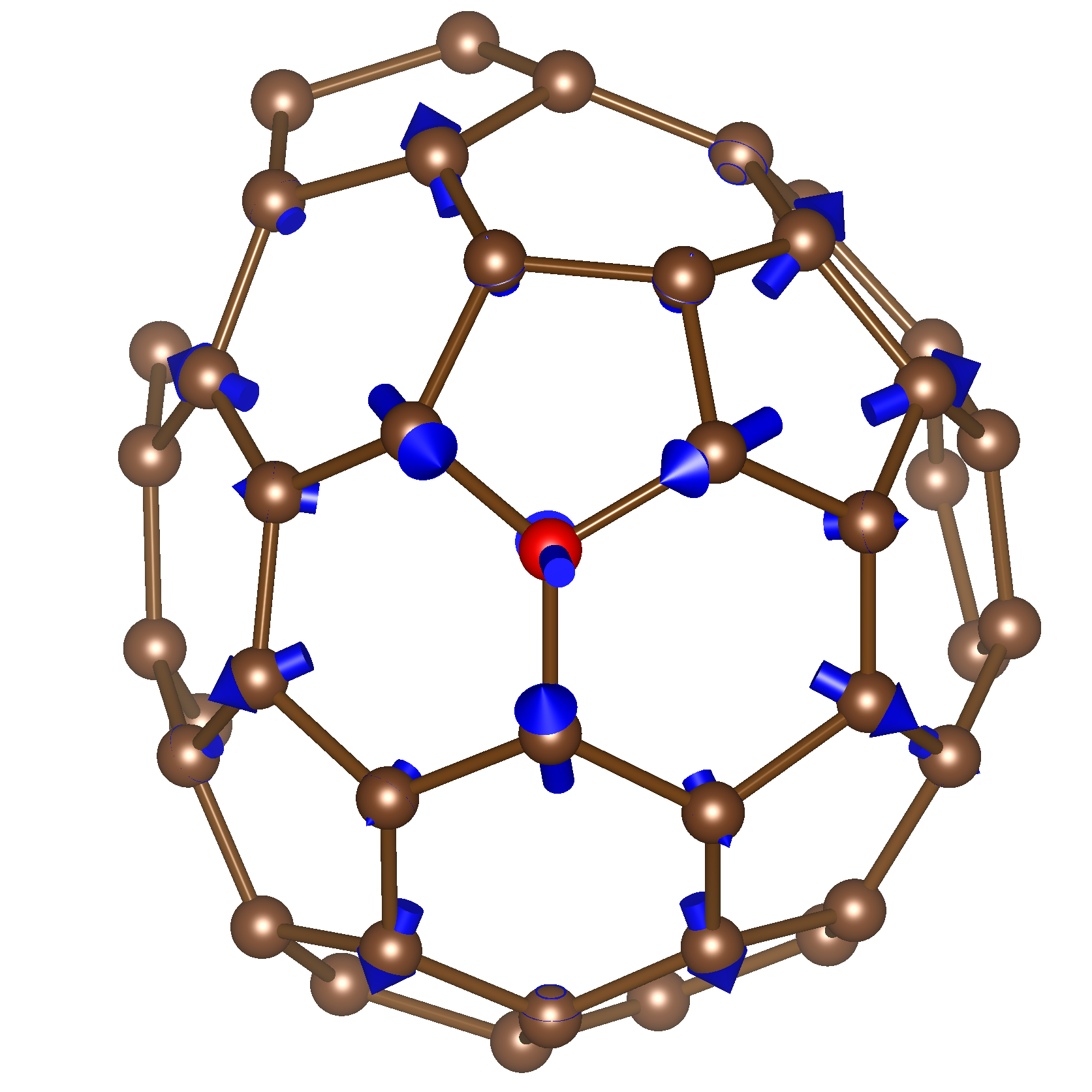}
         \caption{MBSF, $\lambda=0.014$}
         \label{fig:}
     \end{subfigure}
     \begin{subfigure}[b]{0.2\textwidth}
         \centering
         \includegraphics[width=\textwidth]{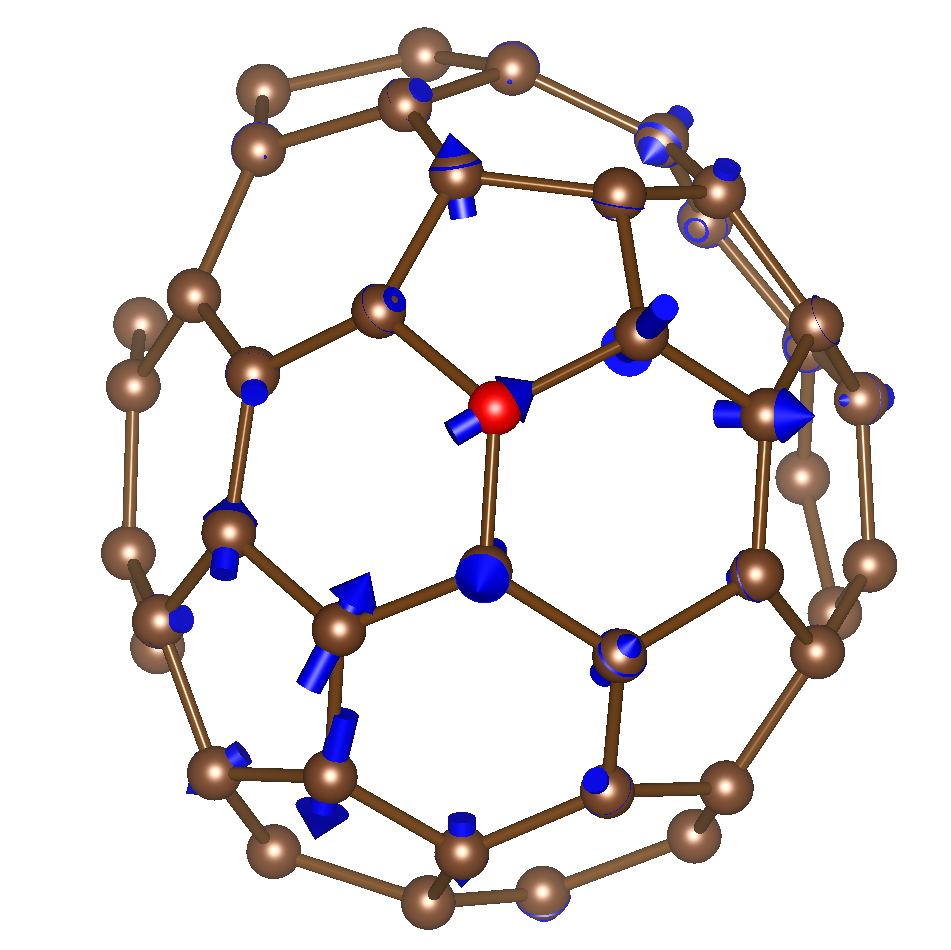}
         \caption{MBSF, $\lambda \sim 1 \times10^{-6}$}
         \label{fig:}
     \end{subfigure}
     \begin{subfigure}[b]{0.2\textwidth}
         \centering
         \includegraphics[width=\textwidth]{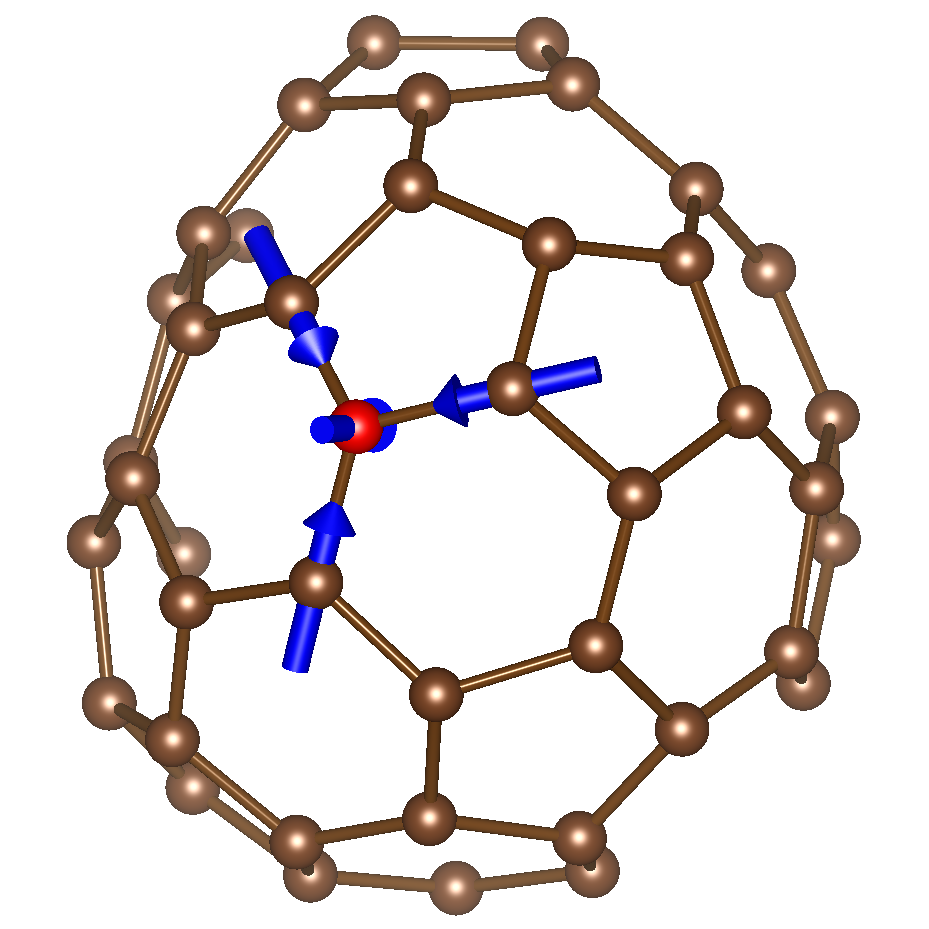}
         \caption{FCHL, $\lambda=1.0$ }
         \label{fig:}
     \end{subfigure}
     \begin{subfigure}[b]{0.2\textwidth}
         \centering
         \includegraphics[width=\textwidth]{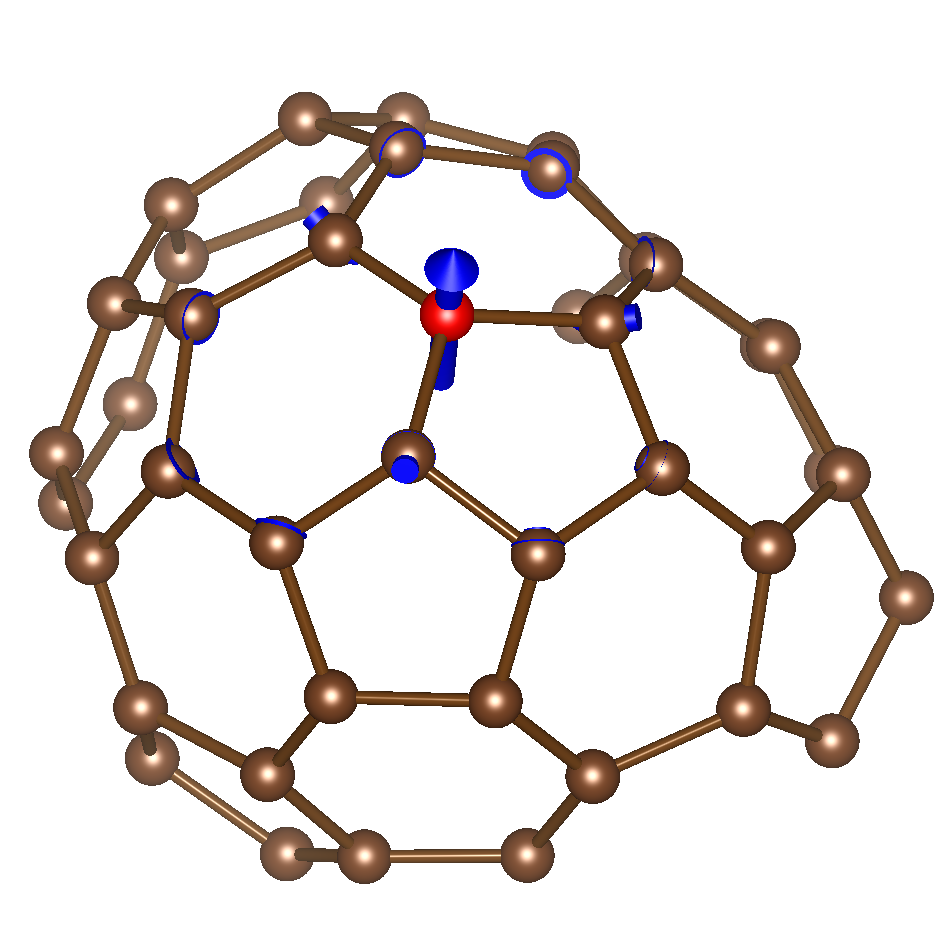}
         \caption{FCHL, $\lambda=0.422$ }
         \label{fig:}
     \end{subfigure}
     \begin{subfigure}[b]{0.2\textwidth}
         \centering
         \includegraphics[width=\textwidth]{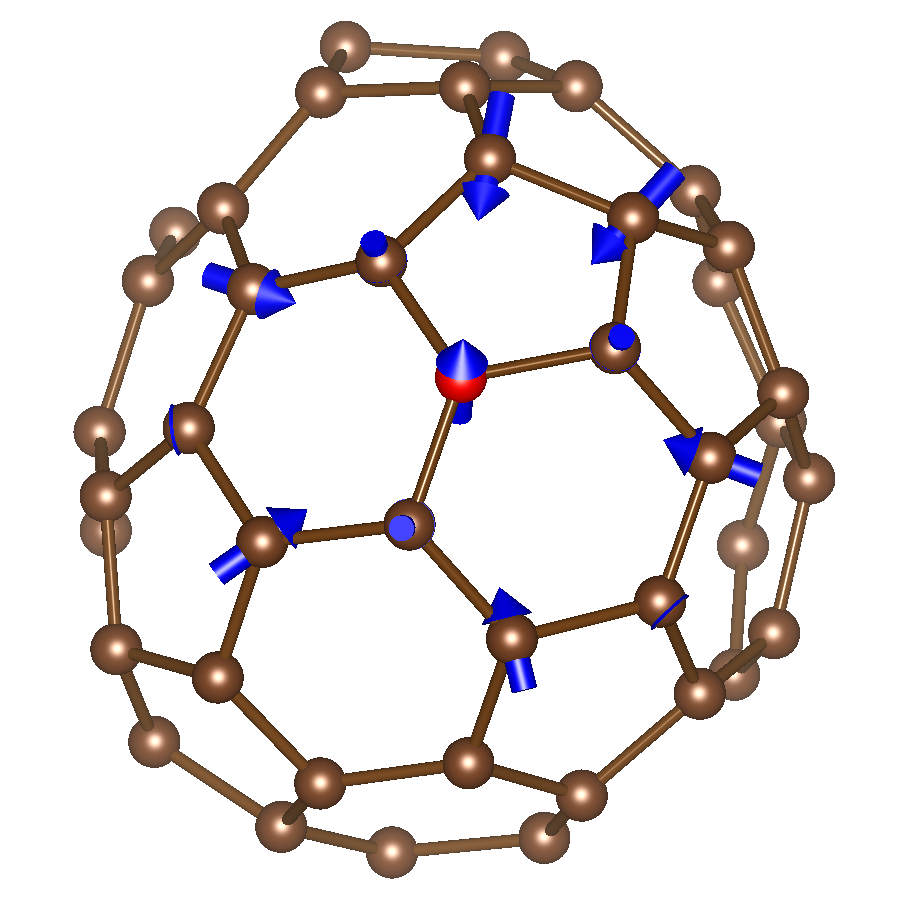}
         \caption{FCHL, $\lambda=0.084$}
         \label{fig:}
     \end{subfigure}
     \begin{subfigure}[b]{0.2\textwidth}
         \centering
         \includegraphics[width=\textwidth]{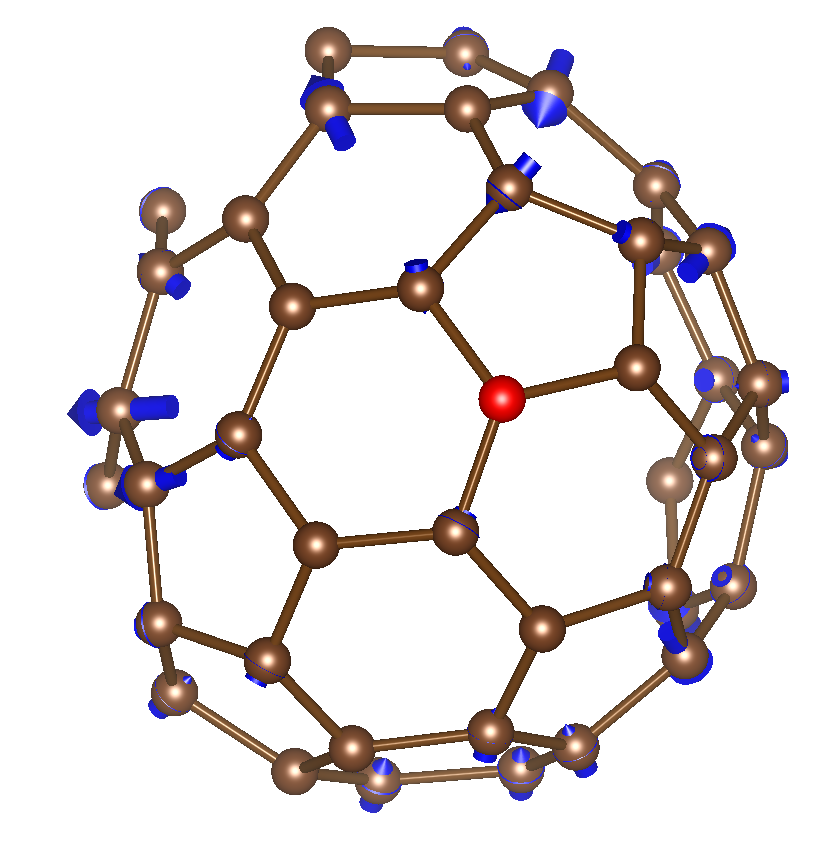}
         \caption{FCHL, $\lambda \sim 1 \times10^{-6}$}
         \label{fig:}
     \end{subfigure}
     
\caption{The eigenvectors belonging to the three largest eigenvalues and one representative small eigenvalue of the sensitivity matrix for the atomic environment in~\ref{fig:conf}. The red atom is again the 
reference atom. The displacement modes given by eigenvectors  are represented by arrows. {\color{black} Only atomic eigenvector components whose length is larger than 0.1 are shown. For this reason 
it is not always visible that all the components exactly sum up to zero.}}
     \label{fig:eigenmodeconf}
\end{figure*}

\begin{figure*}[p!]
     \centering
     \begin{subfigure}[b]{0.2\textwidth}
         \centering
         \includegraphics[width=\textwidth]{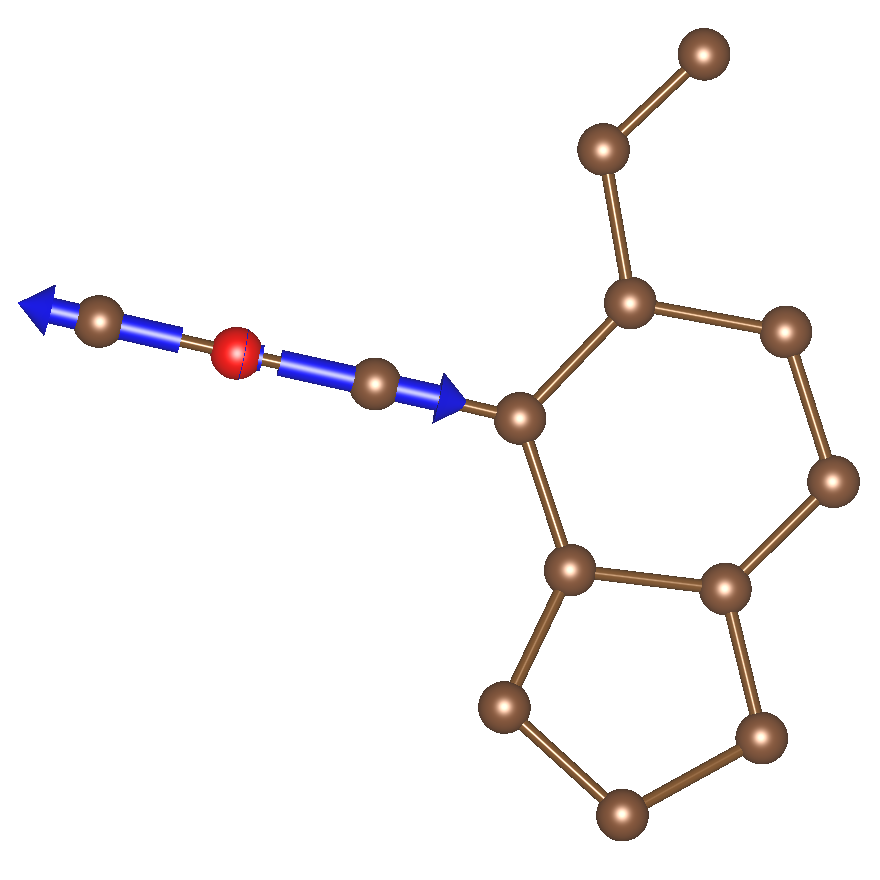}
         \caption{OM[sp], $\lambda=1.0$}
         \label{fig:init1om1}
     \end{subfigure}
      \begin{subfigure}[b]{0.2\textwidth}
         \centering
         \includegraphics[width=\textwidth]{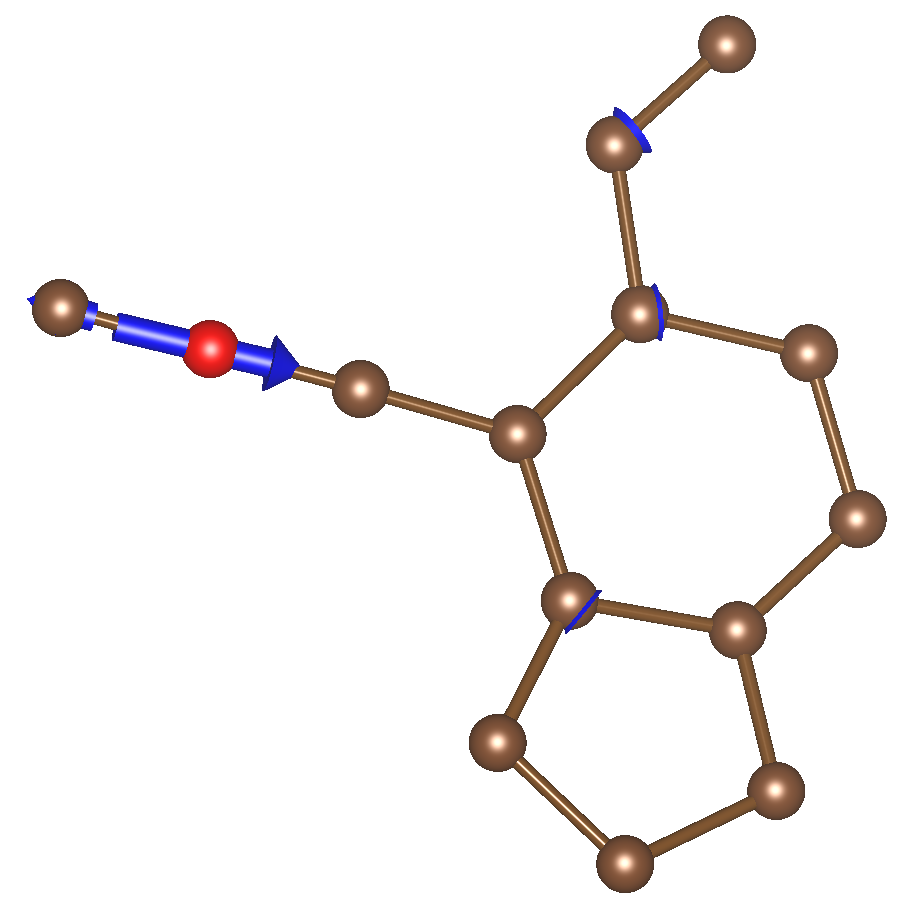}
         \caption{OM[sp], $\lambda=0.594$}
         \label{fig:init1om2}
     \end{subfigure}
     \begin{subfigure}[b]{0.2\textwidth}
         \centering
         \includegraphics[width=\textwidth]{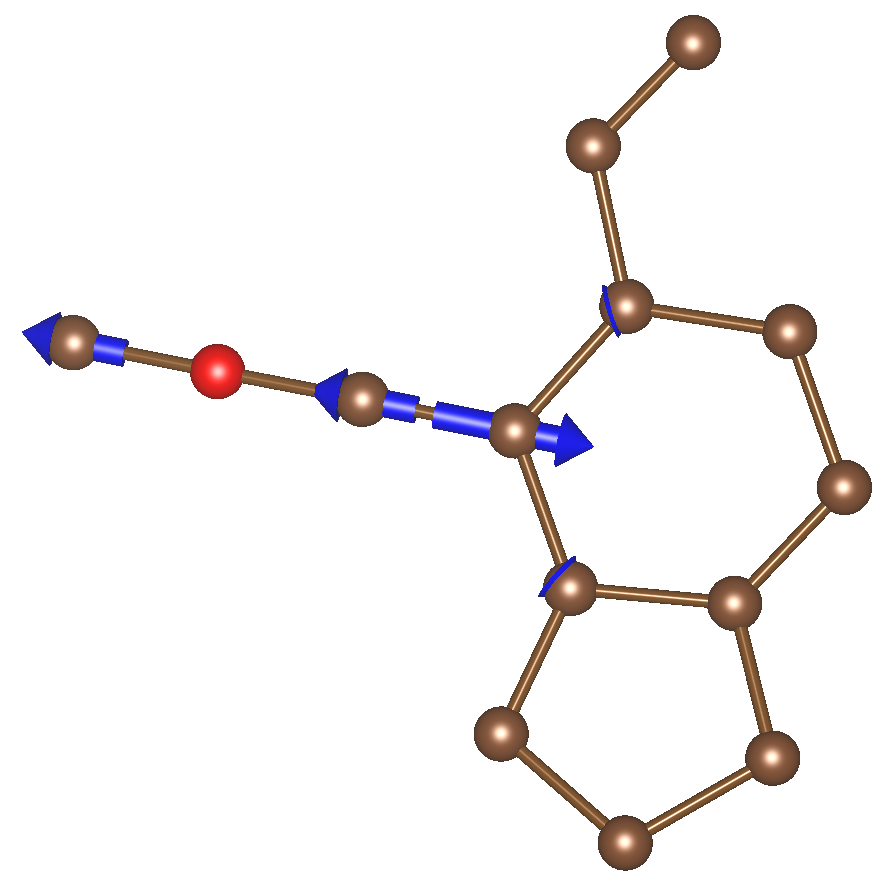}
         \caption{OM[sp], $\lambda=0.231$}
         \label{fig:}
     \end{subfigure}
     \begin{subfigure}[b]{0.2\textwidth}
         \centering
         \includegraphics[width=\textwidth]{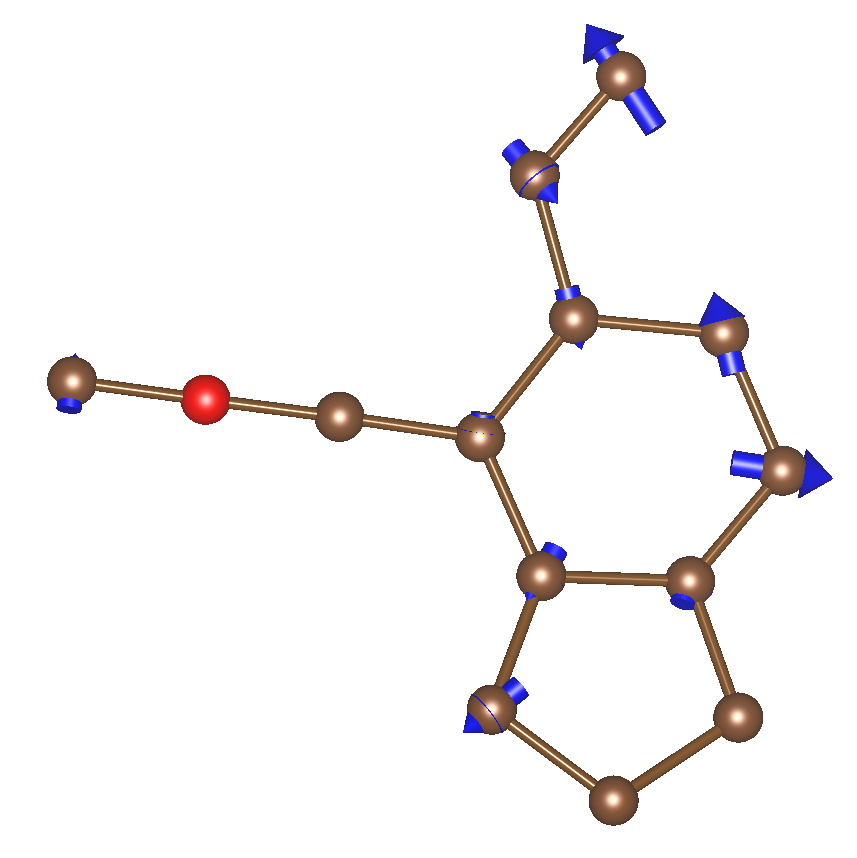}
         \caption{OM[sp], $\lambda \sim 2\times 10^{-6}$}
         \label{fig:}
     \end{subfigure}
     
     \begin{subfigure}[b]{0.2\textwidth}
         \centering
         \includegraphics[width=\textwidth]{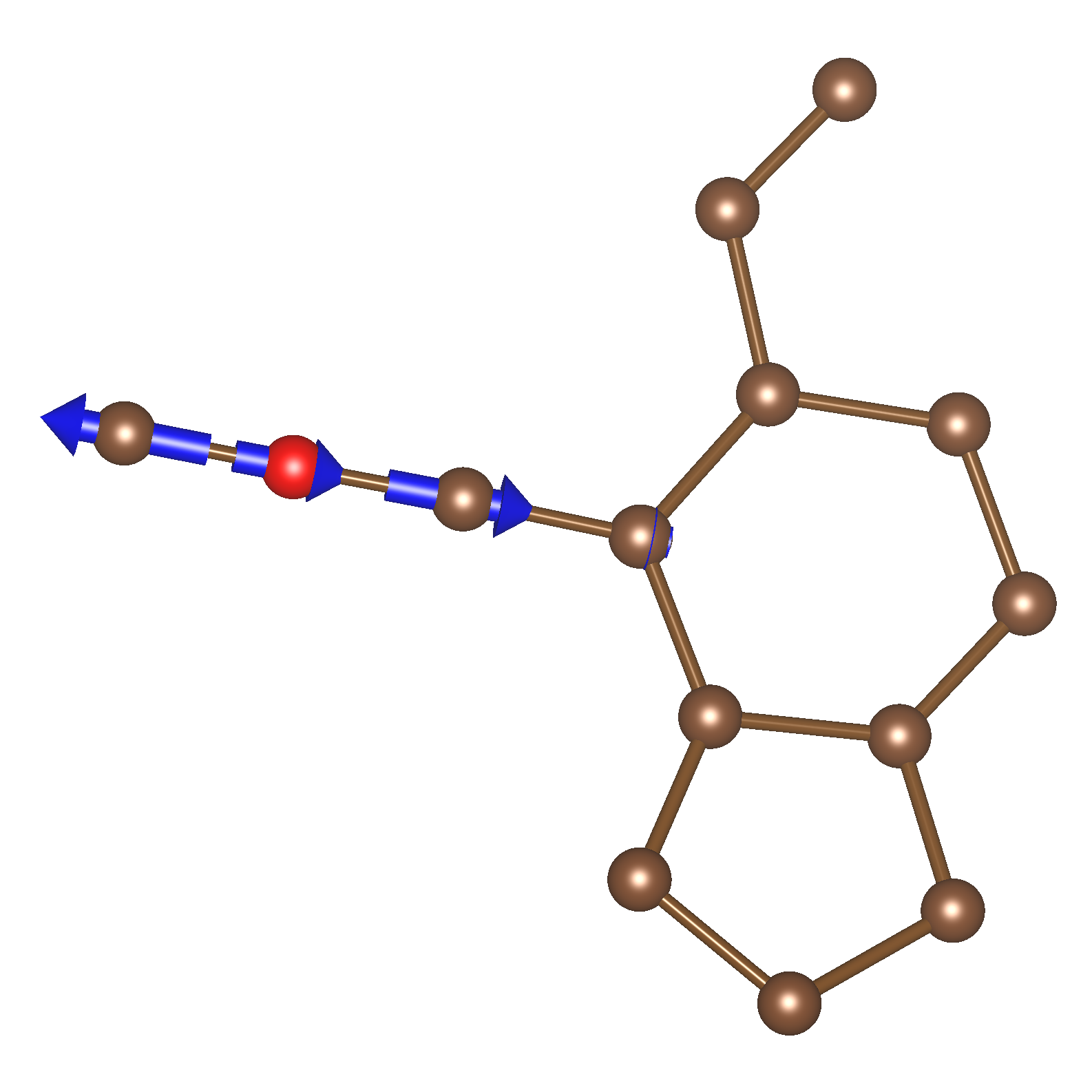}
         \caption{SOAP, $\lambda=1.0$}
         \label{fig:}
     \end{subfigure}
     \begin{subfigure}[b]{0.2\textwidth}
         \centering
         \includegraphics[width=\textwidth]{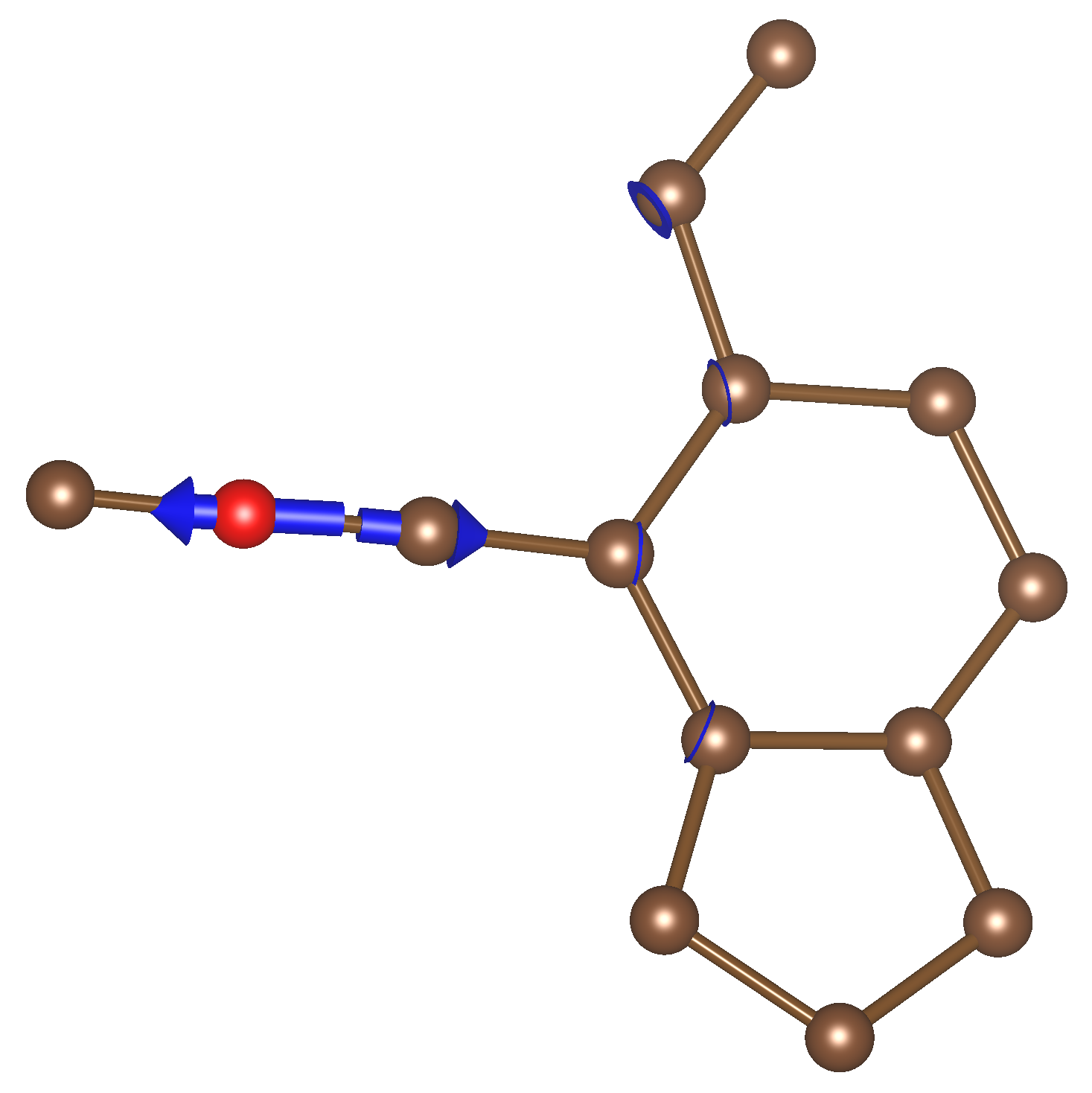}
         \caption{SOAP, $\lambda=0.326$}
         \label{fig:}
     \end{subfigure}
     \begin{subfigure}[b]{0.2\textwidth}
         \centering
         \includegraphics[width=\textwidth]{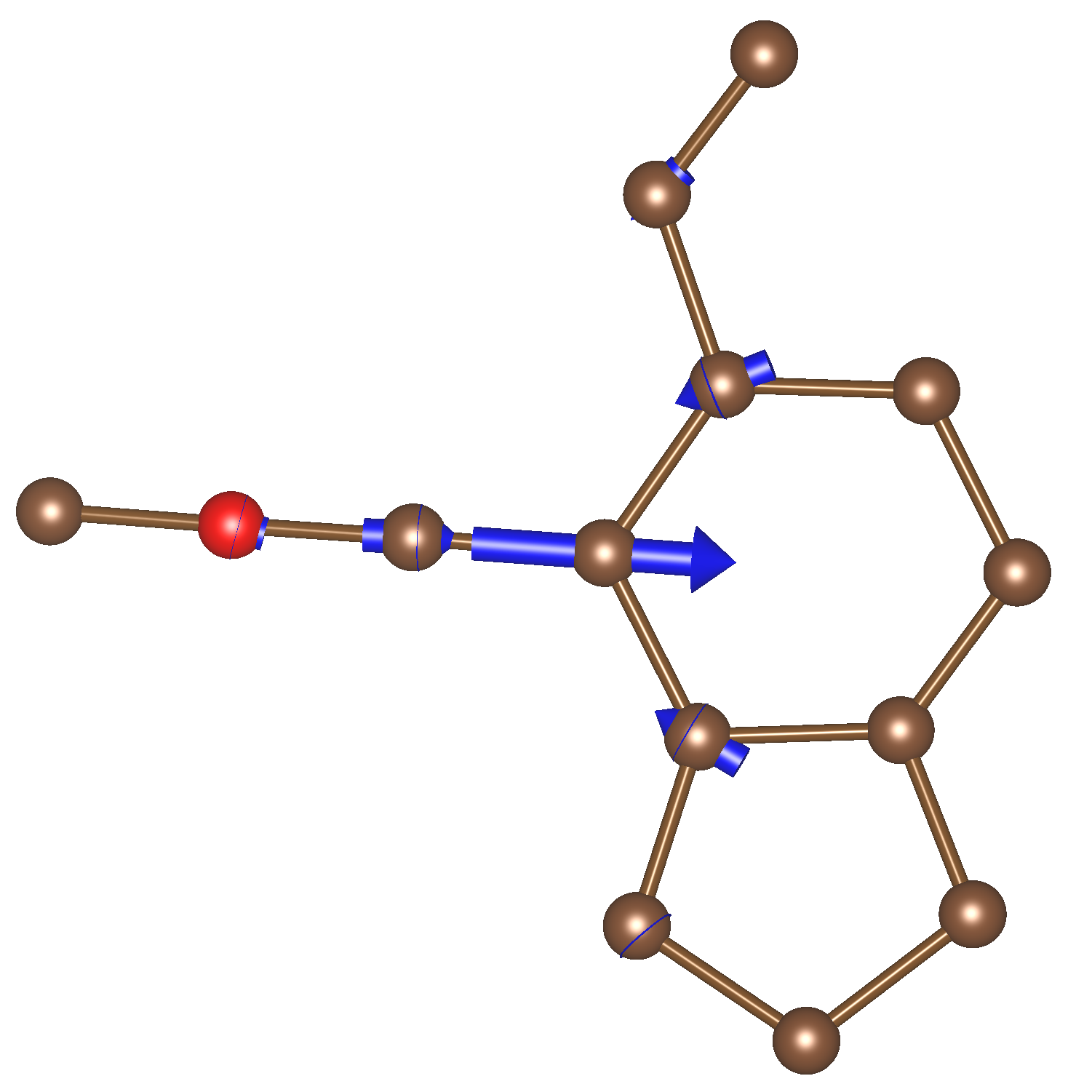}
         \caption{SOAP, $\lambda=0.164$}
         \label{fig:}
     \end{subfigure}
     \begin{subfigure}[b]{0.2\textwidth}
         \centering
         \includegraphics[width=\textwidth]{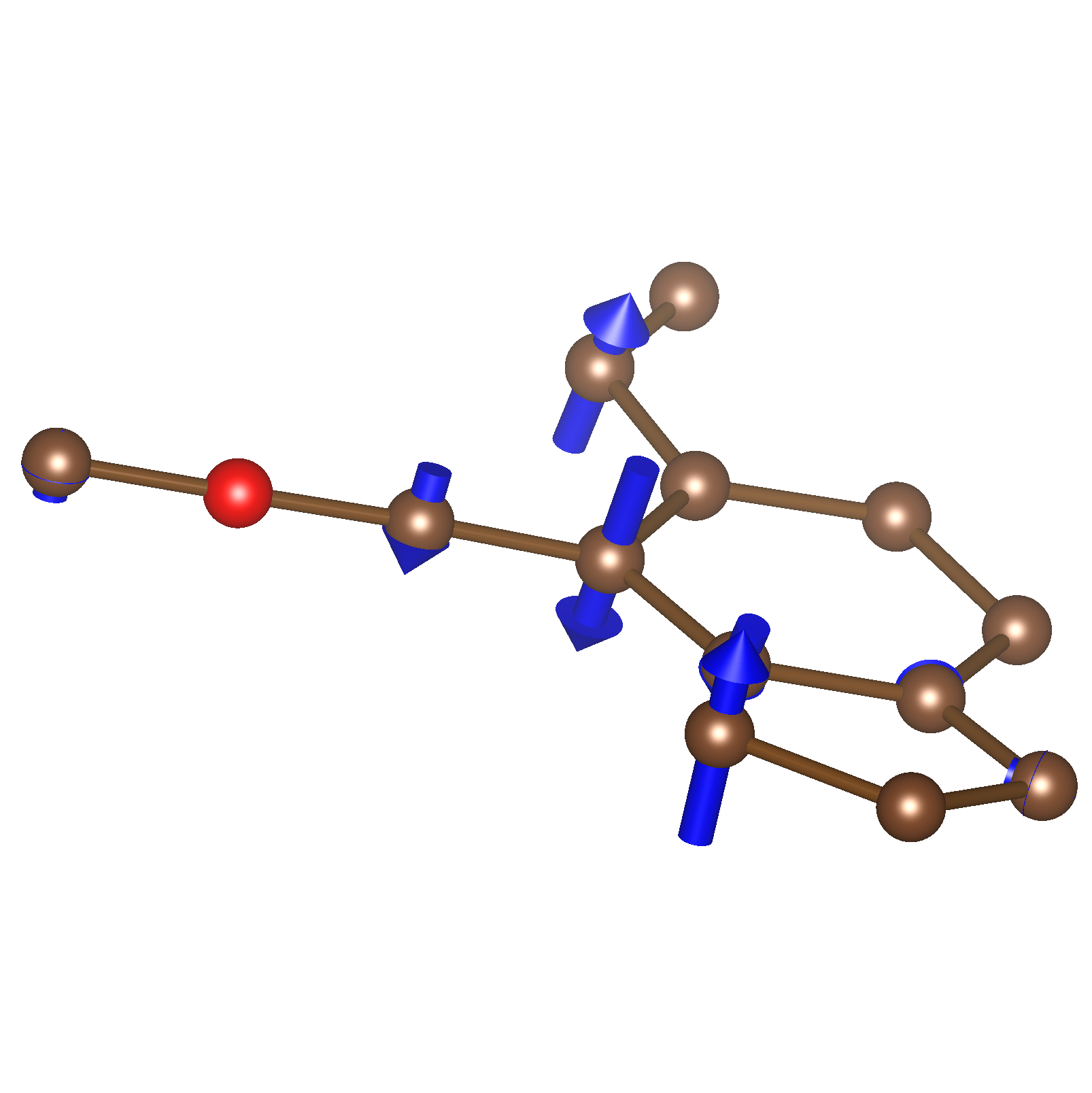}
         \caption{SOAP, $\lambda \sim 2 \times 10^{-6}$}
         \label{fig:}
     \end{subfigure}
     
     \begin{subfigure}[b]{0.2\textwidth}
         \centering
         \includegraphics[width=\textwidth]{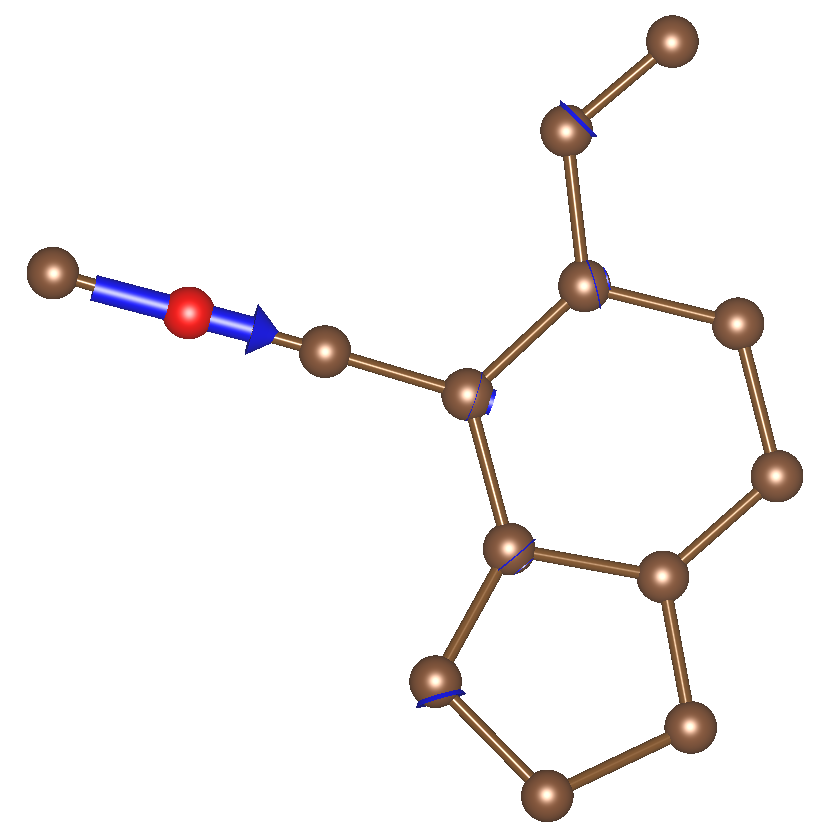}
         \caption{ACSF, $\lambda=1.0$}
         \label{fig:}
     \end{subfigure}
     \begin{subfigure}[b]{0.2\textwidth}
         \centering
         \includegraphics[width=\textwidth]{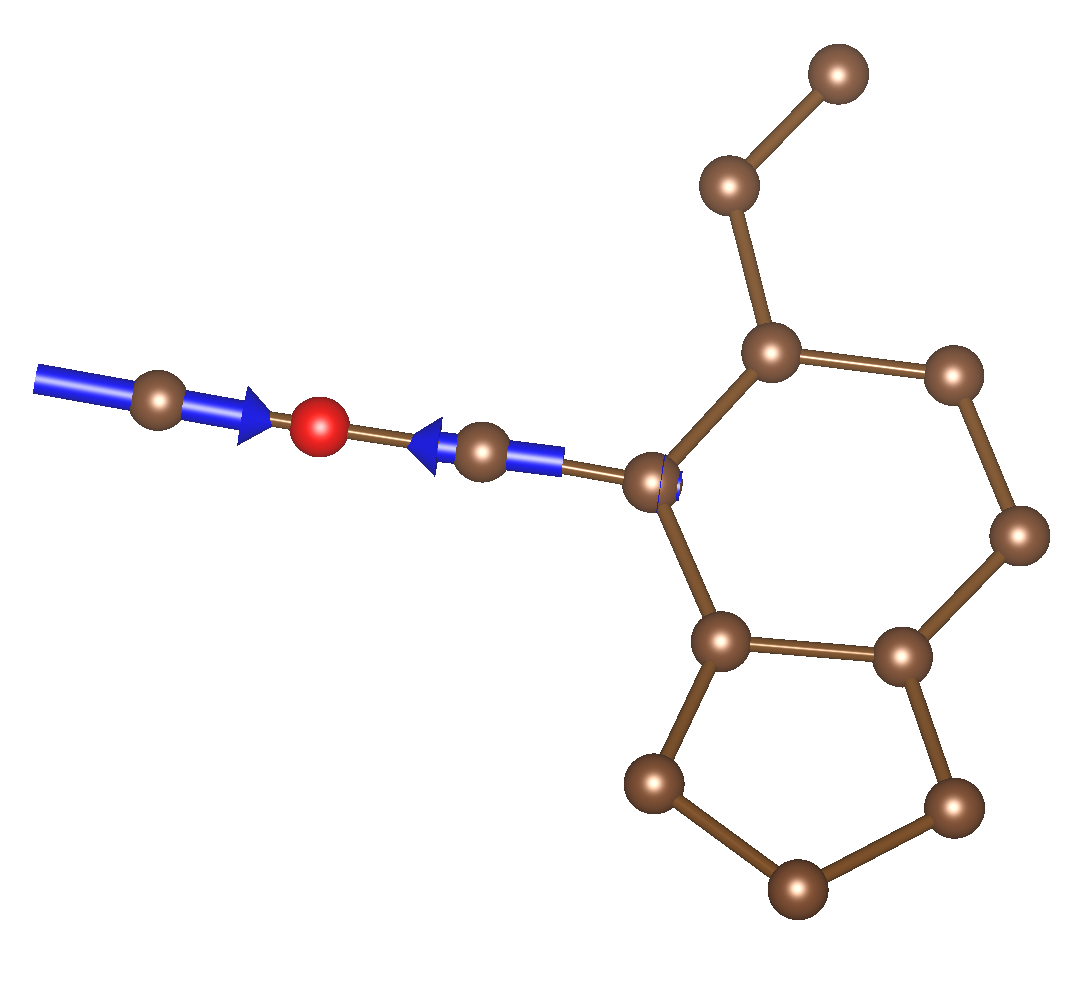}
         \caption{ACSF, $\lambda=0.124$}
         \label{fig:}
     \end{subfigure}
     \begin{subfigure}[b]{0.2\textwidth}
         \centering
         \includegraphics[width=\textwidth]{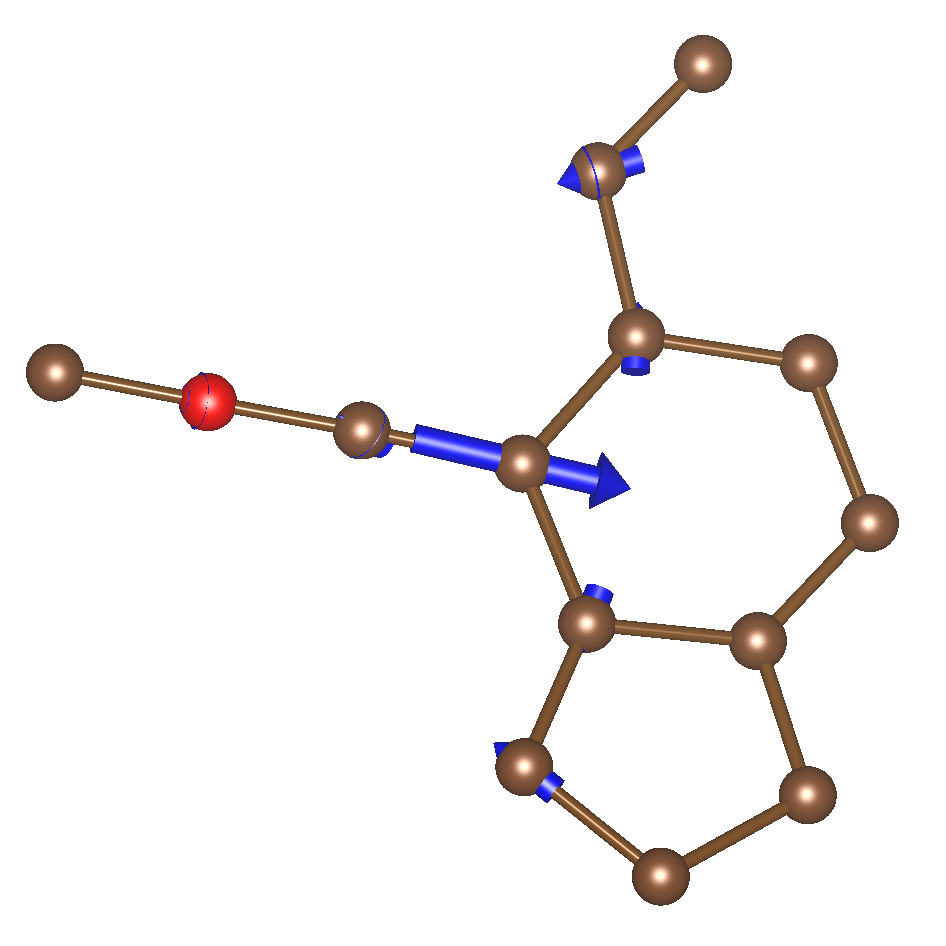}
         \caption{ACSF, $\lambda=0.006$}
         \label{fig:}
     \end{subfigure}
     \begin{subfigure}[b]{0.2\textwidth}
         \centering
         \includegraphics[width=\textwidth]{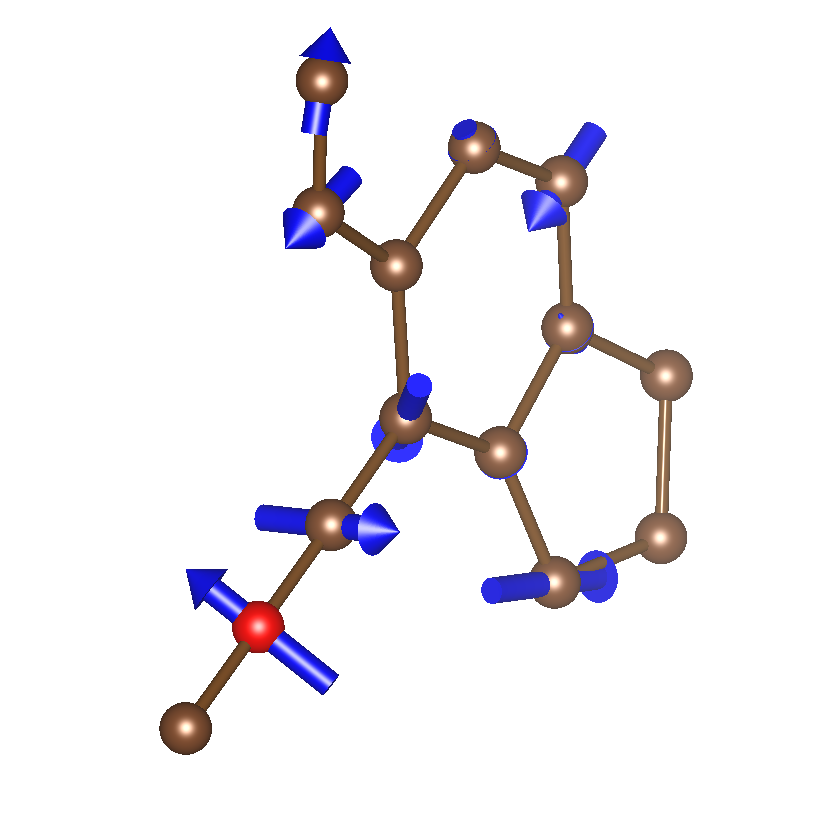}
         \caption{ACSF, $\lambda \sim 2\times 10^{-6}$}
         \label{fig:}
     \end{subfigure}
     
     \begin{subfigure}[b]{0.2\textwidth}
         \centering
         \includegraphics[width=\textwidth]{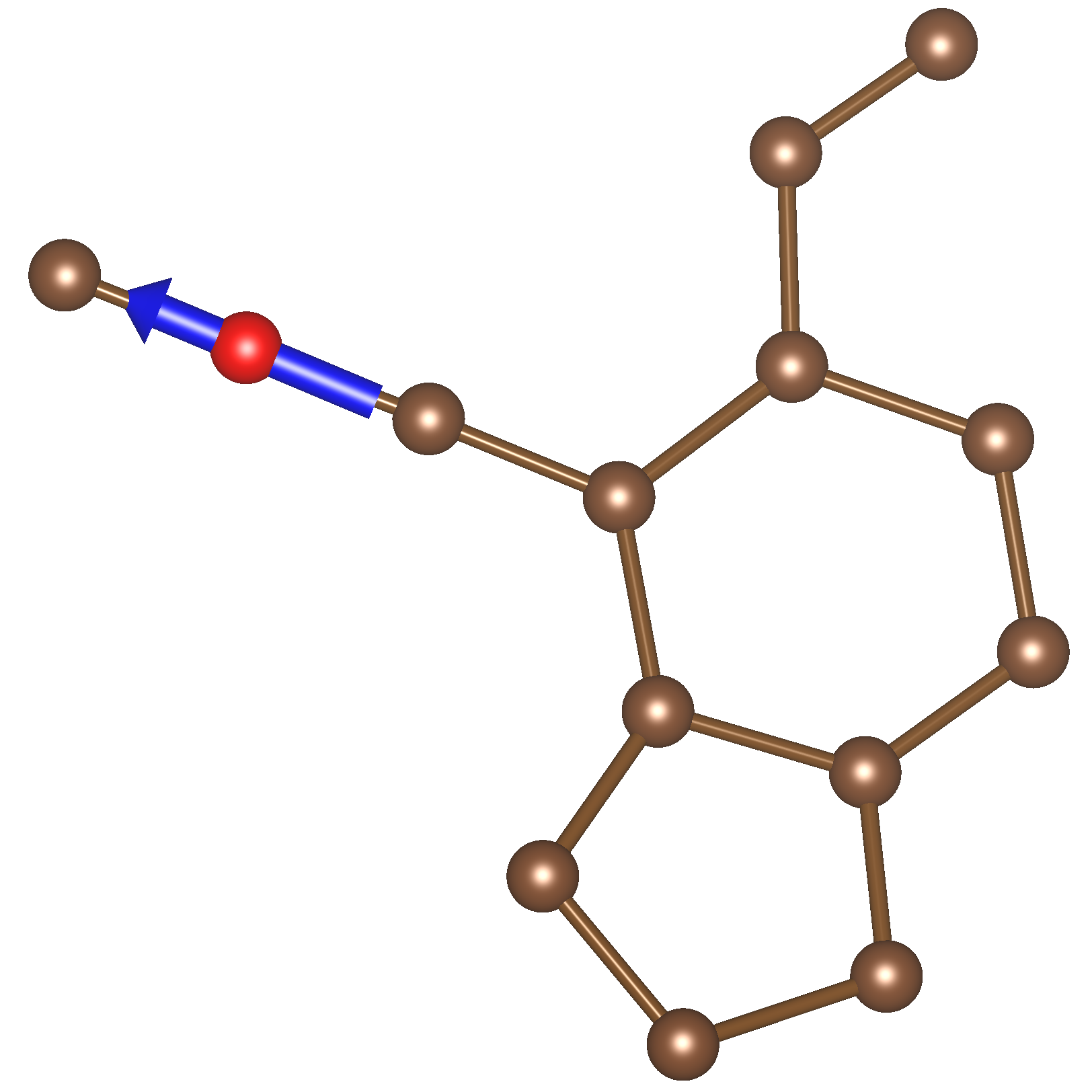}
         \caption{MBSF, $\lambda=1.0$ }
         \label{fig:}
     \end{subfigure}
     \begin{subfigure}[b]{0.2\textwidth}
         \centering
         \includegraphics[width=\textwidth]{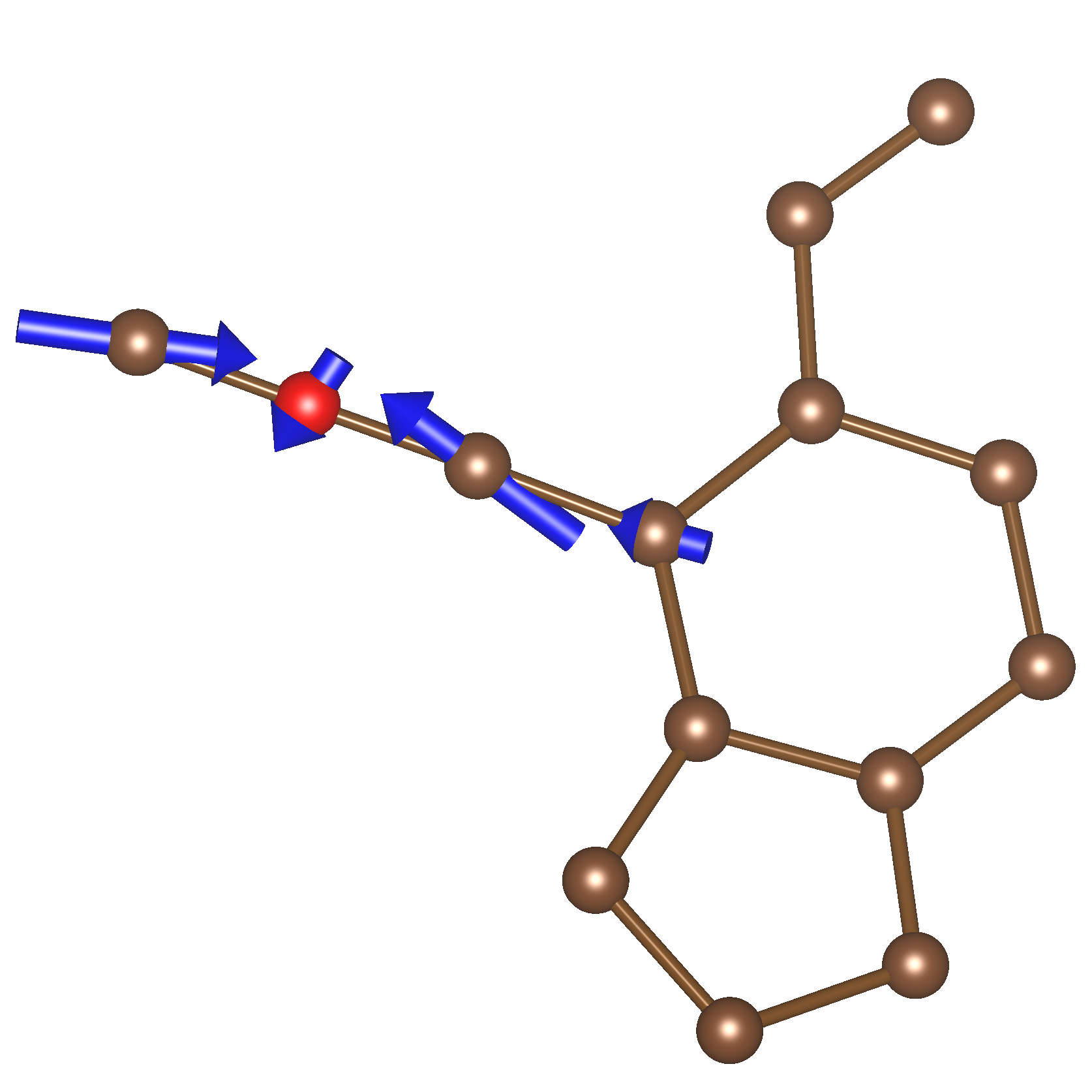}
         \caption{MBSF, $\lambda=0.084$ }
         \label{fig:}
     \end{subfigure}
     \begin{subfigure}[b]{0.2\textwidth}
         \centering
         \includegraphics[width=\textwidth]{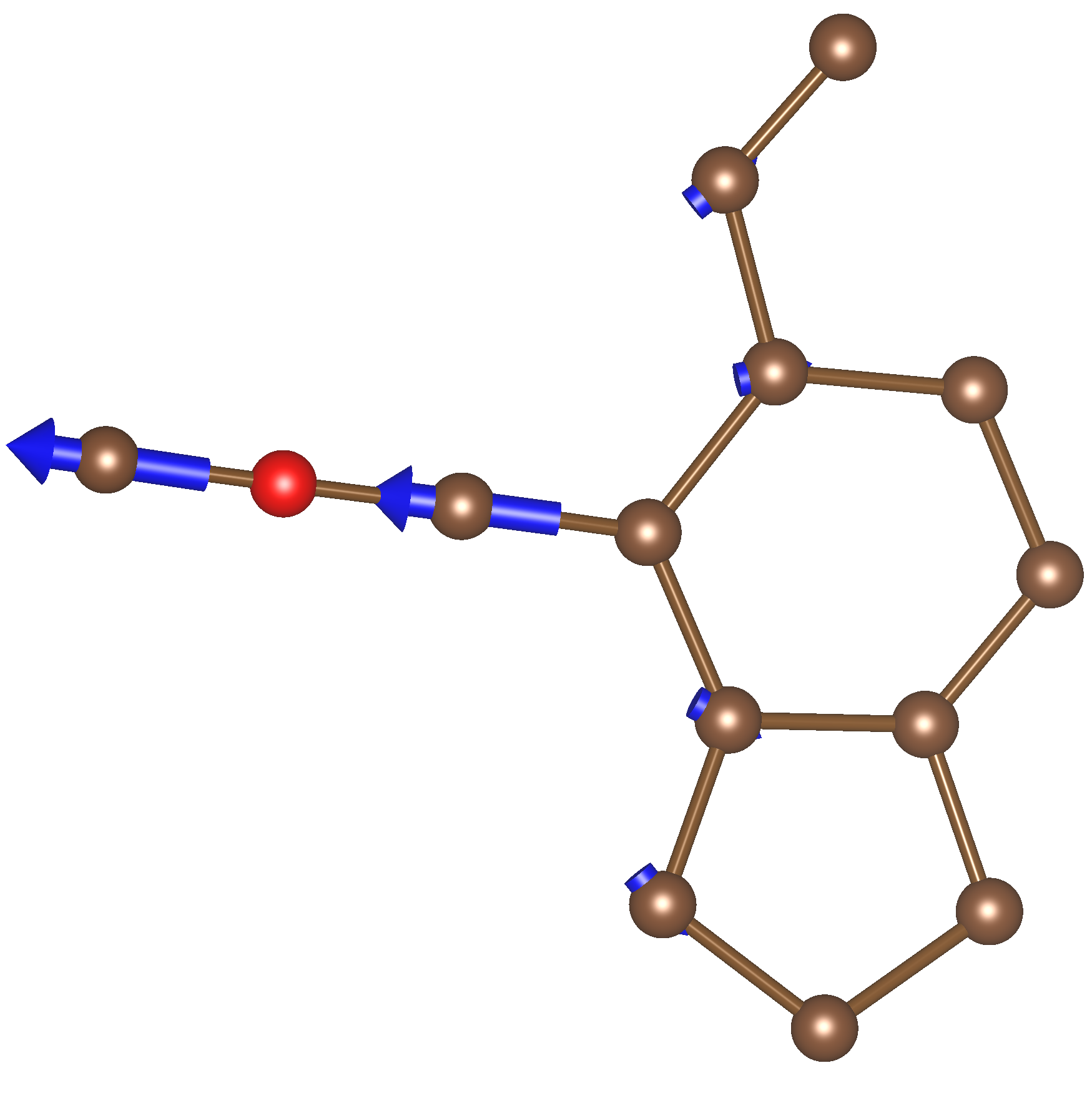}
         \caption{MBSF, $\lambda=0.018$}
         \label{fig:}
     \end{subfigure}
     \begin{subfigure}[b]{0.2\textwidth}
         \centering
         \includegraphics[width=\textwidth]{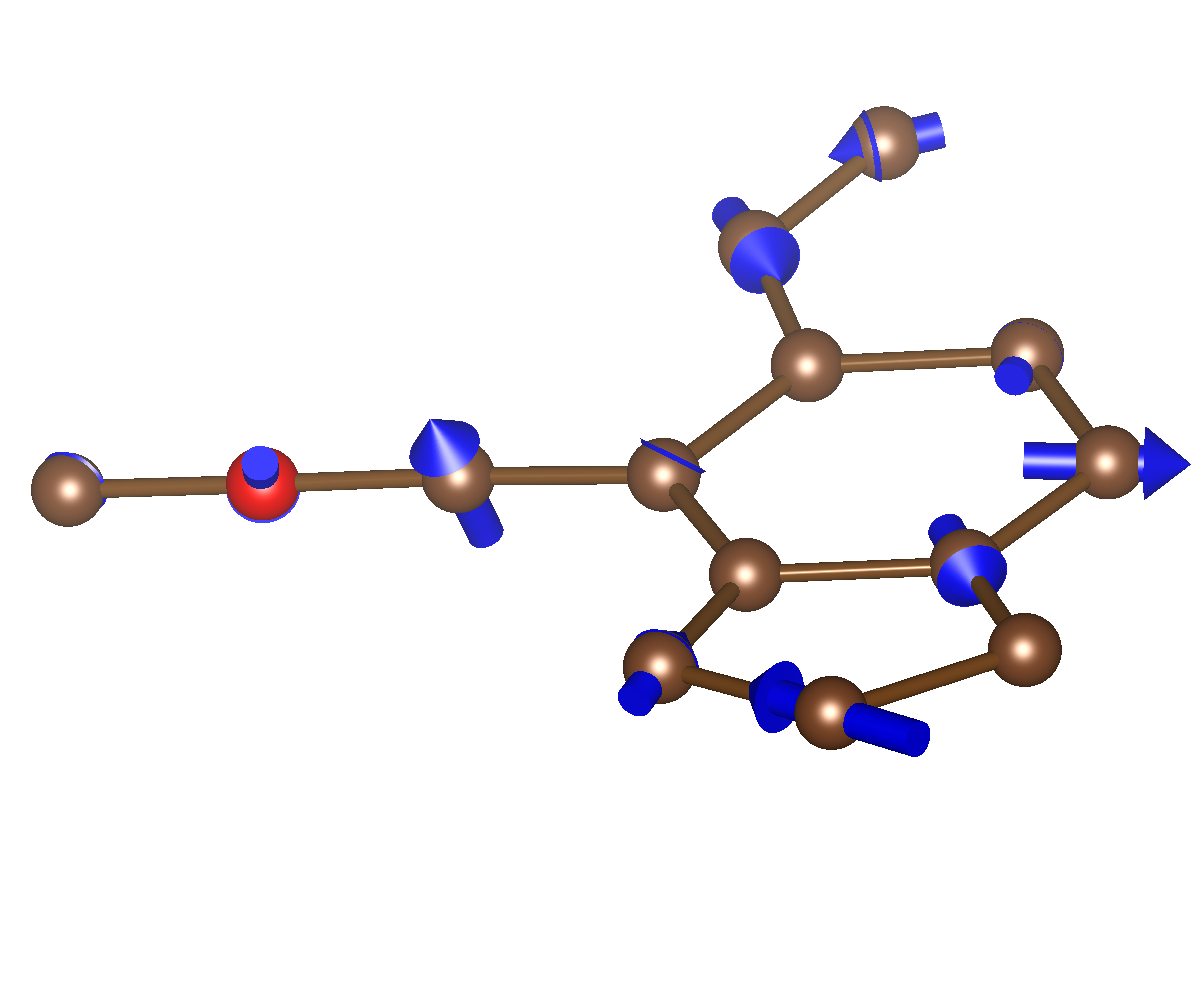}
         \caption{MBSF, $\lambda \sim 2 \times10^{-6}$}
         \label{fig:}
     \end{subfigure}

      \begin{subfigure}[b]{0.2\textwidth}
         \centering
         \includegraphics[width=\textwidth]{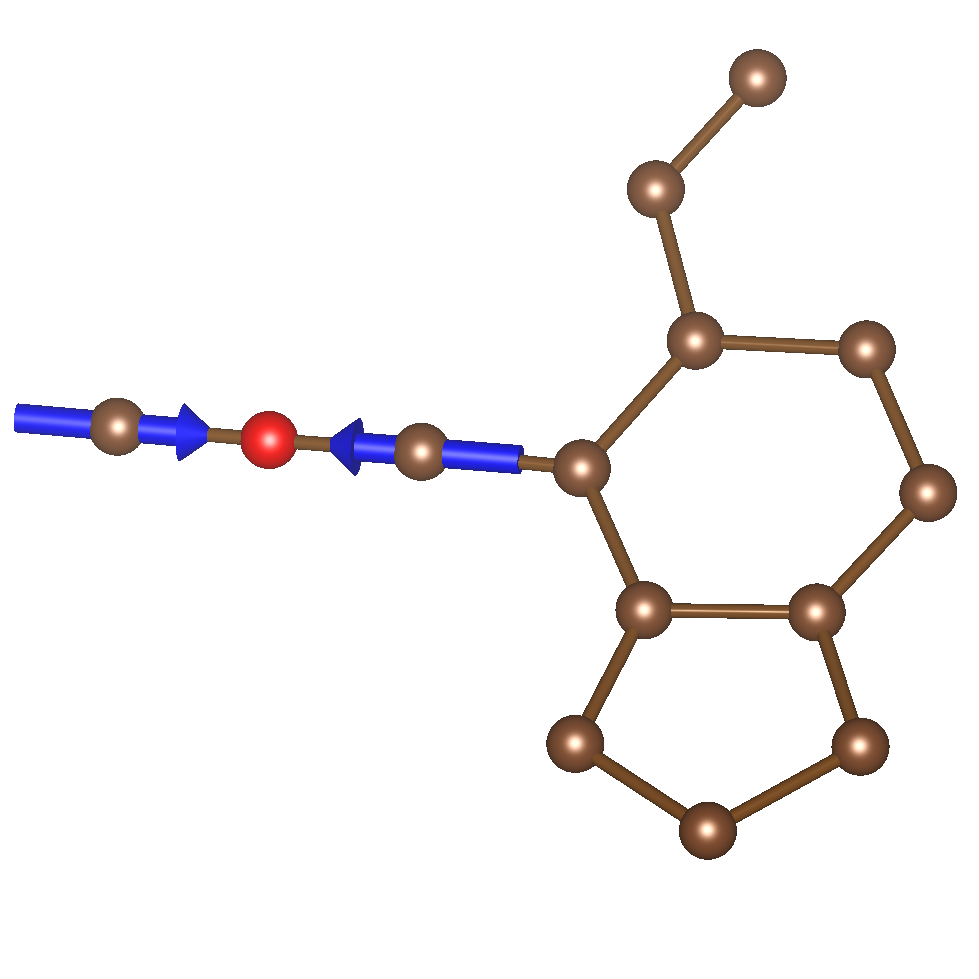}
         \caption{FCHL, $\lambda=1.0$ }
         \label{fig:}
     \end{subfigure}
     \begin{subfigure}[b]{0.2\textwidth}
         \centering
         \includegraphics[width=\textwidth]{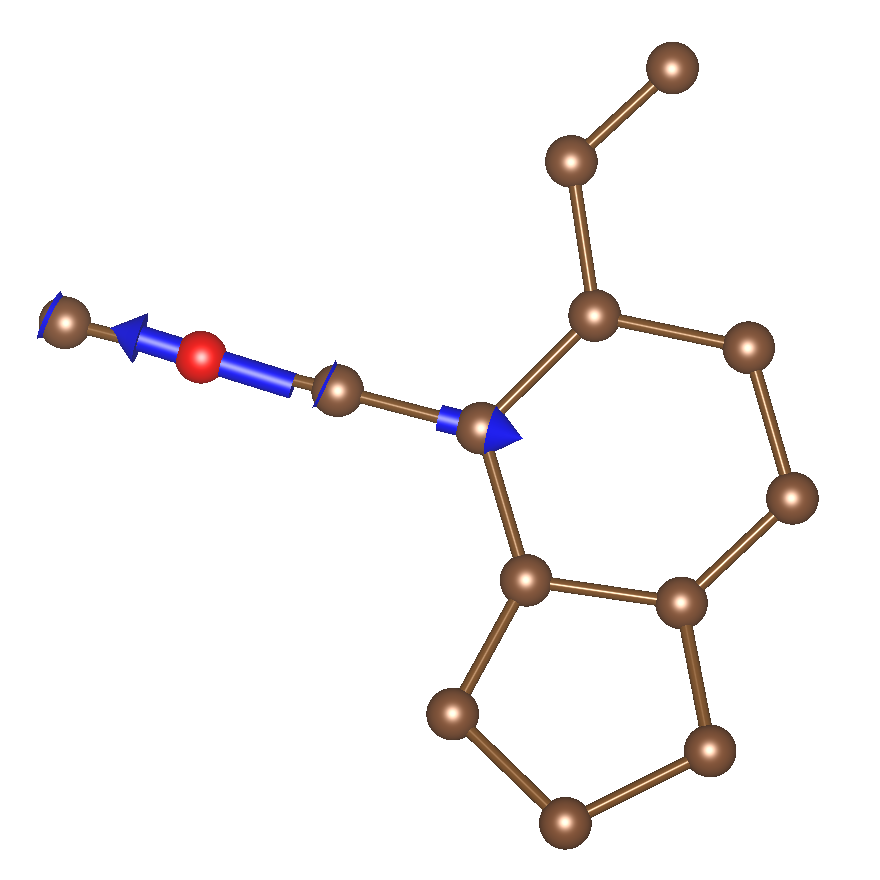}
         \caption{FCHL, $\lambda=0.031$ }
         \label{fig:}
     \end{subfigure}
     \begin{subfigure}[b]{0.2\textwidth}
         \centering
         \includegraphics[width=\textwidth]{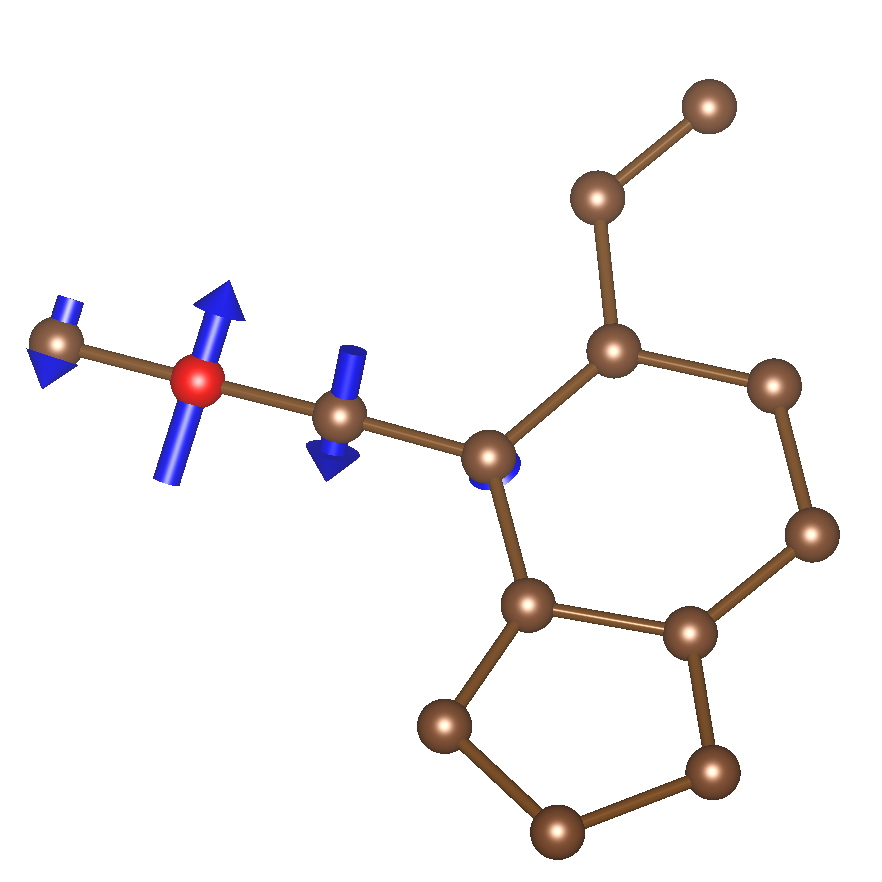}
         \caption{FCHL, $\lambda=0.017$}
         \label{fig:}
     \end{subfigure}
     \begin{subfigure}[b]{0.2\textwidth}
         \centering
         \includegraphics[width=\textwidth]{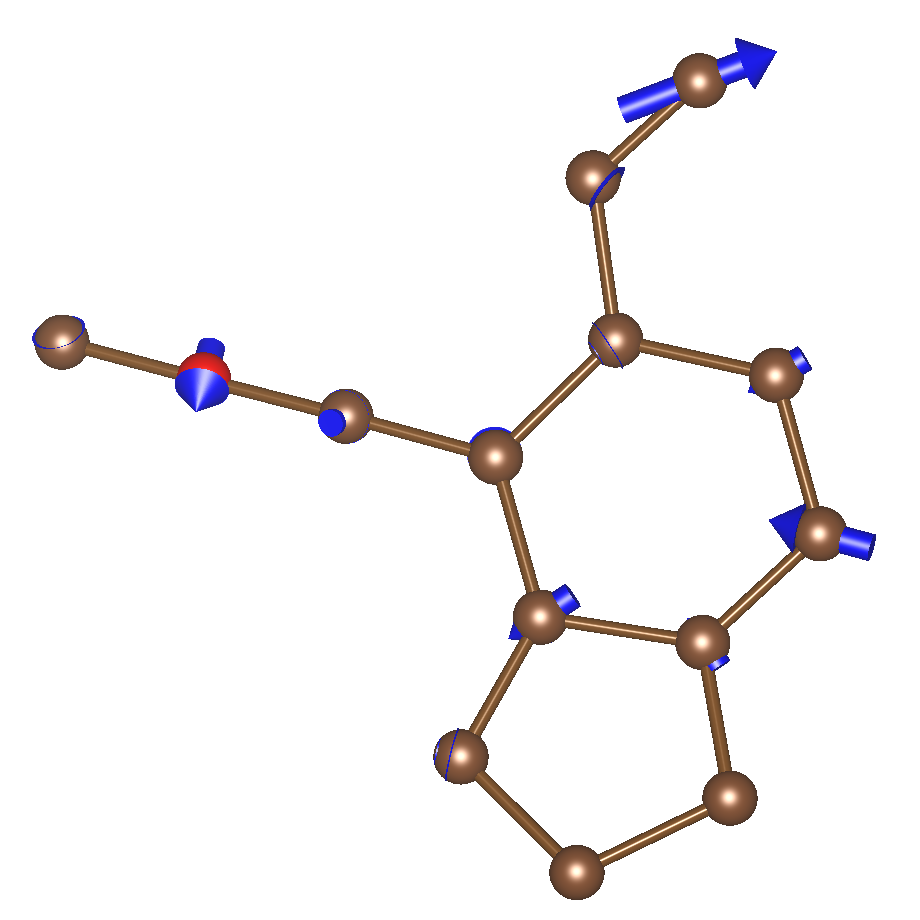}
         \caption{FCHL, $\lambda \sim 2 \times10^{-6}$}
         \label{fig:}
     \end{subfigure}
     
\caption{
Same as~\ref{fig:eigenmodeconf} but for the environment of~\ref{fig:conf22}.}
     \label{fig:eigenmodeconf2}
\end{figure*}

In Fig.~\ref{fig:confeig} we show the eigenvalues of the sensitivity matrix of configuration~\ref{fig:conf} for all the 
fingerprints examined in our study.
The eigenvalues of the sensitivity matrix for ACSF, MBSF, and FCHL decrease much more rapidly 
to zero than the eigenvalues of SOAP and OM[sp]. This means that in ACSF, MBSF, and FCHL, there exist only a few modes that 
have a strong influence on the fingerprint. 
It is also of interest to look at the associated modes   
shown in Fig.~\ref{fig:eigenmodeconf} and~\ref{fig:eigenmodeconf2}. In the context of machine learning one might hope that the
modes that are associated to the largest eigenvalues and will therefore lead to the strongest variation in the fingerprint will also lead to the largest variation of 
physical properties such as forces~\cite{behler2011atom}.
Since movements of atoms close to the central atom 
will in general lead to a strong variation of the environment of the 
reference atom, 
this means 
that modes belonging to large eigenvalues should be localized around the central atom.
The movement that will lead to the strongest variation of the 
energy
for the configurations shown in Fig.~\ref{fig:eigenmodeconf}  is clearly a bond stretching mode where the 3 neighboring atoms either move towards the central atom or away from it (Fig.~\ref{fig:eigenmodeconfom1}, \ref{fig:eigenmodeconfom2}, \ref{fig:eigenmodeconfom3}). 
Then follows a movement where two bonds of the central are compressed and one is 
stretched and finally an out of plane movement of the central atom.
These three modes are exactly the modes 
associated to the 3 largest eigenvalues of the OM sensitivity matrix. SOAP and FCHL also describe 
the physically important modes with reasonably large eigenvalues. In the ACSF and MBSF fingerprints 
however only an out of plane mode has a reasonably large eigenvalue. 
The modes belonging to the few largest eigenvalues are always localized on the 
reference atom and 
a few surrounding  atoms. 
As the eigenvalues become smaller the modes should get more delocalized, and this is indeed true in most cases.
There are however some exceptions such as the modes  of the ACSF shown in the panels l of Fig.~\ref{fig:eigenmodeconf} and   Fig.~\ref{fig:eigenmodeconf2}, the modes of MBSF  
in the panels p of Fig.~\ref{fig:eigenmodeconf} and Fig.~\ref{fig:eigenmodeconf2} and a mode of 
SOAP shown panel h of Fig.~\ref{fig:eigenmodeconf2}.

\begin{figure*}[t!]
     \centering
     \begin{subfigure}[b]{0.4\textwidth}
         \centering
         \includegraphics[width=\textwidth]{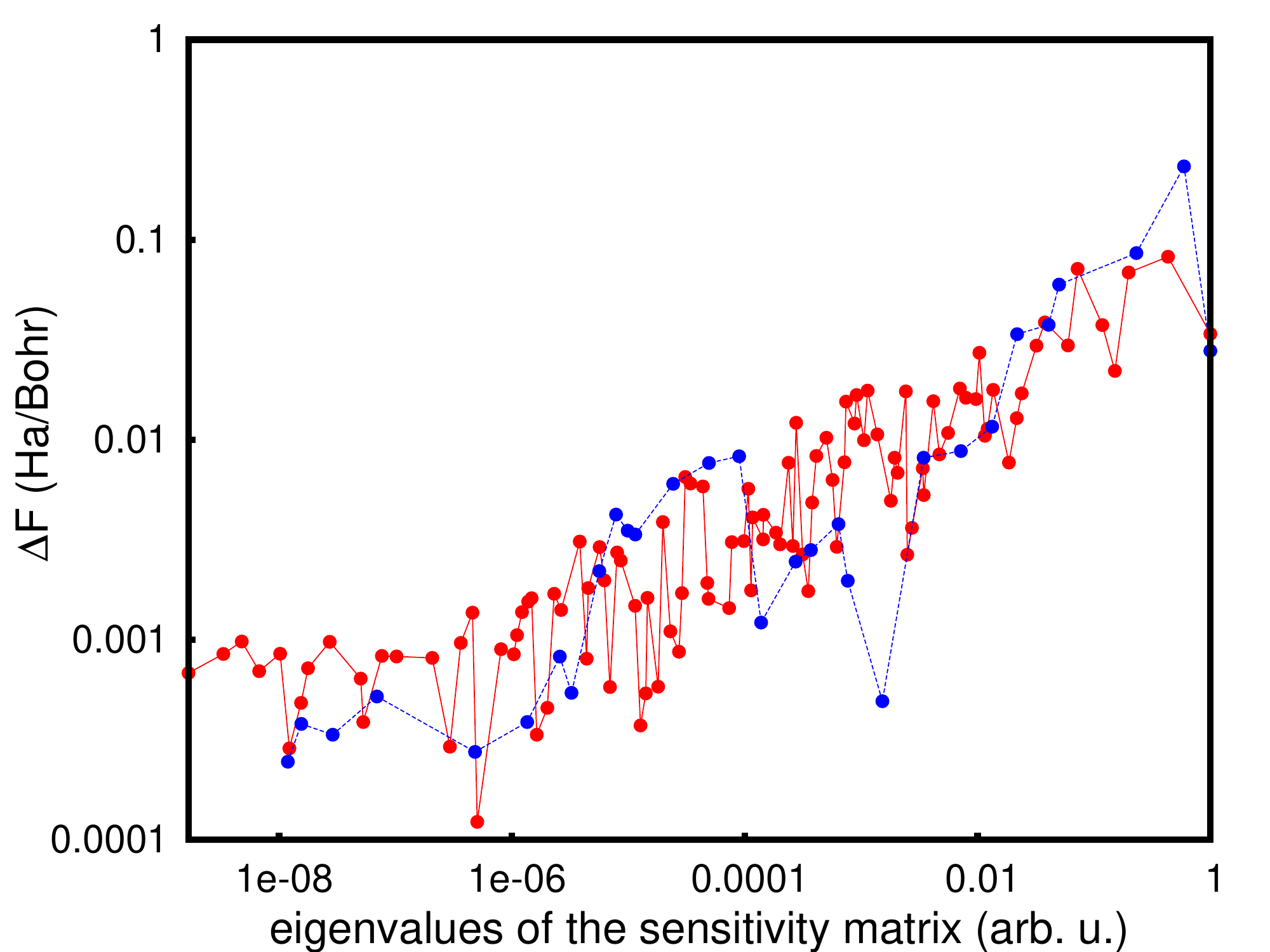}
         \caption{OM[sp]}
         \label{fig:eigvsforcediffom}
     \end{subfigure}
     \begin{subfigure}[b]{0.4\textwidth}
         \centering
         \includegraphics[width=\textwidth]{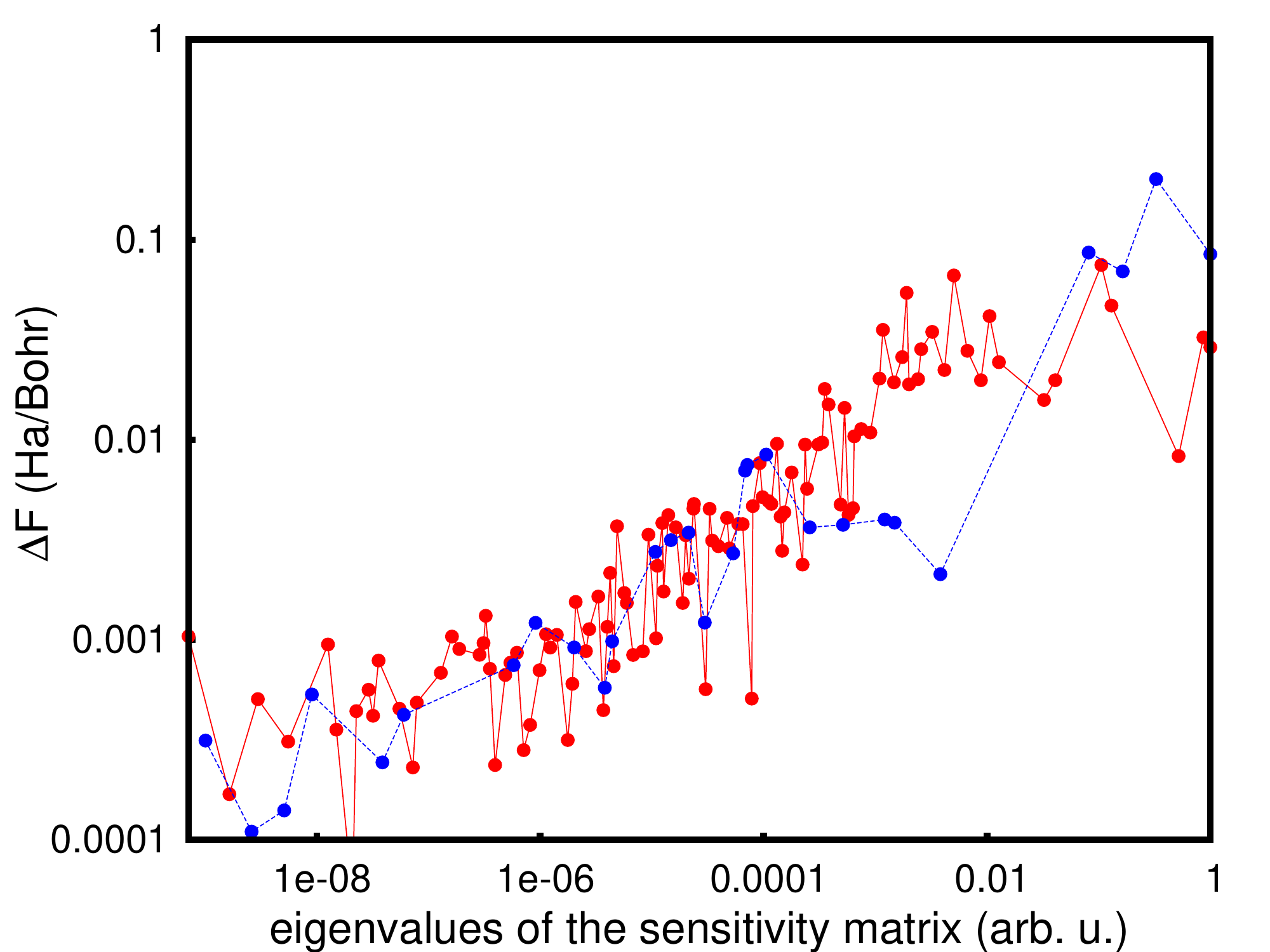}
         \caption{SOAP}
         \label{fig:eigvsforcediffsoap}
     \end{subfigure}
     \begin{subfigure}[b]{0.4\textwidth}
         \centering
         \includegraphics[width=\textwidth]{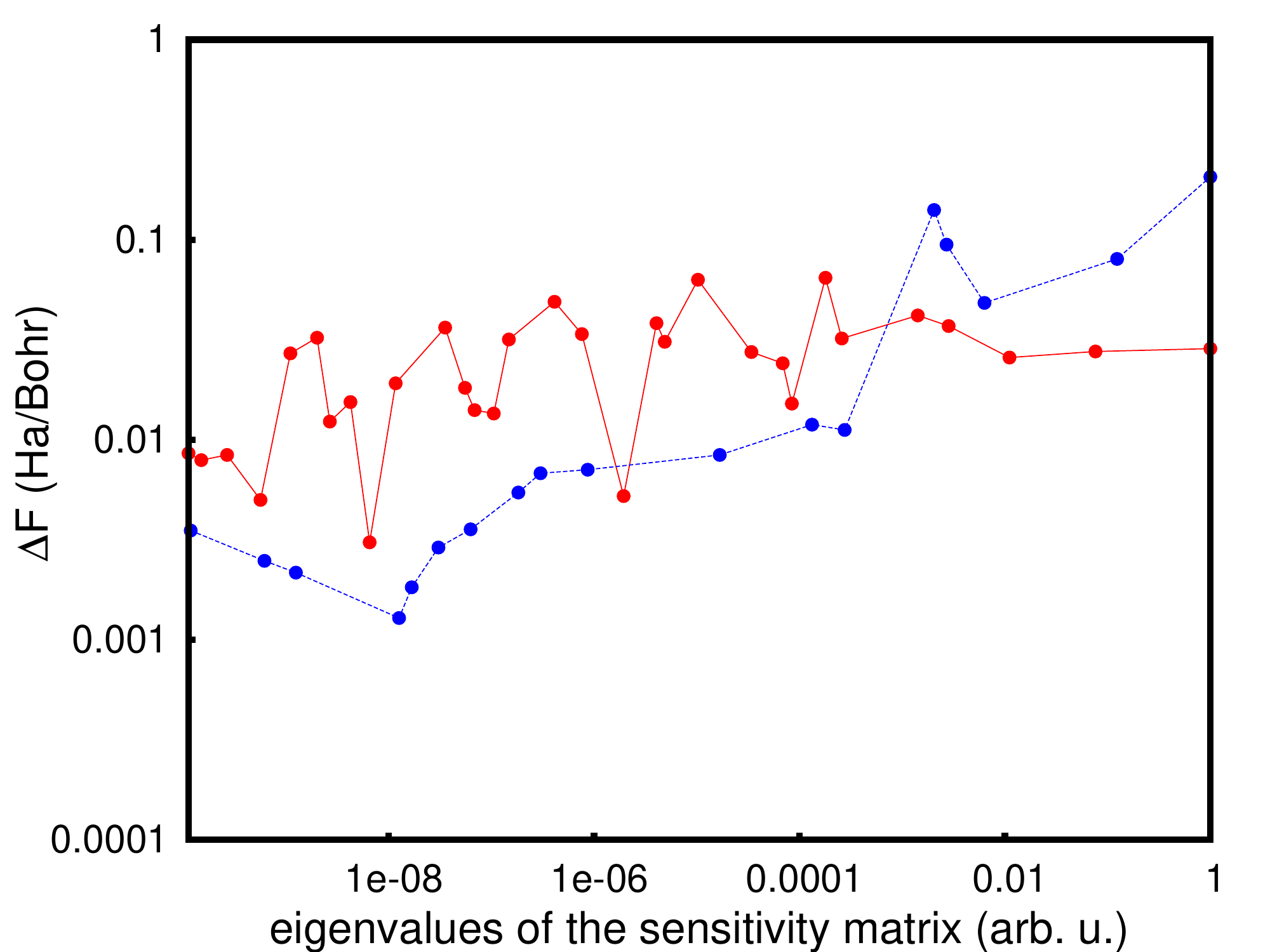}
      \caption{ACSF}
         \label{fig:eigvsforcediffbehler}
     \end{subfigure}
     \begin{subfigure}[b]{0.4\textwidth}
         \centering
         \includegraphics[width=\textwidth]{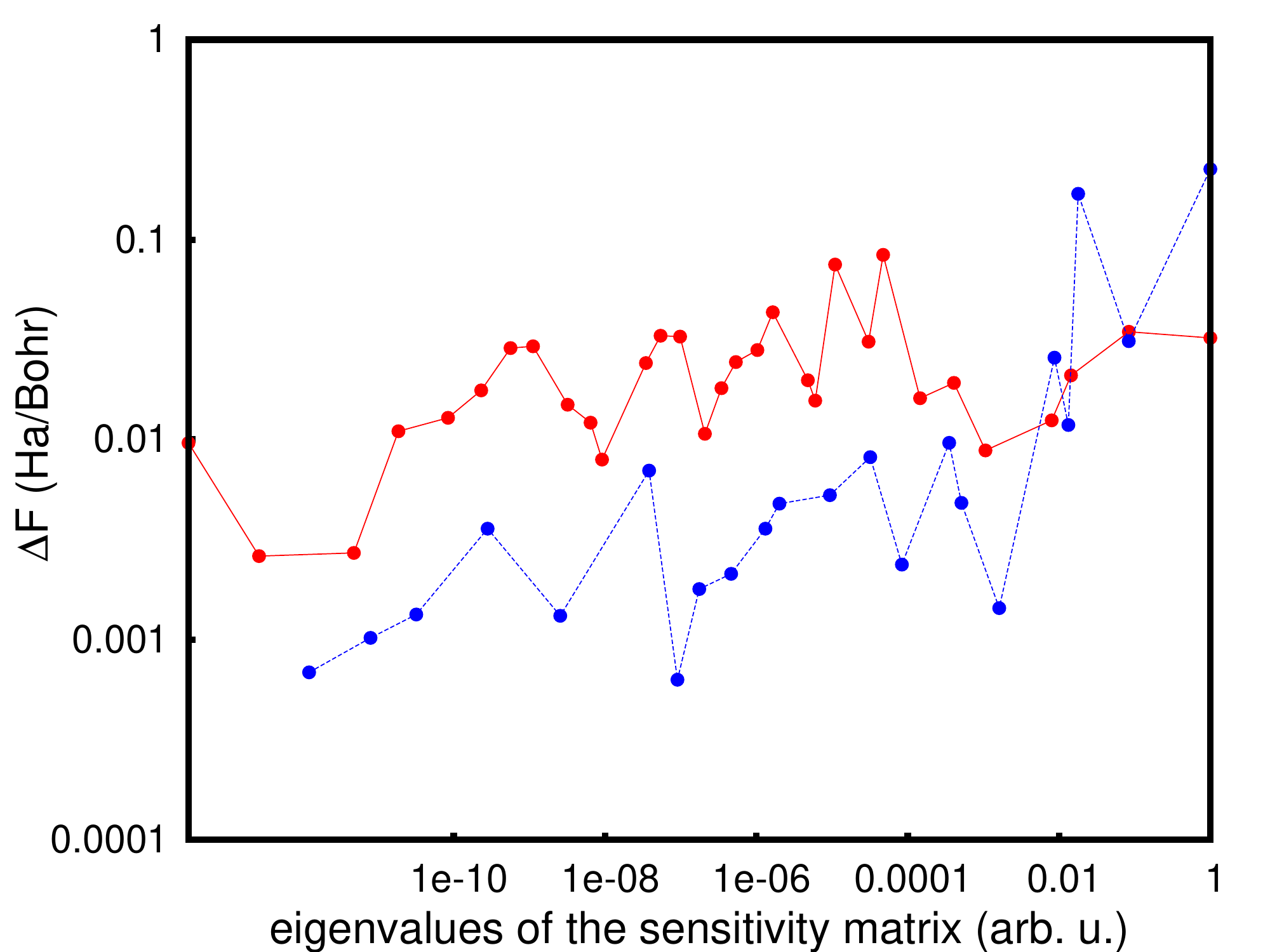}
         \caption{MBSF}
         \label{fig:eigvsforcediffacsf}
     \end{subfigure}
     \begin{subfigure}[b]{0.4\textwidth}
         \centering
         \includegraphics[width=\textwidth]{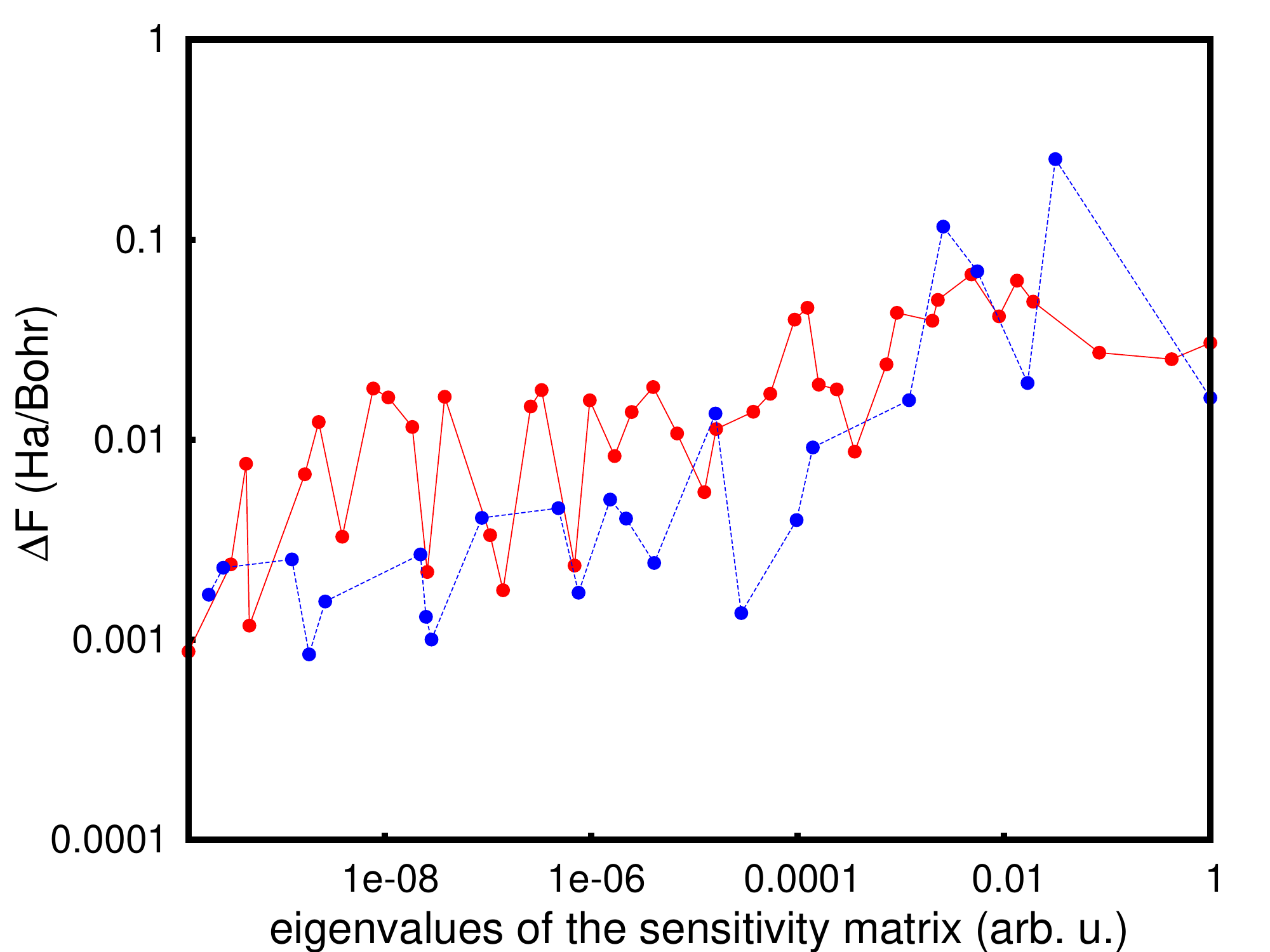}
         \caption{FCHL}
         \label{fig:eigvsforcedifffchl}
     \end{subfigure}
\caption{
Changes of the absolute forces upon displacements along the eigenvectors of the sensitivity matrix vs. its eigenvalues. For each fingerprint the atoms in the system are moved along the respective eigenvectors and the force changes are calculated using DFT~\cite{bigdft,PSP,PBE}. The red and the blue curves belong to the reference atoms in \ref{fig:conf} and \ref{fig:conf22} respectively. 
There is a strong correlation in OM and SOAP since eigenvectors of large eigenvalues are localized around the reference atom and eigenvectors of small eigenvalues are localized on further distances from the reference atom whereas in ACSF, MBSF, and FCHL it is not the case (there is no preferred spatial order of the components, which is why a clear correlation cannot be seen).
} 
     \label{fig:forcediffvseigs}
\end{figure*}

This discussion, which was based on some physical insight into which modes are important, can also be 
made more quantitative. We do this by plotting the change in the force acting on the central atom when the system  is moved along the different modes against the eigenvalue of this mode.
This is shown in Fig.~\ref{fig:forcediffvseigs}. 
A 
clear correlation is found for OM and SOAP, while for ACSF, MBSF and FCHL the correlation is substantially weaker, with FCHL showing at least the correct trend.
This  means that movements along modes associated to large and small eigenvalues have almost the same influence on the force on the 
reference atom. 

Even though the environment of Fig.~\ref{fig:conf22} is quite different, the performance of the fingerprints is quite similar. Only  OM and SOAP detect the physically important modes (Fig.~\ref{fig:conf2eig}), i.e. assign a large eigenvalue to these modes.
They are also the only two fingerprints that give a good correlation between the eigenvalues and the 
change in the force (Fig.~\ref{fig:forcediffvseigs}).
 
While SOAP is performing well in our test case where many atoms are contained in the sphere, 
it was recently shown~\cite{pozdnyakov2020completeness} that for a methane molecule there are movements that leave the SOAP 
fingerprint of the carbon invariant.  We detected the same 
deficiency also for ACSF, MBSF and FCHL. 
We have also tested the OM fingerprint for these configurations and did 
not find any small or even zero eigenvalues. This is to be expected since the OM fingerprint is based on a matrix diagonalization 
scheme that is similar to the diagonalization of the Hamiltonian matrix in a quantum-mechanical calculation. 
Hence the scheme is not restricted to the information obtained only from the radial and angular distribution of the atoms in the sphere.

\subsection{Correlation of fingerprint distances}\label{powerres}

\begin{figure*}[p!]
    \centering
    \includegraphics[width=0.7\textwidth]{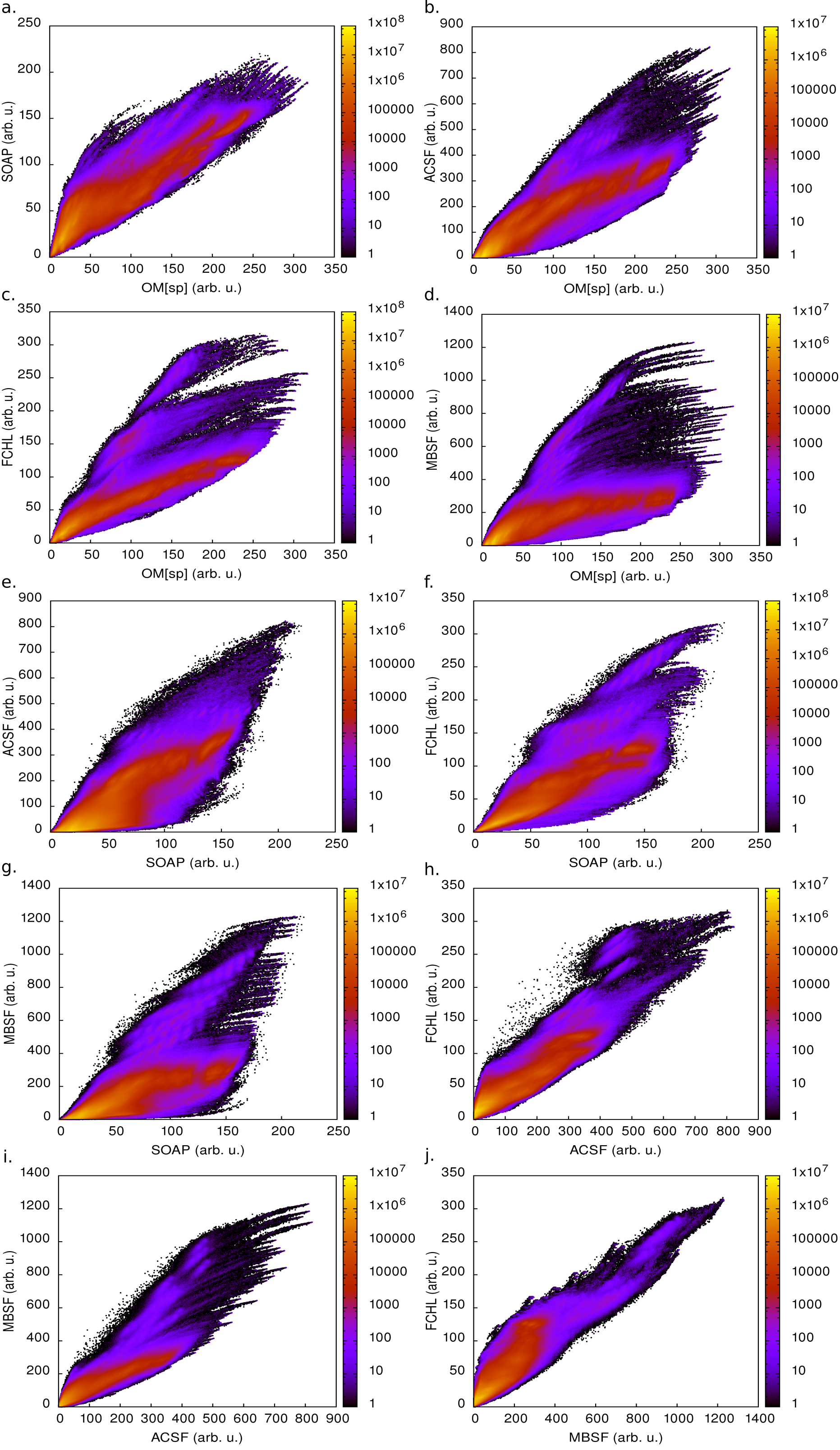}
    \caption{The correlation intensity plot for a) OM vs. SOAP; b) OM vs ACSF; c) OM vs. FCHL; d) OM vs. MBSF; e) SOAP vs. ACSF; f) SOAP vs. FCHL; g) SOAP vs. MBSF; h) ACSF vs. FCHL; i) ACSF vs. MBSF; j) MBSF vs. FCHL.}
    \label{fig:intplots}
\end{figure*}

In this section, we are going to compare the resolution power of different fingerprints, i.e. 
their 
numerical sensitivity to
small  dissimilarities between atomic environments. 
To perform the tests we have generated a set of 1000 C$_{60}$ structures using minima hopping~\cite{MH} coupled to DFTB~\cite{aradi2007dftb+}.
In this way we have obtained $60\times1000$ environments arising from  a large variety of structural motifs such as chains,  
planar structures and cages. We will in the following correlate all the $\frac{60000\times(60000-1)}{2}$ pairwise atomic fingerprint distances 
obtained from different 
fingerprint types.
Obviously large fingerprint distances should be obtained for environments that are quite different whereas 
small distances correspond to similar environments. Since the absolute value of a fingerprint distance 
is arbitrary, we scale all our fingerprint distances such that a distance of one corresponds to the noise level. We define the noise level as the fingerprint distance between identical structures, whose 
atoms were randomly displaced by an amount of up to $ \pm 0.02 $ \AA. 
 Since the number of environment pairs is huge we would not be able to resolve each pair in 
 a simple correlation plot where we would plot the fingerprint distances $\Delta^A_{I,J}$ according to fingerprint A versus the distance  $\Delta^B_{I,J}$  according to fingerprint B. However this large number of data allows us to generate a histogram. This histogram tells us how many 
 environments have fingerprint distances  $\Delta^A_{I,J}$ and  $\Delta^B_{I,J}$. These two distances are plotted along the x and y axis and the height of the bins of the histogram is indicated by the color in this plot shown in Fig.~\ref{fig:intplots}.

 As can be seen in Fig. \ref{fig:intplots}, in  most cases, the intensity is peaked around the diagonal which implies that both fingerprints agree on the degree of similarity or dissimilarity between the environment pairs. It can not be expected that all the points lie directly on the diagonal since different fingerprints weight different types of similarity or dissimilarity in different ways.
 There is however a problem if a point lies exactly on or very close to the x or y axis which means that the $\Delta$ is either zero or very small.
 This means that one fingerprint categorizes this pair of environments as identical whereas the other fingerprint can detect differences, i.e. it's $\Delta $ value is large. In Table~\ref{tab:compare0} we show several pairs of 
 environments that correspond to such problematic points in the correlation plot.

 
 In Table~\ref{tab:compare0} \textbf{a} we show the two most distinct  environments in the data according to OM[sp]. 
One environment is at the end of a chain and the other is 3-fold coordinated. 
So OM recognizes the atoms with the highest and lowest coordination 
 number found in this data set as being the most different. 
The fingerprint distance is $\Delta ^{OM[sp]}=317$. 
Diamond-like environments were not in our MH generated data set.  
Due to their large number of surface dangling bonds such structures are considerably higher in energy 
than the structures arising from sp2 and sp1 hybridized carbon atoms. 
However,  when we add by hand such a diamond derived cluster, OM predicts the central 4 fold coordinated atom of this cluster 
together with the previous atom at the end of the chain as the most distinct atoms.
So again it classifies  the two environments with the highest and lowest coordination as the most different ones. 
ACSF, SOAP, FCHL, and MBSF predict the environments in Table~\ref{tab:compare0} \textbf{b} and \textbf{c} to be the most distinct environments in the data.
The fingerprint distances are $\Delta^{SOAP}=214$, $\Delta^{FCHL}=315$, $\Delta^{ACSF}=822$, and $\Delta^{MBSF}=1224$ respectively. 
This is not in agreement with our basic chemical concepts  of what structural differences are important.
According to these concepts the coordination number is the most important quantity in the chemistry of carbon, 
since it is related to the hybridization 
state.
When adding the four fold coordinated carbon from the diamond-like cluster, then  ACSF, MBSF and FCHL correctly identify this 
fourfold coordinated environment and the one from the end of the chain as the most different ones.
The assignment of the largest fingerprint distance in SOAP is however unchanged by the addition of this fourfold 
coordinated environment. So the assignments of the symmetry-function-related fingerprints are at least partly 
compatible
with chemical concepts, whereas for SOAP this is not the case.
It is unclear whether a fingerprint that is compatible with chemical concepts gives better performance in machine learning schemes.
By choosing a shorter $r_\delta$ in the case of SOAP and shorter cutoff radii for ACSF-related fingerprints, it is however expected that the immediate environment gets more weight and that then the other fingerprints can also better distinguish different coordinations. We note that also for the cutoff employed in the present work individual components of the fingerprint vectors in ACSF-related fingerprints adopt different values for varying coordinations, while this effect is much less visible in the combined fingerprint distances.
    In the following we look at the correlation plots of fingerprint distances obtained with different fingerprints.
	We check whether some fingerprints can not recognize structural differences. 
\begin{table*}[p!]
   \begin{center}
   \begin{tabular}{|c|c|}
   
       \hline
       \begin{tabular}{cc}   
          \includegraphics[width=0.23\linewidth]{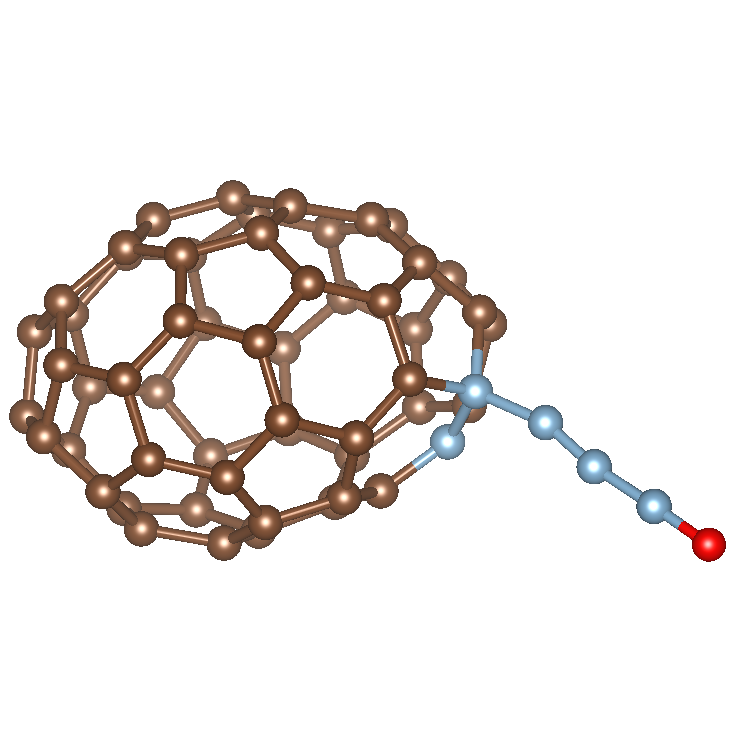} &
          \includegraphics[width=0.23\linewidth]{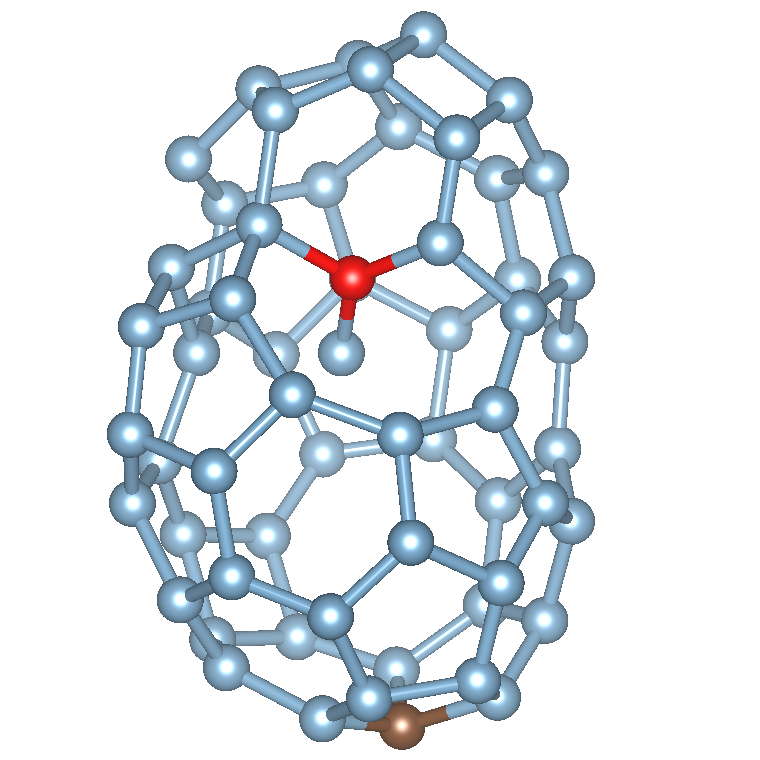} \\
      \end{tabular} &
       \begin{tabular}{cc}   
          \includegraphics[width=0.23\linewidth]{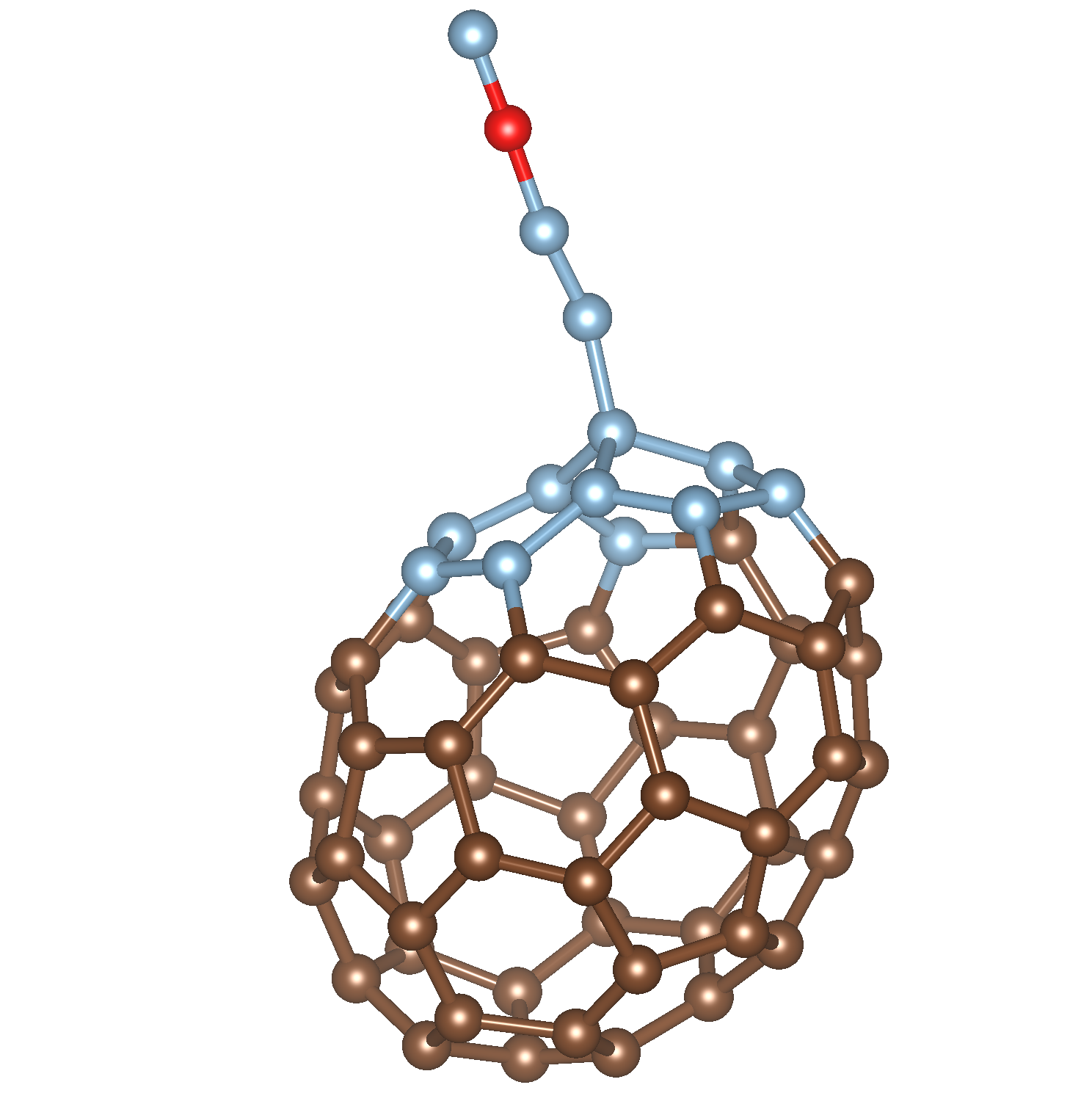}&
          \includegraphics[width=0.23\linewidth]{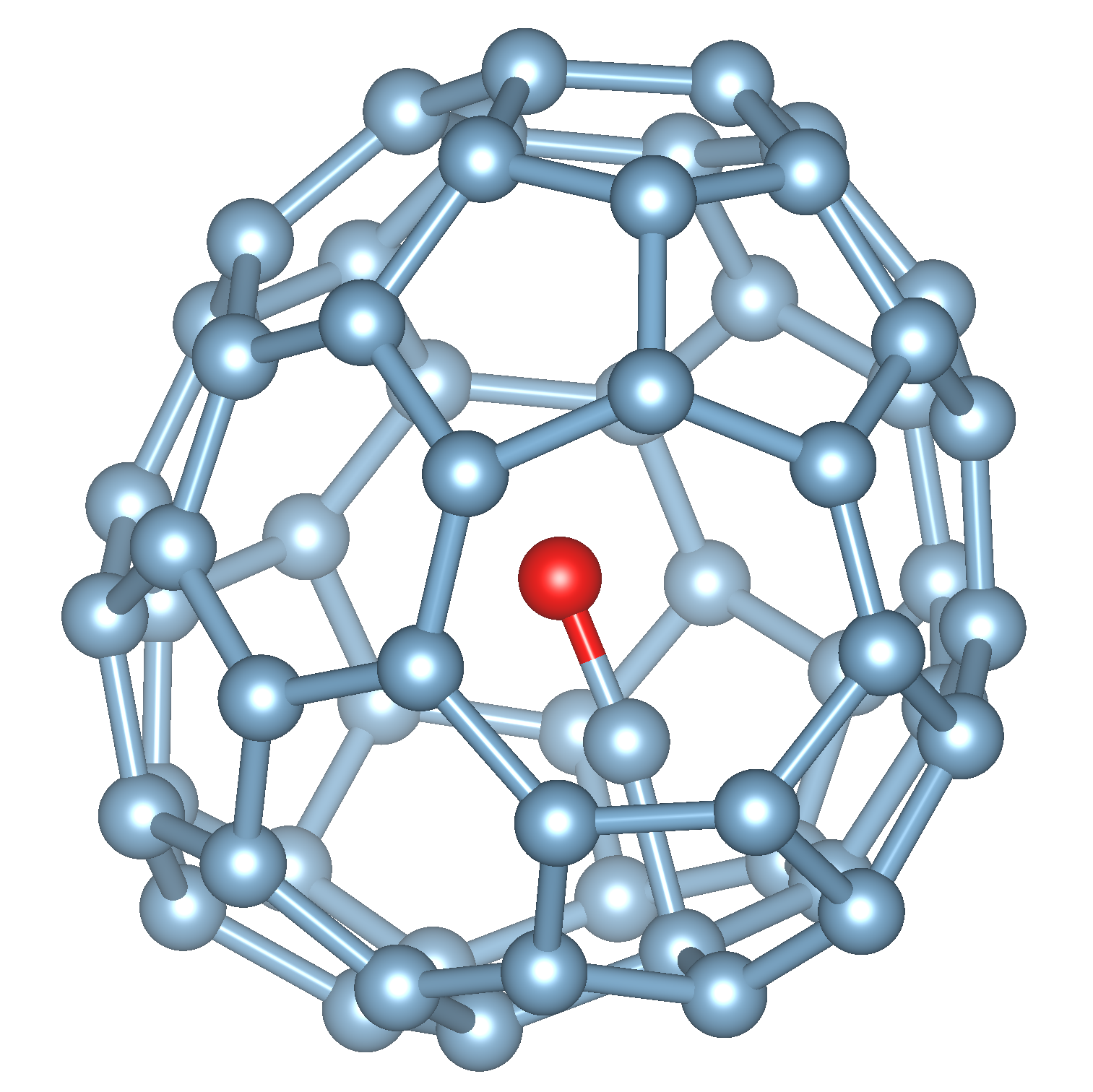} \\
      \end{tabular} \\
      \begin{tabular}{cc}  
         \textbf{a}) $\Delta^{OM[sp]}=317$(1.0);&
          $\Delta^{SOAP}=189$(0.88);\\
          $\Delta^{ACSF}=738$(0.90);&
          $\Delta^{FCHL}=256$(0.82);\\
          $\Delta^{MBSF}=844$(0.69)\\
      \end{tabular} &
      \begin{tabular}{cc} 
          \textbf{b}) $\Delta^{OM[sp]}=251$(0.79);&
          $\Delta^{SOAP}=214$(1.0);\\
          $\Delta^{ACSF}=802$(0.98);&
          $\Delta^{FCHL}=315$(1.0);\\
          $\Delta^{MBSF}=1224$(1.0)\\
      \end{tabular} \\
       \hline
       \begin{tabular}{cc}  
          \includegraphics[width=0.23\linewidth]{Tables_56_0219all.png}&
          \includegraphics[width=0.23\linewidth]{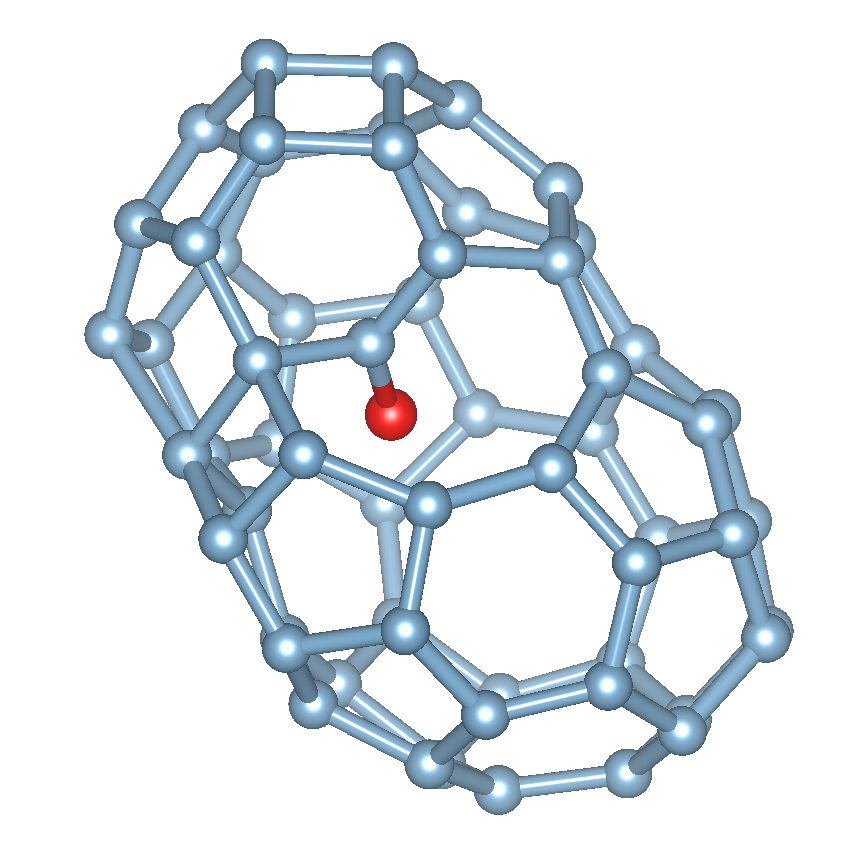} \\
      \end{tabular} &
      \begin{tabular}{cc} 
          \includegraphics[width=0.23\linewidth]{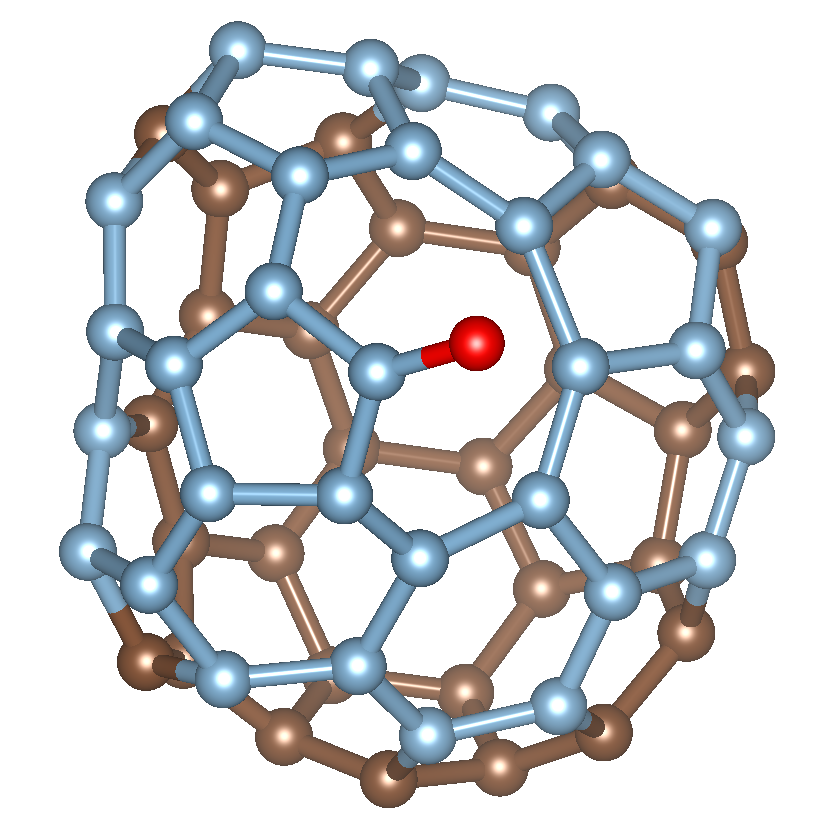}&
          \includegraphics[width=0.23\linewidth]{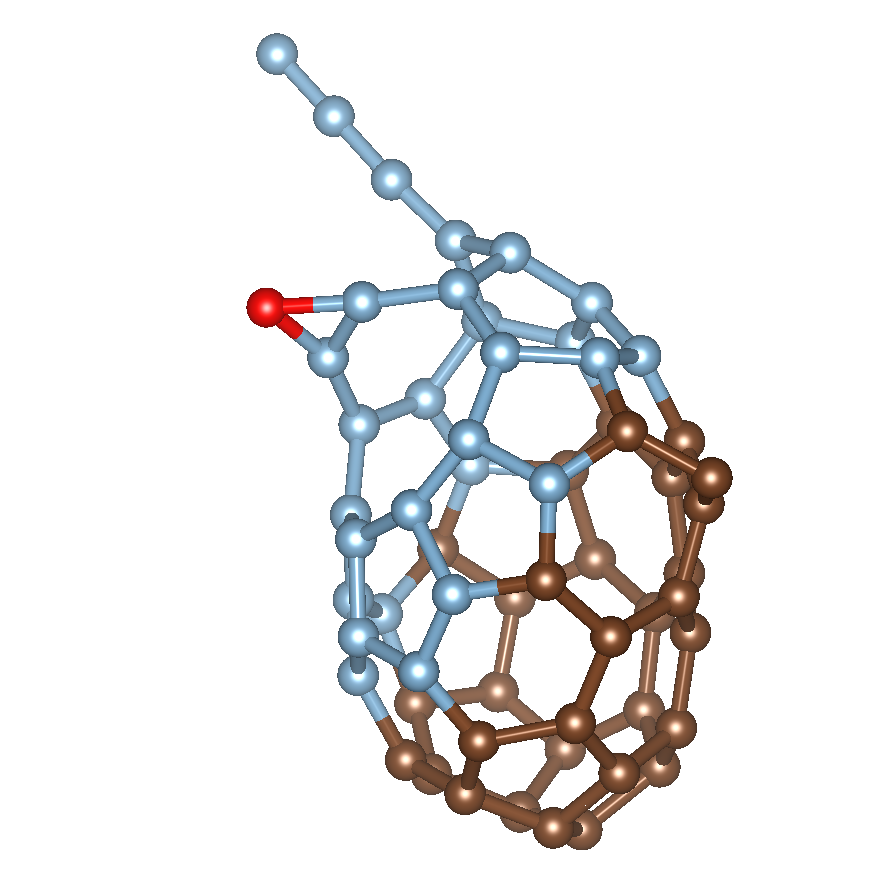} \\
      \end{tabular} \\
      \begin{tabular}{cc}  
          \textbf{c}) $\Delta^{OM[sp]}=292$(0.92);&
          $\Delta^{SOAP}=206$(0.96);\\
          $\Delta^{ACSF}=822$(1.0);&
          $\Delta^{FCHL}=292$(0.93);\\
          $\Delta^{MBSF}=1119$(0.91)\\
      \end{tabular} &
       \begin{tabular}{cc}  
         \textbf{d}) $\Delta^{OM[sp]}=38$(0.12);&
          $\Delta^{SOAP}=67$(0.32);\\
          $\Delta^{ACSF}=3$(0.0);&
          $\Delta^{FCHL}=33$(0.11);\\
          $\Delta^{MBSF}=5$(0.0)\\
      \end{tabular} \\
      \hline

      \begin{tabular}{cc} 
          \includegraphics[width=0.23\linewidth]{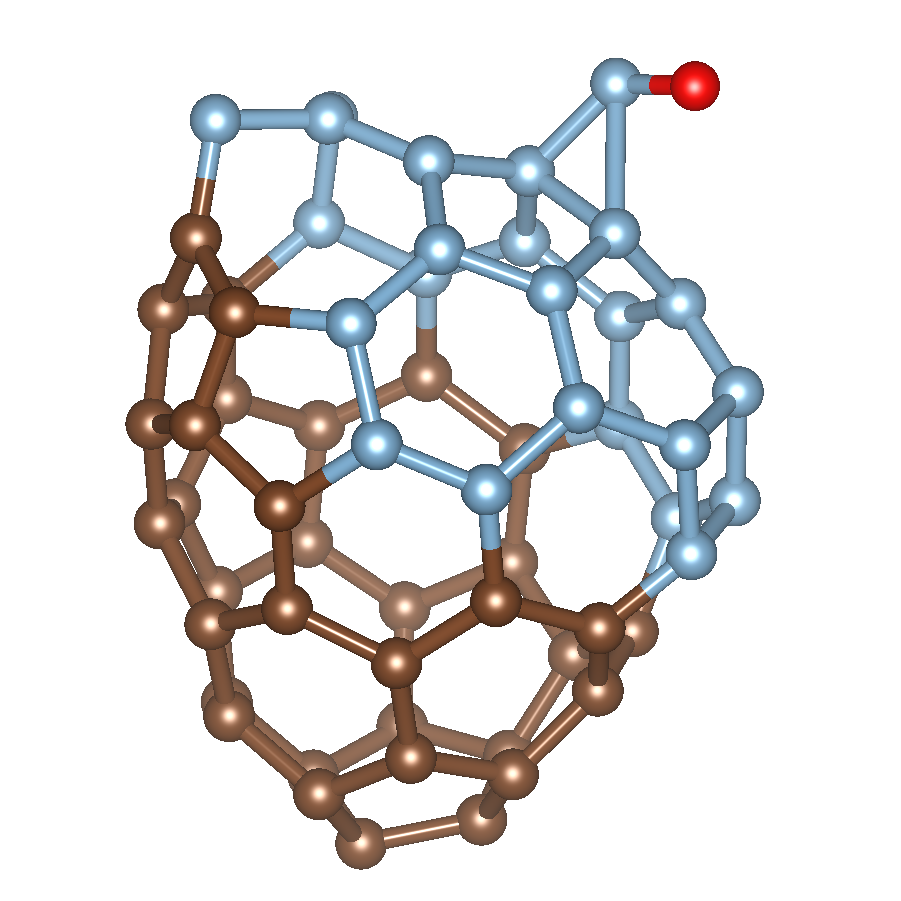} &
          \includegraphics[width=0.23\linewidth]{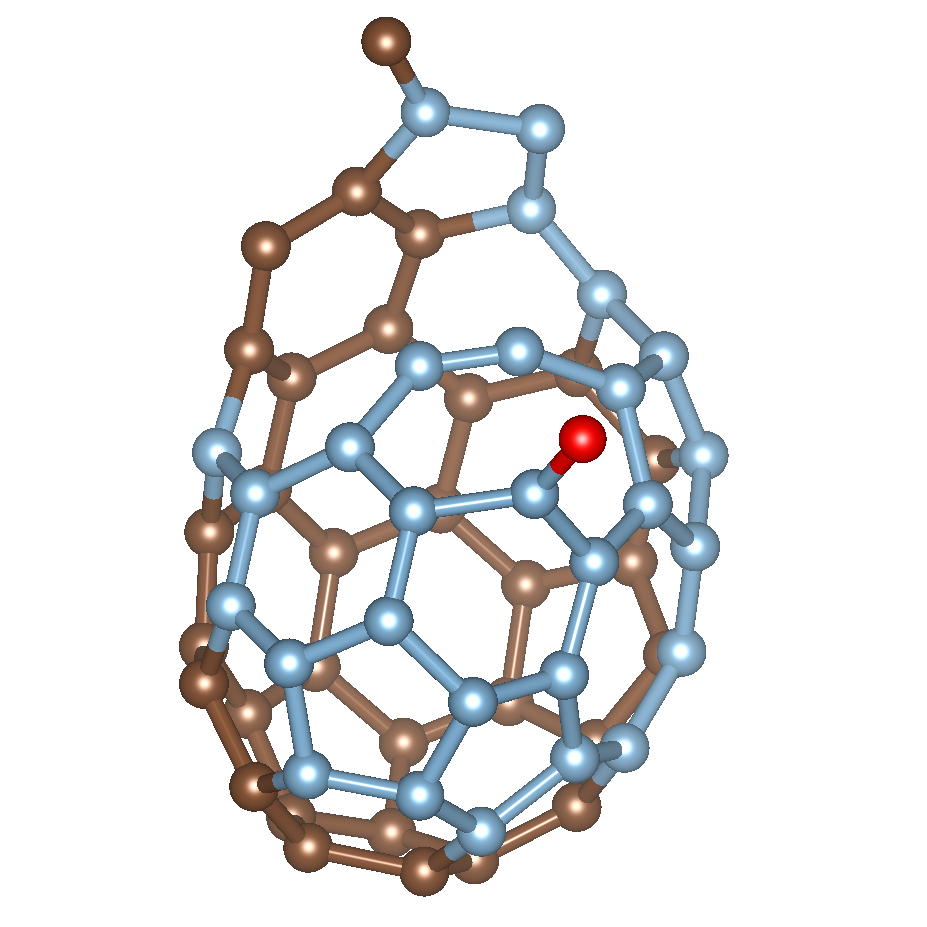}  \\
      \end{tabular} &
        \begin{tabular}{cc} 
         \includegraphics[width=0.23\linewidth]{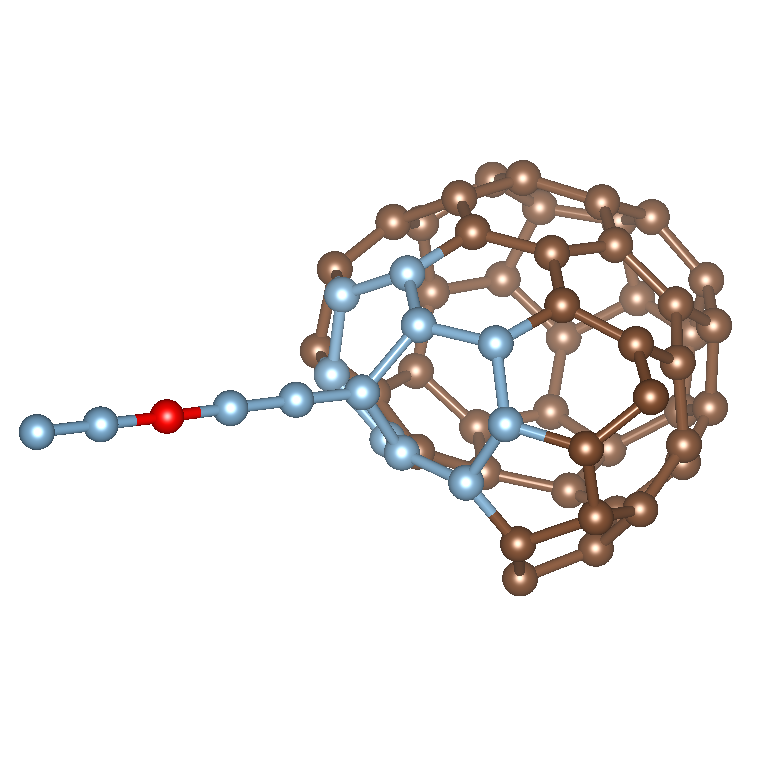} &
          \includegraphics[width=0.23\linewidth]{Tables_56_0219all.png} \\
          \end{tabular} \\
      \begin{tabular}{cc}
          \textbf{e}) $\Delta^{OM[sp]}=34$(0.11);&
          $\Delta^{SOAP}=43$(0.2);\\
          $\Delta^{ACSF}= 2$(0.0);&
          $\Delta^{FCHL}=17$(0.05);\\
          $\Delta^{MBSF}=8$(0.01)\\
      \end{tabular} &
      \begin{tabular}{cc}
          \textbf{f}) $\Delta^{OM[sp]}=79$(0.25);&
          $\Delta^{SOAP}=66$(0.31);\\
          $\Delta^{ACSF}=34$(0.04);&
          $\Delta^{FCHL}=46$(0.15);\\
          $\Delta^{MBSF}=13$(0.01)\\
      \end{tabular} \\
       \hline
        
      \begin{tabular}{cc} 
          \includegraphics[width=0.23\linewidth]{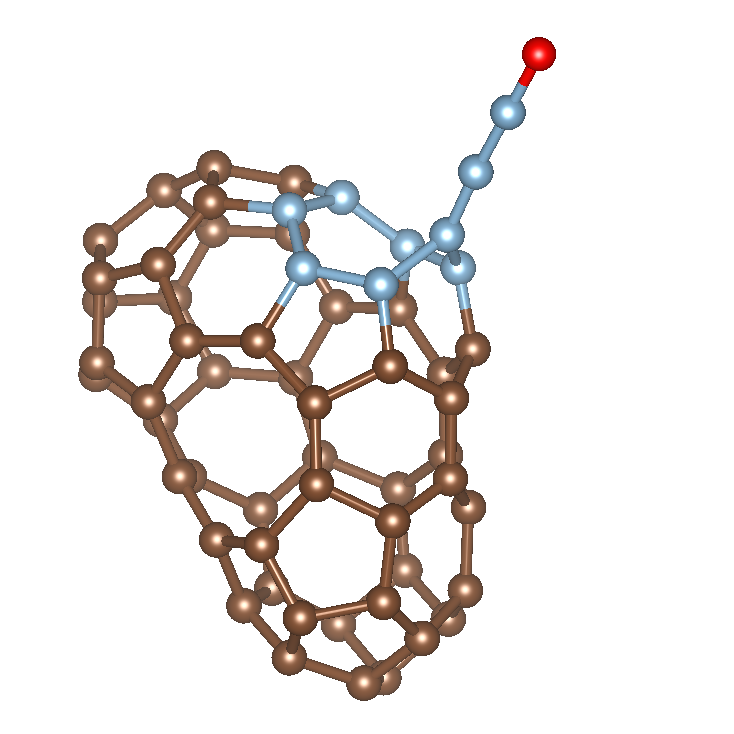}&
          \includegraphics[width=0.23\linewidth]{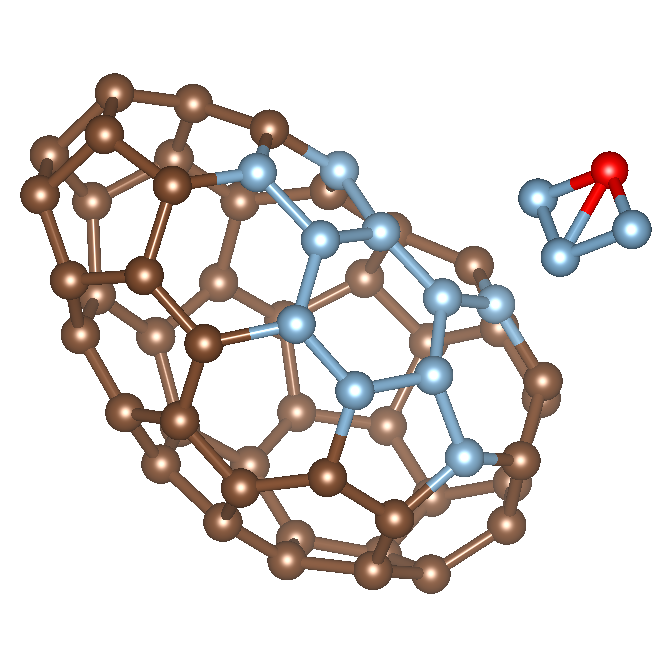} \\
      \end{tabular} &
       \begin{tabular}{cc} 
          \includegraphics[width=0.23\linewidth]{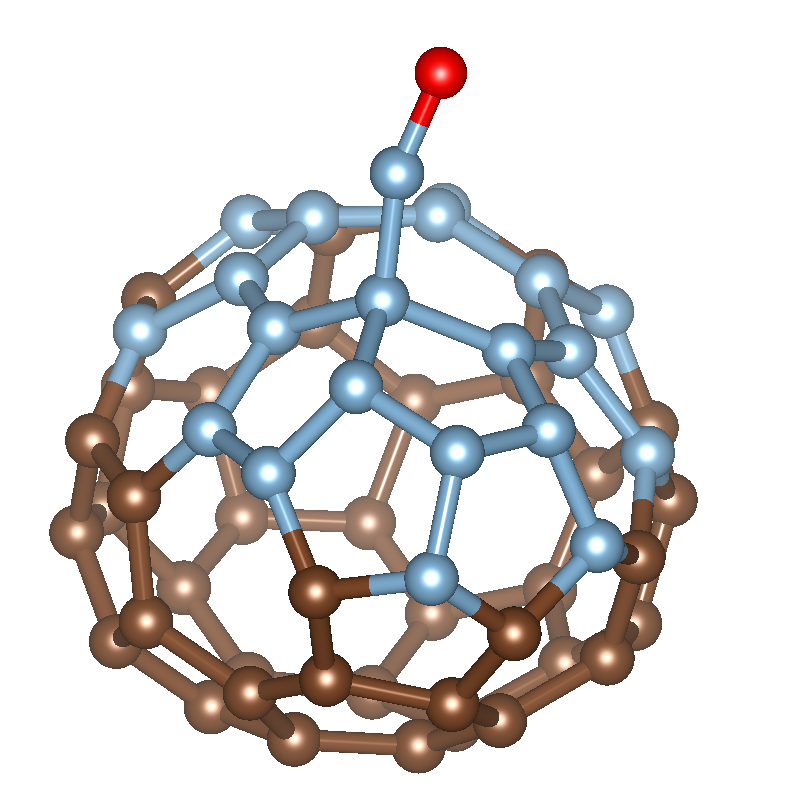} &
          \includegraphics[width=0.23\linewidth]{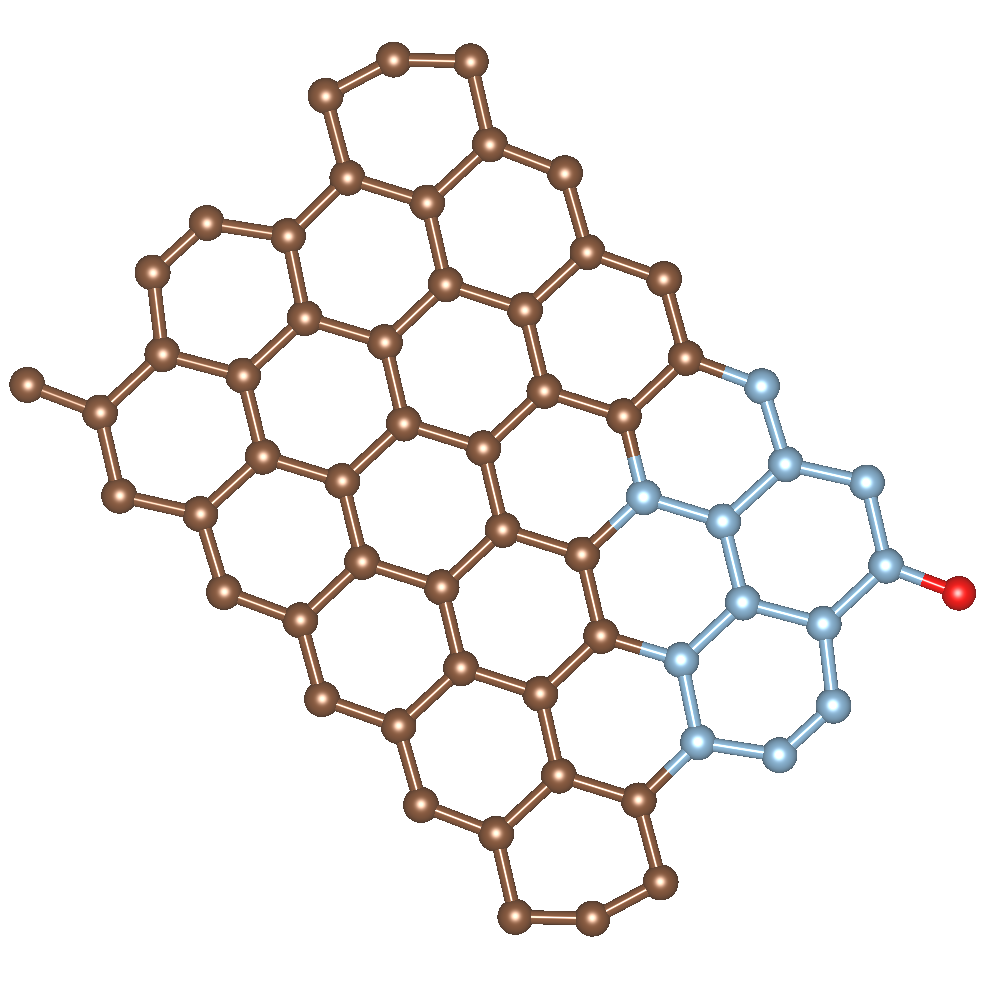} \\
      \end{tabular} \\ 
      \begin{tabular}{cc}
         \textbf{g}) $\Delta^{OM[sp]}=78$(0.25);&
          $\Delta^{SOAP}=79$(0.37);\\
          $\Delta^{ACSF}=22$(0.03);&
          $\Delta^{FCHL}=60$(0.19);\\
          $\Delta^{MBSF}=11$(0.01)\\
      \end{tabular} &
      \begin{tabular}{cc}
         \textbf{h}) $\Delta^{OM[sp]}=37$(0.12);&
          $\Delta^{SOAP}=74$(0.35);\\
          $\Delta^{ACSF}=7$(0.01);&
          $\Delta^{FCHL}=23$(0.07);\\
          $\Delta^{MBSF}=14$(0.01)\\
      \end{tabular} \\
      \hline
     
   
  \end{tabular}
  
  \caption{The most distinct atomic environments according to \textbf{a}) OM; \textbf{b}) SOAP, FCHL, and MBSF; and \textbf{c}) ACSF. The rest of the panels are problematic atomic environments in which one fingerprint predicts a large fingerprint distance whereas the other fingerprint predicts a small one.  The first number is the absolute 
  fingerprint distance whereas the number in parenthesis is the percentage of the largest 
  distance. The 
reference atom whose environment we want to describe, is red colored, 
  the atoms in the vicinity of the 
 reference atom are blue colored and the remaining atoms 
  in the structure which are outside of the cutoff sphere and do not affect the fingerprint are shown 
  in brown.}
  \label{tab:compare0}
  \end{center}
  
\end{table*}

\begin{table*}[p!]
   \begin{center}

   \begin{tabular}{|c|c|}
   
      \hline
      
      \begin{tabular}{cc} 
          \includegraphics[width=0.23\linewidth]{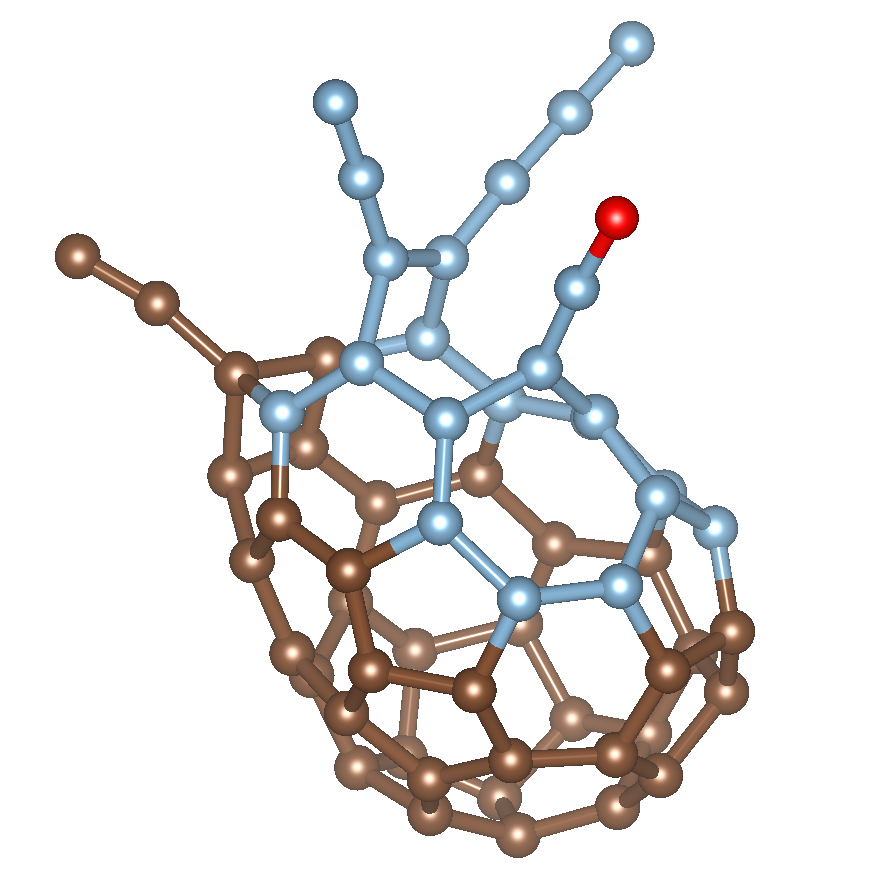}&
          \includegraphics[width=0.23\linewidth]{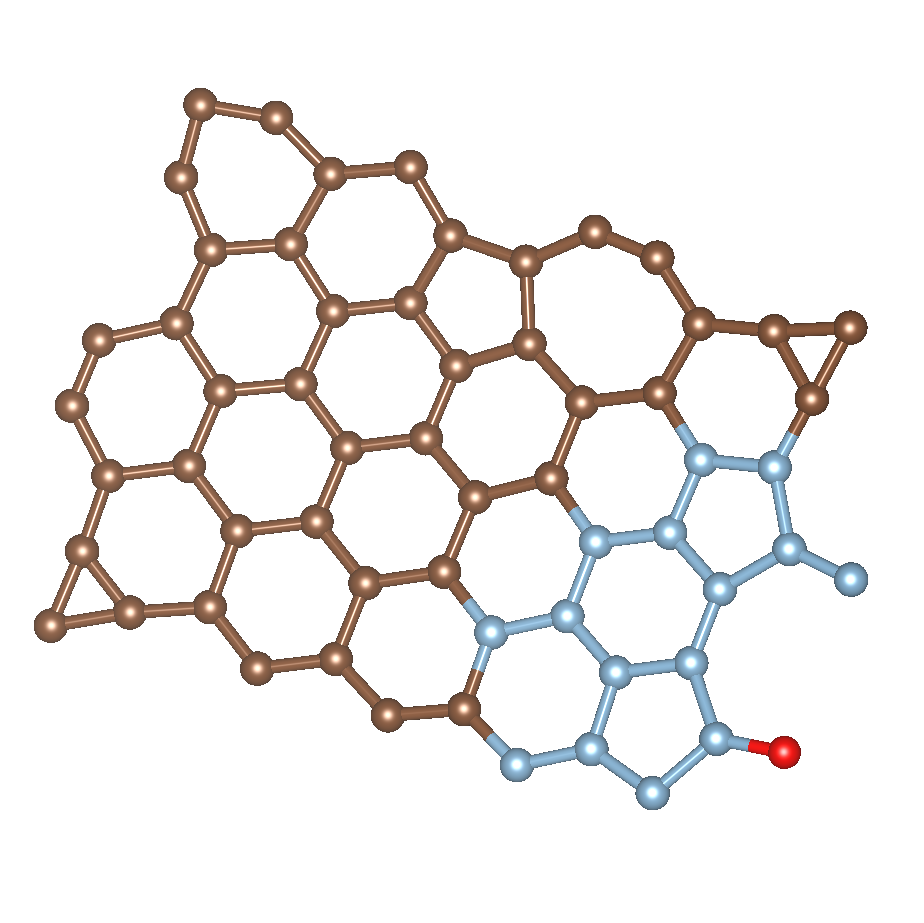} \\
      \end{tabular} &
        \begin{tabular}{cc} 
          \includegraphics[width=0.23\linewidth]{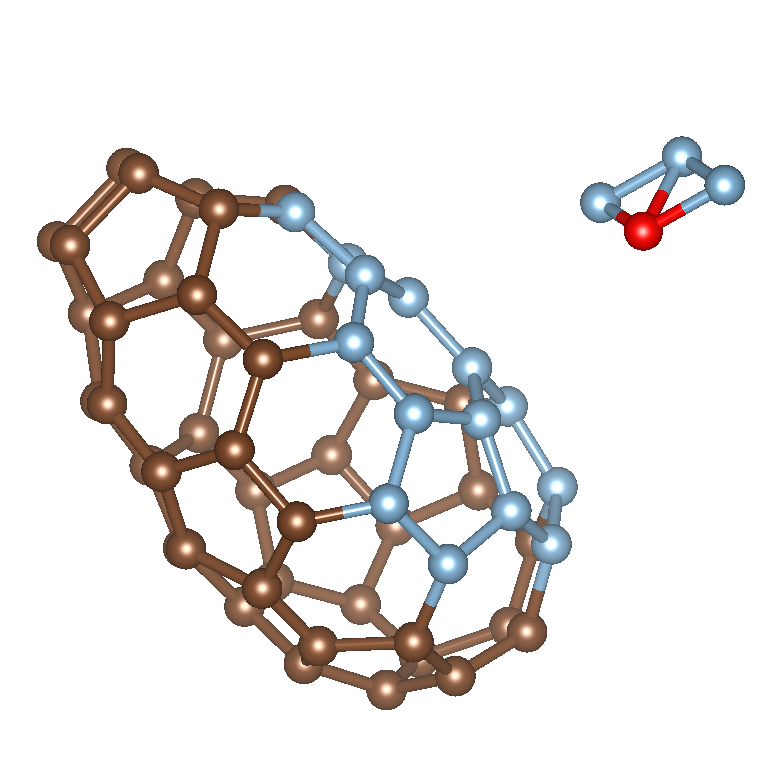} &
          \includegraphics[width=0.23\linewidth]{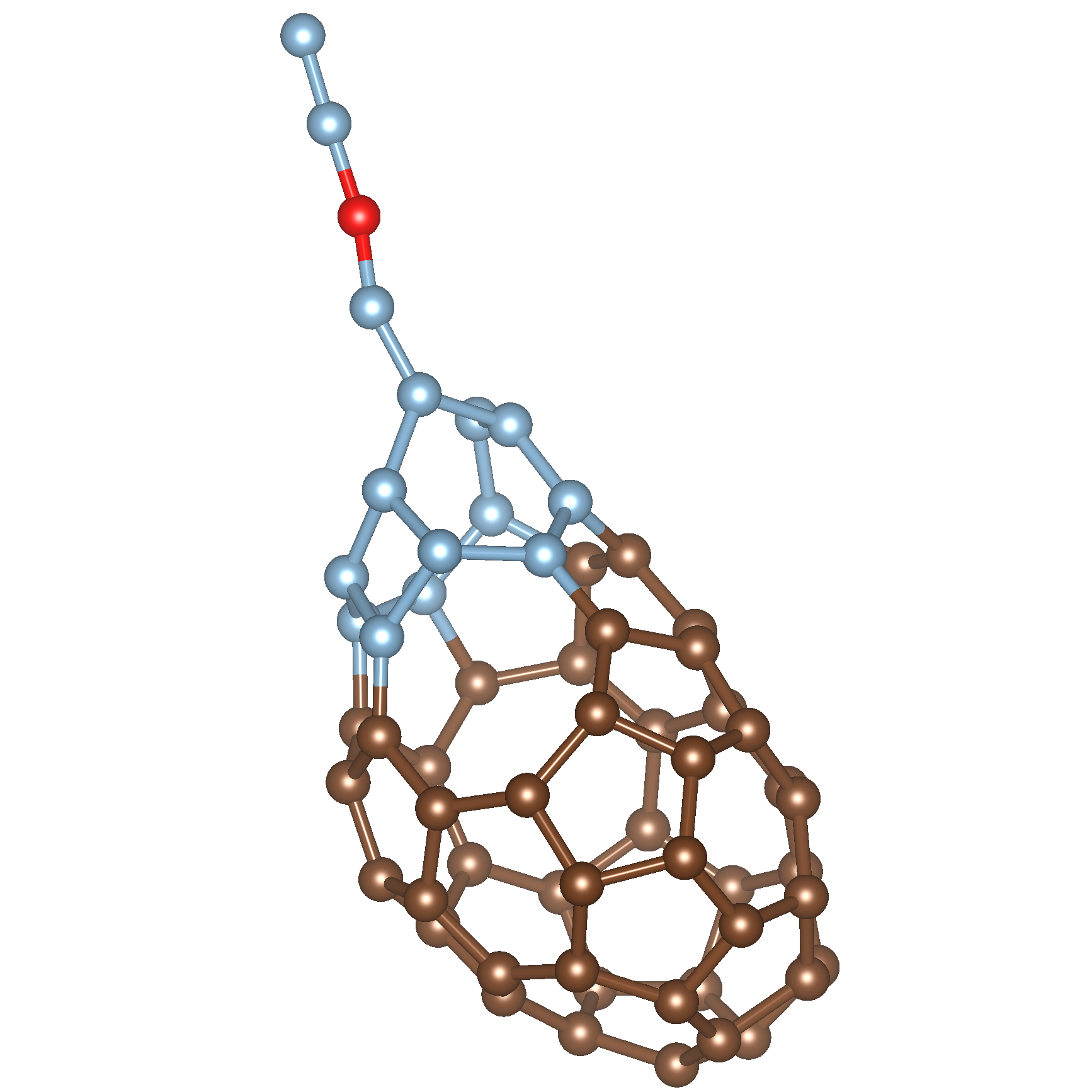} \\
      \end{tabular} \\
      \begin{tabular}{cc}
          \textbf{a}) $\Delta^{OM[sp]}=34$(0.11);&
          $\Delta^{SOAP}=74$(0.35);\\
          $\Delta^{ACSF}=6$(0.01);&
          $\Delta^{FCHL}=18$(0.06);\\
          $\Delta^{MBSF}=18$(0.01)\\
      \end{tabular} &
      \begin{tabular}{cc}
          \textbf{b}) $\Delta^{OM[sp]}=55$(0.18);&
          $\Delta^{SOAP}=85$(0.40);\\
          $\Delta^{ACSF}=37$(0.05);&
          $\Delta^{FCHL}=35$(0.11);\\
          $\Delta^{MBSF}=7$(0.01)\\
      \end{tabular} \\
   
       \hline
  
      \begin{tabular}{cc} 
          \includegraphics[width=0.23\linewidth]{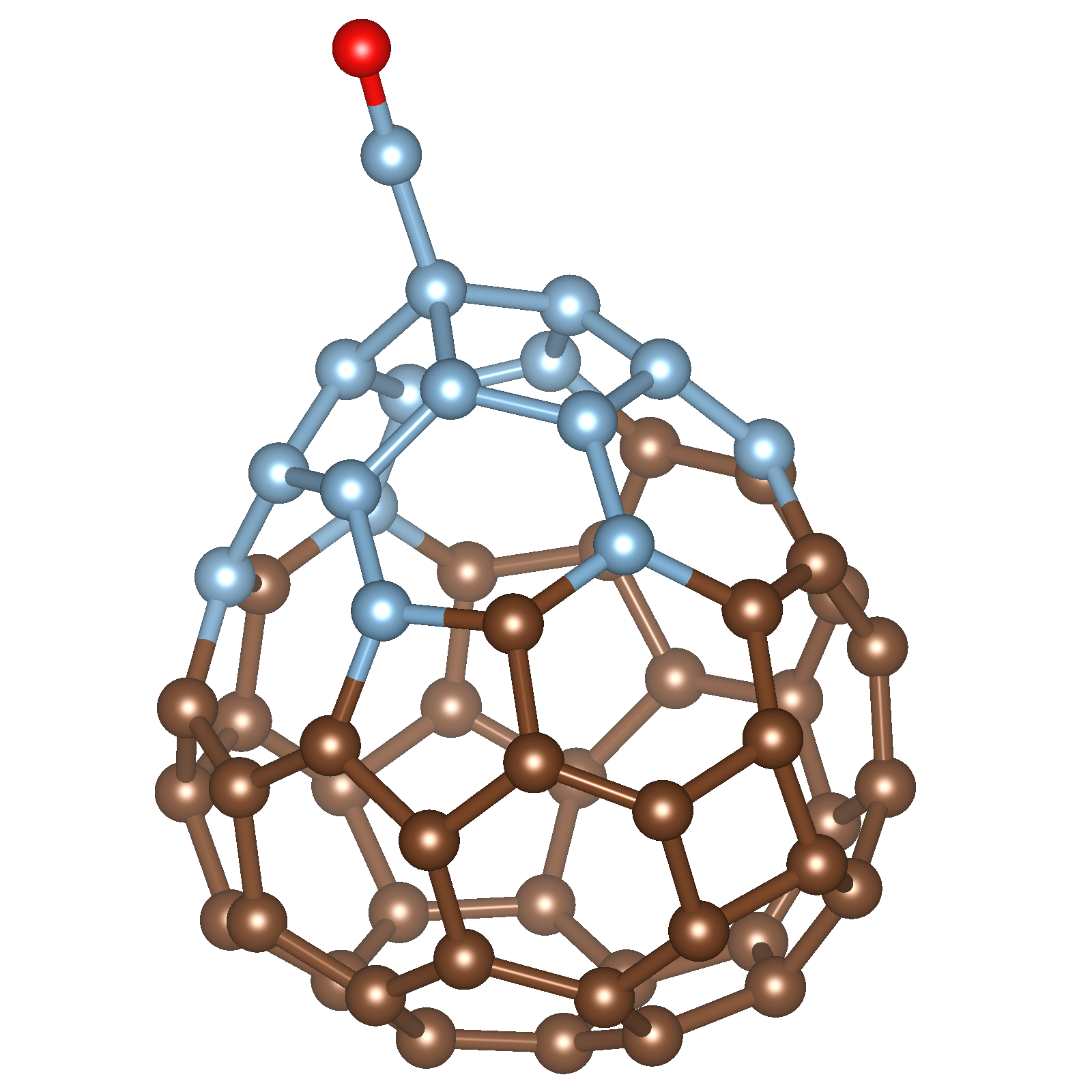} &
          \includegraphics[width=0.23\linewidth]{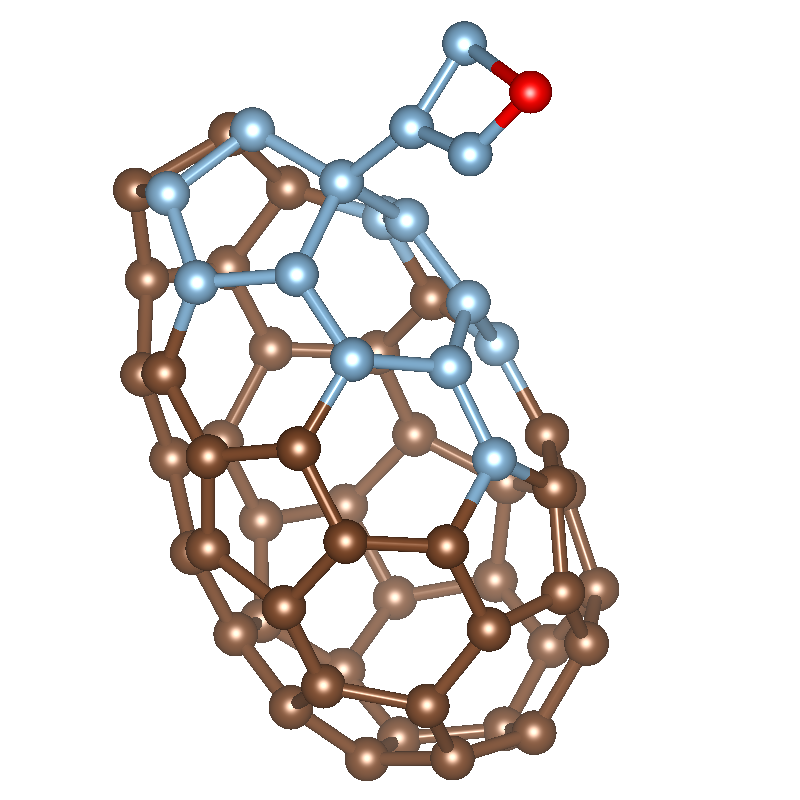} \\
      \end{tabular} &
       \begin{tabular}{cc}  
          \includegraphics[width=0.23\linewidth]{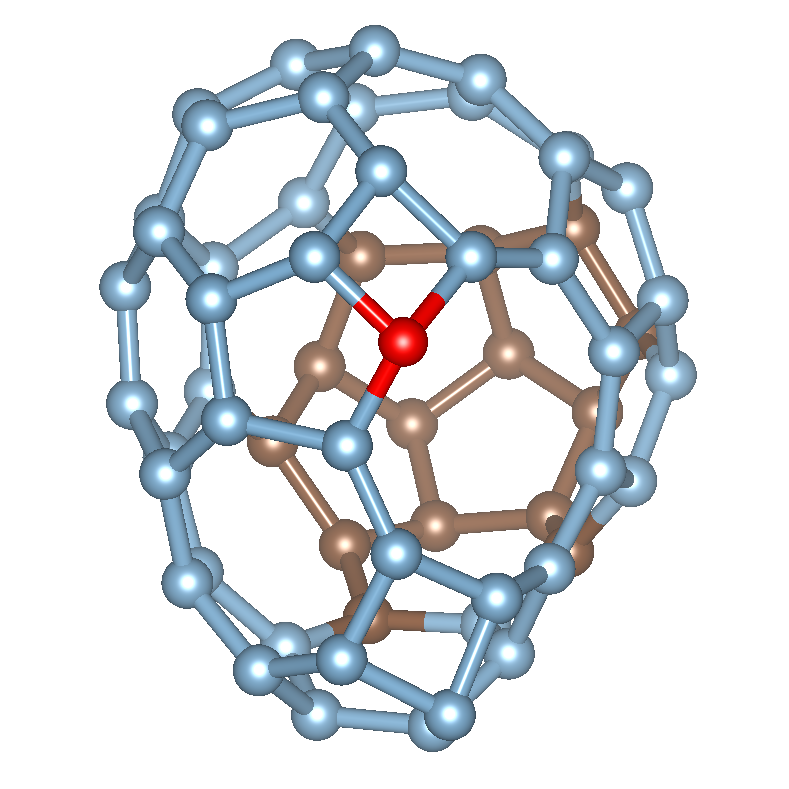} &
          \includegraphics[width=0.23\linewidth]{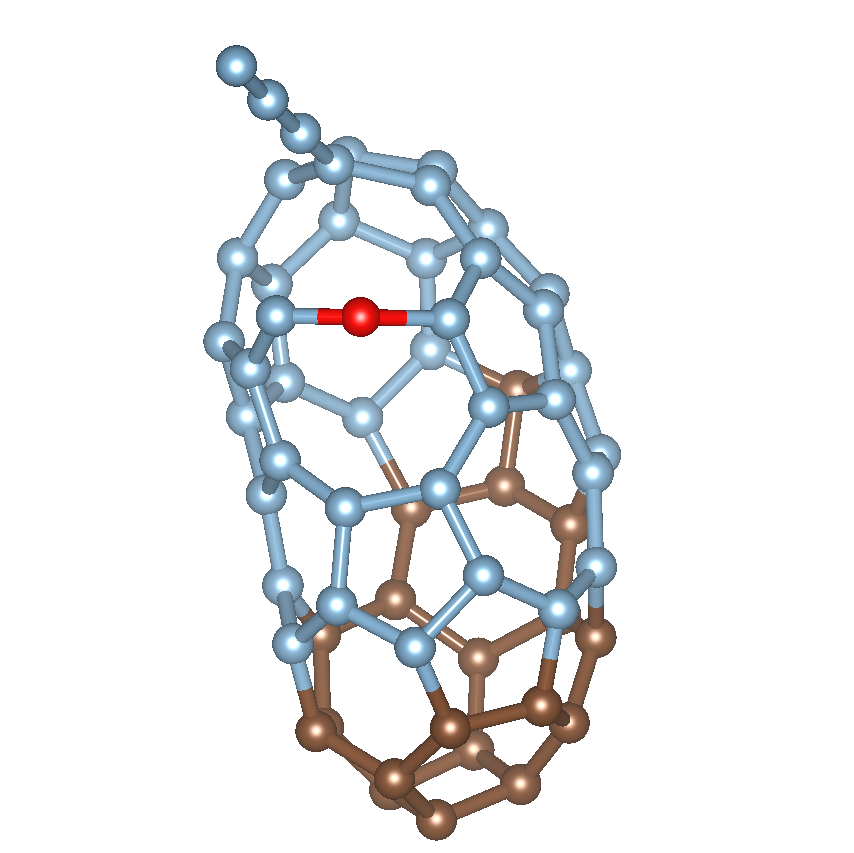} \\
      \end{tabular} \\
      \begin{tabular}{cc}
        \textbf{c}) $\Delta^{OM[sp]}=37$(0.12);&
          $\Delta^{SOAP}=73$(0.34);\\
          $\Delta^{ACSF}=16$(0.02);&
          $\Delta^{FCHL}=26$(0.08);\\
          $\Delta^{MBSF}=7$(0.01)\\
      \end{tabular} &
      \begin{tabular}{cc}
         \textbf{d}) $\Delta^{OM[sp]}=46$(0.15);&
          $\Delta^{SOAP}=60$(0.28);\\
          $\Delta^{ACSF}=8$(0.01);&
          $\Delta^{FCHL}=46$(0.15);\\
          $\Delta^{MBSF}=25$(0.02)\\
      \end{tabular} \\
       \hline
      
     
      \begin{tabular}{cc}  
          \includegraphics[width=0.23\linewidth]{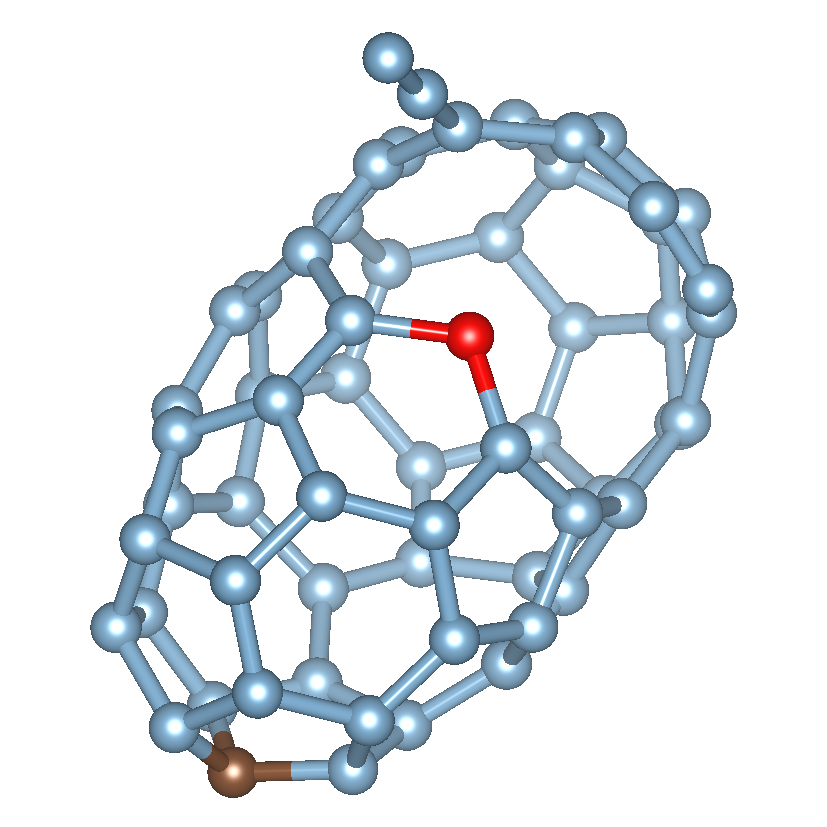}&
          \includegraphics[width=0.23\linewidth]{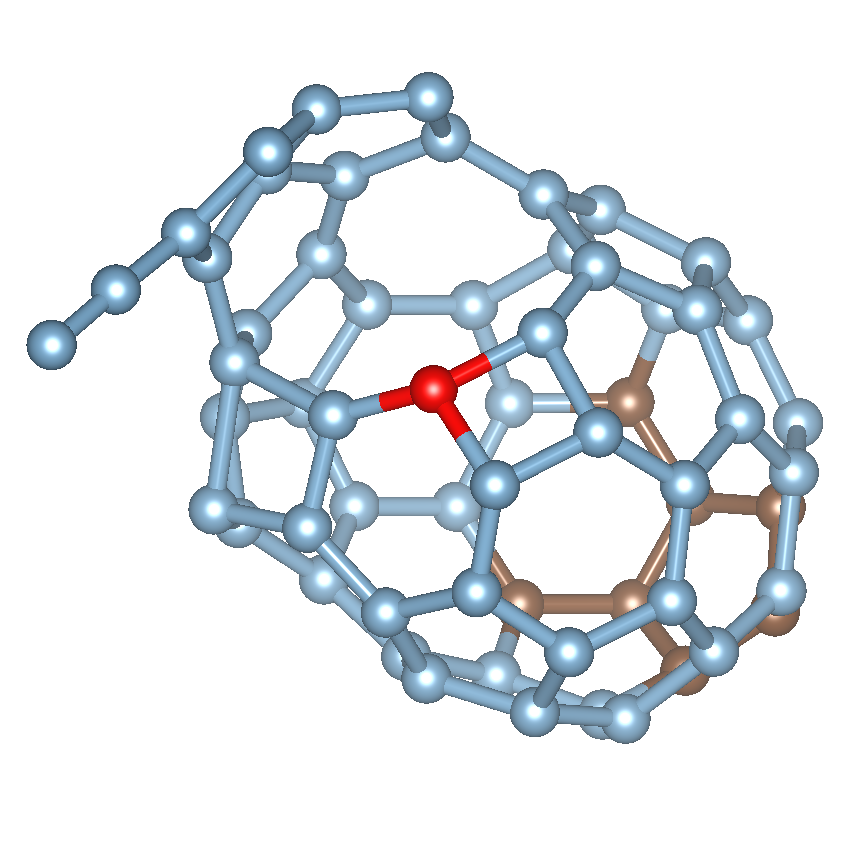} \\
      \end{tabular} &
       \begin{tabular}{cc}  
          \includegraphics[width=0.23\linewidth]{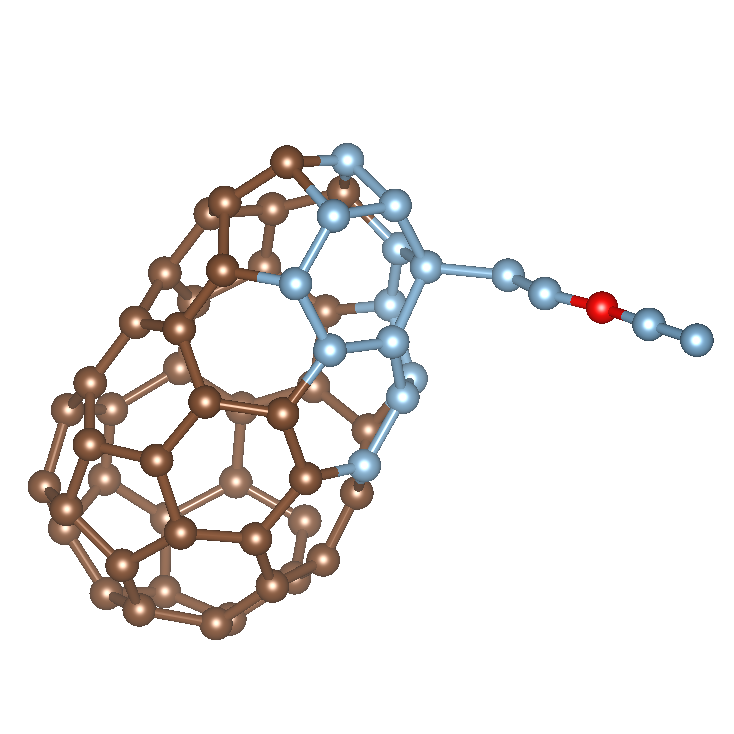} &
          \includegraphics[width=0.23\linewidth]{Tables_27_0844all.png} \\
      \end{tabular} \\
      \begin{tabular}{cc}
         \textbf{e}) $\Delta^{OM[sp]}=36$(0.12);&
          $\Delta^{SOAP}=50$(0.24);\\
          $\Delta^{ACSF}=8$(0.01);&
          $\Delta^{FCHL}=44$(0.14);\\
          $\Delta^{MBSF}=29$(0.02)\\
      \end{tabular} &
        \begin{tabular}{cc}
          \textbf{f}) $\Delta^{OM[sp]}=52$(0.16)&
          $\Delta^{SOAP}=76$(0.36);\\
          $\Delta^{ACSF}=28$(0.4);&
          $\Delta^{FCHL}=33$(0.11);\\
          $\Delta^{MBSF}=5$(0.0)\\
      \end{tabular} \\
      \hline
      
      \begin{tabular}{cc}   
          \includegraphics[width=0.23\linewidth]{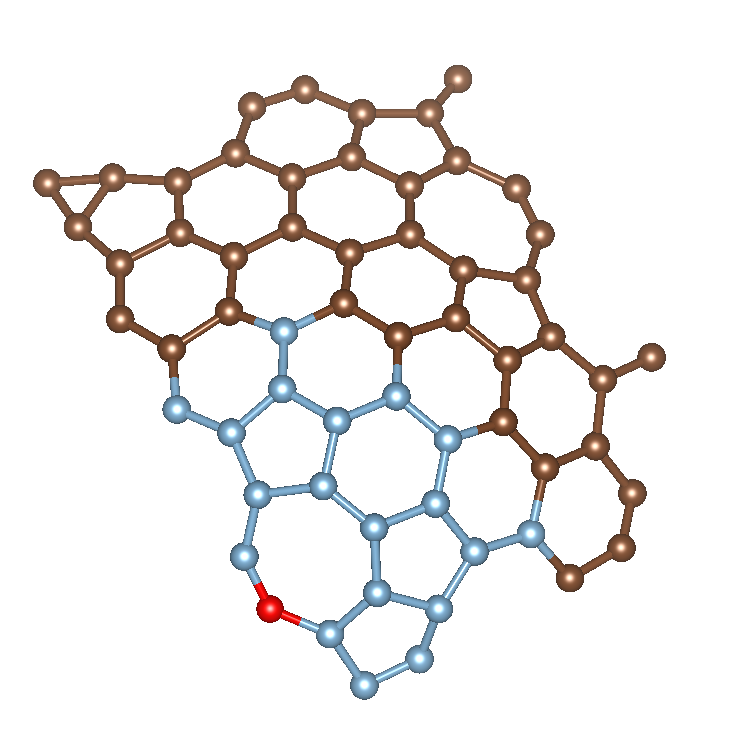} &
          \includegraphics[width=0.23\linewidth]{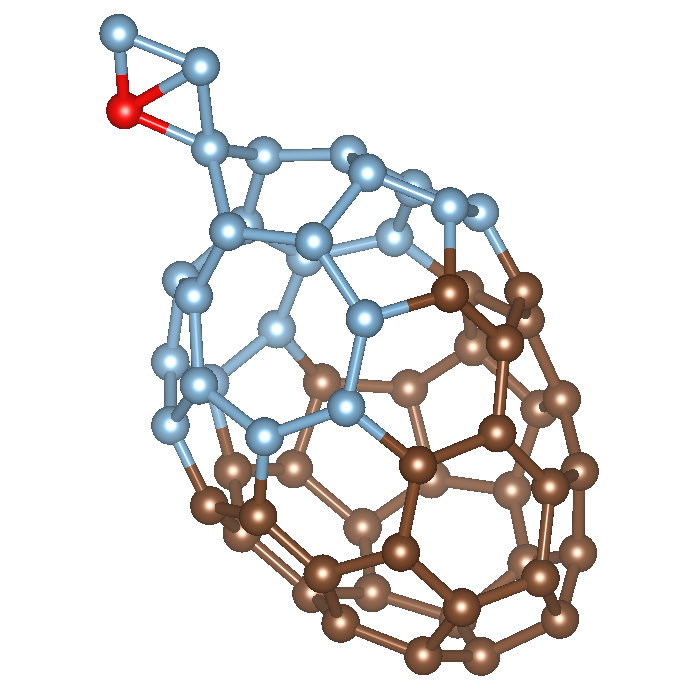} \\
      \end{tabular} &
     \begin{tabular}{cc}  
          \includegraphics[width=0.23\linewidth]{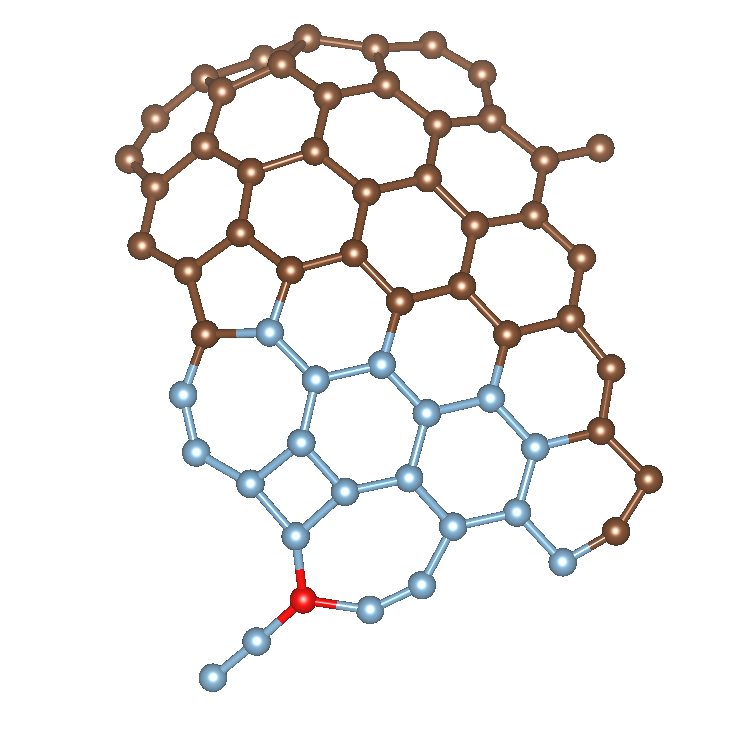} &
          \includegraphics[width=0.23\linewidth]{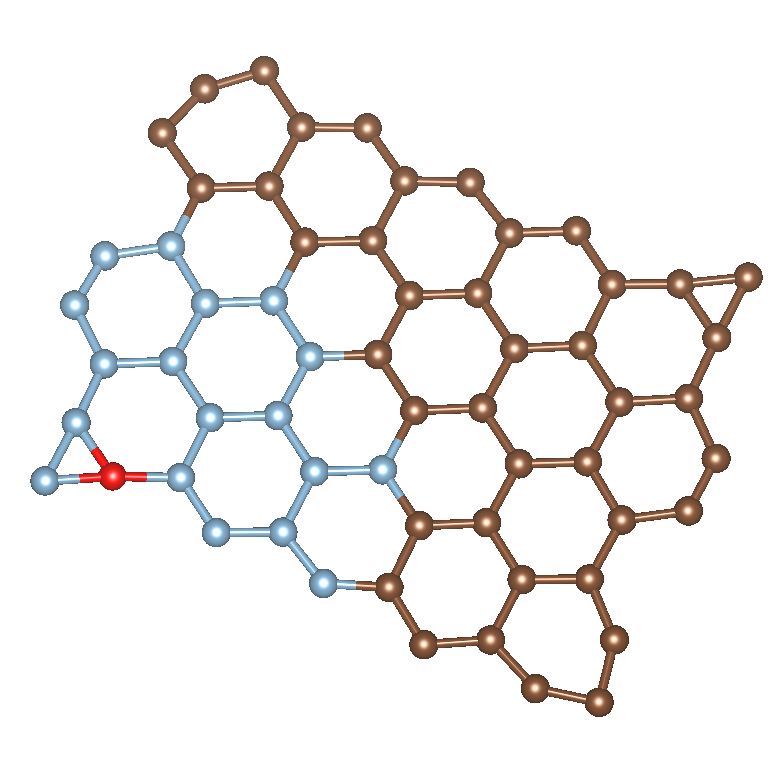} \\
      \end{tabular} \\
      \begin{tabular}{cc}
         \textbf{g}) $\Delta^{OM[sp]}=34$(0.11);&
          $\Delta^{SOAP}=43$(0.20);\\
          $\Delta^{ACSF}=14$(0.02);&
          $\Delta^{FCHL}=31$(0.10);\\
          $\Delta^{MBSF}=5$(0.0)\\
      \end{tabular} &
      \begin{tabular}{cc}
          \textbf{h}) $\Delta^{OM[sp]}=21$(0.07);&
          $\Delta^{SOAP}=34$(0.16);\\
          $\Delta^{ACSF}=15$(0.02);&
          $\Delta^{FCHL}=29$(0.09);\\
          $\Delta^{MBSF}=5$(0.0)\\
      \end{tabular} \\
       \hline
   
  \end{tabular}
  
  \caption{Further problematic environments.}
  
  \label{tab:compare1}
  \end{center}
  
\end{table*}

 Fig.~\ref{fig:intplots} \textbf{a} shows the resolution plot between the OM and SOAP fingerprints. 
	In this case, both OM[sp] and SOAP fingerprints agree quite well on similarities and dissimilarities between the environments.  
 
 Fig.~\ref{fig:intplots} \textbf{b} shows the resolution intensity plot between OM[sp] and ACSF.
 There exist some points with significant values on the OM[sp] axis. These points represent different environments where ACSF 
 cannot resolve the differences between them since the ACSF FP distance is close to zero. 
 In Table~\ref{tab:compare0} \textbf{d} we show two atomic environments which are obviously quite different, but whose 
 ACSF distance is very small. The two environments are very different since the central atom in the left panel makes one bond 
 with its nearest neighbor while the central atom in the right panel is two-fold coordinated. 
 In Table~\ref{tab:compare0} \textbf{e} we also show another example where the difference vectors of the ACSF are rather small.

 Fig.~\ref{fig:intplots} \textbf{c} shows the correlation intensity plot between OM[sp] and FCHL. There is not any point on the axes with significant values. So both fingerprints agree on similarities.
 
 Fig.~\ref{fig:intplots} \textbf{d} shows the correlation plot between OM[sp] and MBSF.
In Table~\ref{tab:compare0} \textbf{f} and  \textbf{g} 
we show two examples in which the MBSF does not recognize the differences between the two environments. 
In Table~\ref{tab:compare0} \textbf{f} left, the central environment is in the middle of the chain and has two nearest neighbors while on right, it is at the end of the chain and has one nearest neighbor. In Table~\ref{tab:compare0} \textbf{g} left, the 
reference atom is again at the end of a chain while on right it is three-fold coordinated. 

 Fig.~\ref{fig:intplots} \textbf{e} shows the correlation intensity between SOAP and ACSF. We can also see problematic points where 
 the fingerprint distance is very small according to ACSF but not according to SOAP. 
 In Table~\ref{tab:compare0} \textbf{h} 
 we show an example of two different environments where ACSF 
 predicts 
 a very small fingerprint distance. Although the central atom in both cases have one nearest neighbour, but the second and third shells are different.
 Table~\ref{tab:compare1} \textbf{a} shows another example in which ACSF does not recognize the differences in the local environment. 
 
 The correlation intensity between SOAP and FCHL is shown in Fig.~\ref{fig:intplots} \textbf{f}. There isn't any point on either axes 
 with significant values and both fingerprints therefore agree on similarities and differences between environments.  
 
 Correlation intensity between SOAP and the MBSF is shown in Fig.~\ref{fig:intplots} \textbf{g}. 
 There exist again some problematic points on the SOAP axis which indicates that there are some different environments 
 that MBSF predicts to be the same or very similar. 
 In Table~\ref{tab:compare1} \textbf{b} and \textbf{c} we show two such examples.


 The correlation intensity between ACSF and FCHL is shown in Fig~\ref{fig:intplots} \textbf{h}. There are also some points 
 lying on and very close to the FCHL axis (points with fingerprint distances up to 50 near the FCHL axis). These points indicate environments which are different according to FCHL and very 
 similar according to ACSF. In Table~\ref{tab:compare1} \textbf{d} and \textbf{e} we show two such examples where the two environments are different while fingerprint distance according to ACSF is very small. The 
 reference atom is in one case two-fold coordinated while it is three-fold coordinated in the other case.
 
 In Fig.~\ref{fig:intplots} \textbf{i} we show the correlation intensity between ACSF and the MBSF. 
 The two fingerprint agree on most similarities and there are no points on axes with significant values. 
 
 As a last illustration we show the correlation plot between the MBSF and FCHL in Fig.~\ref{fig:intplots} \textbf{j}. 
 In Table~\ref{tab:compare1} \textbf{f}, \textbf{g}, and \textbf{h} we show examples where the MBSF does not recognize differences between the local environments and predicts very small fingerprint distances compared to FCHL. 
{\color{black} To summarize, our analysis of the eigen modes of the sensitivity matrix 
shows that ACSF, MBSF, and partly FCHL are quite insensitive to certain displacements of the neighbouring atoms and have in this way an unsatisfactory structural resolution power. 
SOAP and OM perform significantly better in this respect. }

\section{Correlation between molecular fingerprints and global physical properties}

\begin{figure*}[ht!]
     \centering
     \begin{subfigure}[b]{\columnwidth}
         \centering
         \includegraphics[width=\textwidth]{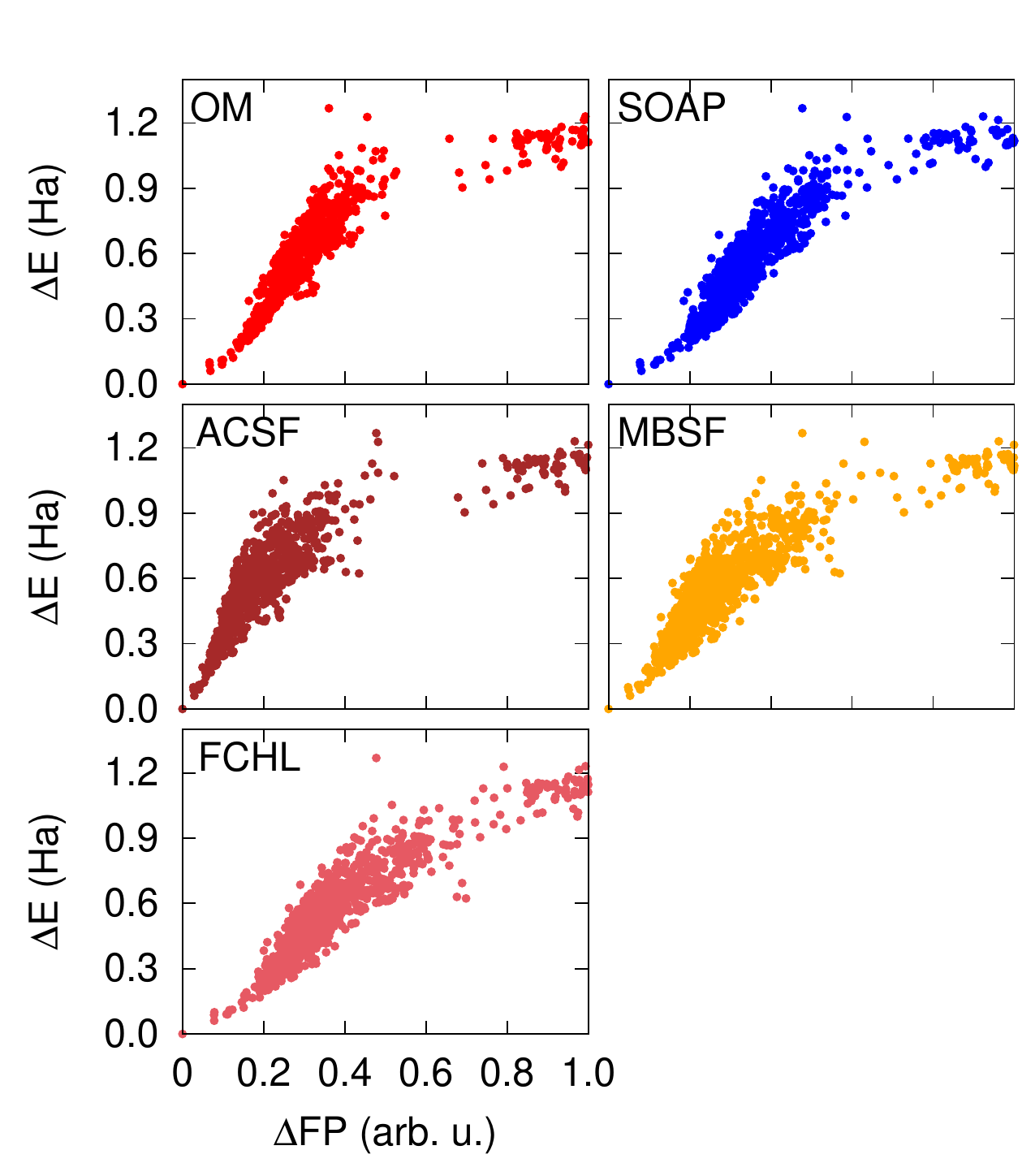}
         \caption{$\Delta FP$ vs. $\Delta E$}
         \label{fig:fp_vs_e}
     \end{subfigure}
      \begin{subfigure}[b]{\columnwidth}
         \centering
         \includegraphics[width=\textwidth]{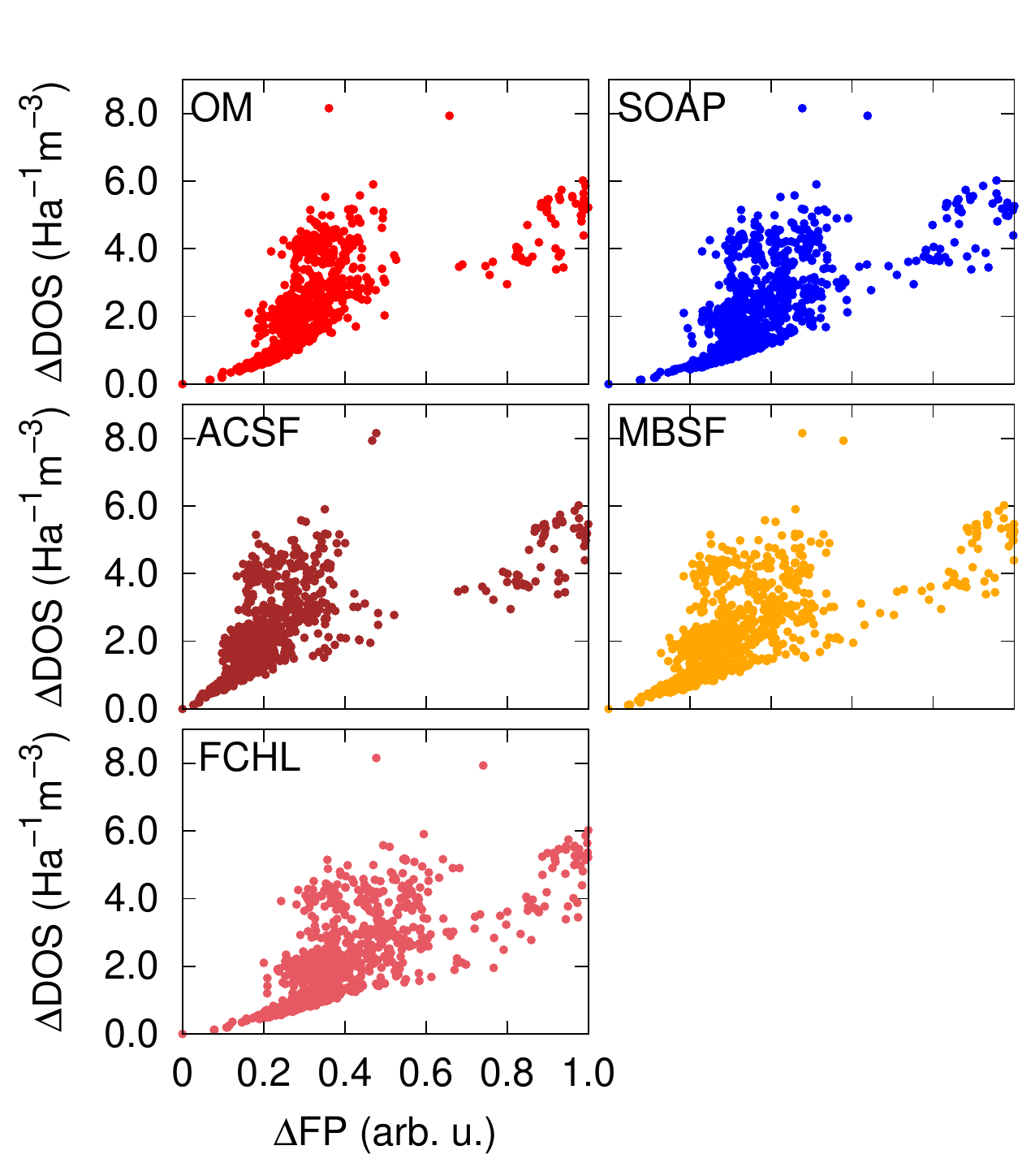}
         \caption{$\Delta FP$ vs. $\Delta DOS$}
         \label{fig:fp_vs_dos}
     \end{subfigure}
      \caption{The correlation between molecular fingerprint distance and $\Delta E$ (left hand side) and $\Delta DOS$ (right hand side) for OM[sp], SOAP, ACSF, FCHL, and MBSF. The global minimum of $C_{60}$ is taken as the reference structure. The fingerprint distances are scaled such that the maximum fingerprint distance for each fingerprint is 1.0}
     \label{fig:fig7and8}
\end{figure*}

According to our analysis reported above several fingerprints that are widely and successfully used for instance in machine learning schemes are apparently sometimes unable
to distinguish between different chemical environments. One would thus expect that this gives rise to errors in the prediction of 
physical properties. 
One typical application that in principle could be affected is the development of machine learning potentials~\cite{P4885}, which predict the energy and forces as a function of the atomic positions. Most of these ML potentials rely on a construction of the total energy as a sum of environment-dependent atomic energies~\cite{behler2007generalized,bartok2010gaussian,P4862} and thus should be sensitive to deficiencies in the discrimination of these environments. In this section we will discuss possible implications of our findings with respect to such applications of ML.

For our investigation, we need to distinguish between local and global properties. While local properties like forces are observables that can be uniquely assigned to individual atoms, the total energy of the system is not an observable, and there is no physically unique definition of atomic energies. While ML potentials are supposed to represent both, forces and energies, with high accuracy and consistently, their analysis requires different approaches.

We will now investigate the role of the total energy as a global property. It has been shown for instance for the distribution of atomic energies within extended systems~\cite{P5537}, that atomic energies determined by ML can compensate each other to yield the correct total energy if there is enough flexibility in the system. For many systems this flexibility can be reduced by adding constraints on the energy distribution in form of different stoichiometries~\cite{P5537}, but in general there is no way to extract unique atomic energies for arbitrary systems using ML. This finding is independent of the ability of the fingerprint vectors to  distinguish chemically inequivalent atomic environments.

Here, we now go one step further and investigate if even
a few "wrong" environment descriptions, which cannot resolve some structural differences as reported above, might be tolerable as the total energy could still be well represented due to some error cancellation. 
To check the correlation of global properties  with various atomic fingerprints we first have to construct a global, i.e. 
molecular fingerprint from our local atomic fingerprints. 
We do this by finding the optimal matching between all the atomic environments in the two structures~\cite{zhu2016fingerprint}, {\color{black} i.e. the matching that minimizes the root-mean-square distance (RMSD) between the two molecules~\cite{sadeghi2013metrics}}. 
In this approach the fingerprint distance between two molecules $p$ and $q$ is defined  as
\begin{equation}
	\Delta^{p,q}= \min_{P}{ \left( \sum_{i}^{N}{|\mathbf{F}_p^i-\mathbf{F}_q^{P(i)}}|^2 \right)} ^{1/2} \: 
\end{equation}
where $\mathbf{F}_p^i$ is the fingerprint vector for atom $i$ in configuration $p$  and 
$\mathbf{F}_q^{P(i)}$ is the fingerprint of the best matching atom $P(i)$ in configuration $q$. 
The permutation function $P$ which gives the best overall match is found with the Hungarian 
algorithm~\cite{kuhn1955hungarian} in polynomial time. We note, however, that this construction of a global molecular fingerprint is different from the procedure that is usually applied in the construction of ML potentials, and here we use it primarily as a tool to detect correlations between global properties and the entire structure of a system.
 
 While the atomic fingerprint distance shows how different two atomic environments are, the molecular fingerprint distance indicates 
 the difference between two entire  molecules. In the next step, we calculate the correlation between molecular fingerprints  
 and two global properties, namely the total energy and the density of states (DOS). 
 If two molecules have different energies or DOS's, they have to be different and so the fingerprint distance should be non-zero.
On the other hand, if two molecules have nearly the same energies or  DOS they can be similar (in case of degeneracy) or different. 
So the fingerprint distance does not need to be necessarily non-zero.

The density of states for molecule $p$, $D_p(\epsilon)$ is
\begin{equation}
  D_p(\epsilon)=\sum_{i}{\delta(\epsilon-\epsilon_i^p)}
\end{equation}
where $\epsilon_i^p$ are the Kohn-Sham eigenvalues for molecule $p$.
We replace $\delta(\epsilon-\epsilon_i^p)$ with $\frac{1}{\sqrt{2\pi \sigma^2}} \exp \left(\frac{-(\epsilon-\epsilon_i^p)^2}{2\sigma^2}\right)$ with $\sigma$ some smearing parameter. We define the difference between the density of states to be:
\begin{equation}
  \Delta DOS_{p,q}=\sqrt{\int d\epsilon  (D_p(\epsilon)-D_q(\epsilon))^2}
\end{equation}
Taking advantage of the properties of Gaussian functions, we can calculate the integral analytically. Hence, $\Delta DOS_{p,q}$ can be calculated as 
\begin{widetext}
 \begin{equation}
   \Delta DOS_{p,q}=\sqrt{\sum_{i,j} {\left( e^{-(\epsilon_i^p-\epsilon_j^p)^2/4\sigma^2}+ e^{-(\epsilon_i^q-\epsilon_j^q)^2/4\sigma^2} \right. \left. -e^{-(\epsilon_i^p-\epsilon_j^q)^2/4\sigma^2}-e^{-(\epsilon_i^q-\epsilon_j^p)^2/4\sigma^2} \right) }}
 \end{equation}
\end{widetext}

We chose $\sigma=0.01$~Ha in this work.
The molecule with the lowest energy is taken as reference structure and fingerprint distances and energy differences are 
calculated with respect to it. In Fig.~\ref{fig:fig7and8} we see the correlation between the molecular fingerprint distance $\Delta FP$ and 
$\Delta E$ and $\Delta DOS$ with respect to the global minimum for OM[sp], SOAP, 
 ACSF, FCHL, and MBSF. 

Remarkably, all fingerprints show a quite similar behavior in these tests. 
In particular we could  not find any pair of molecules 
that has a very small molecular fingerprint distance, but different energy or DOS. 
As also noted in a study highlighting difficulties in the structural description of methane~\cite{pozdnyakov2020completeness}, the fingerprints 
of neighboring atoms usually change under displacements 
even if the fingerprint of the central atom remains
invariant. Through this effect machine learning schemes 
may compensate the deficiencies 
of a fingerprint, and the quality of the machine learning results for global quantities
based on different fingerprints can become very similar in practice.

\begin{figure*}[pth!]
    \centering
     \begin{subfigure}[b]{0.48\textwidth}
         \centering
         \includegraphics[width=\textwidth]{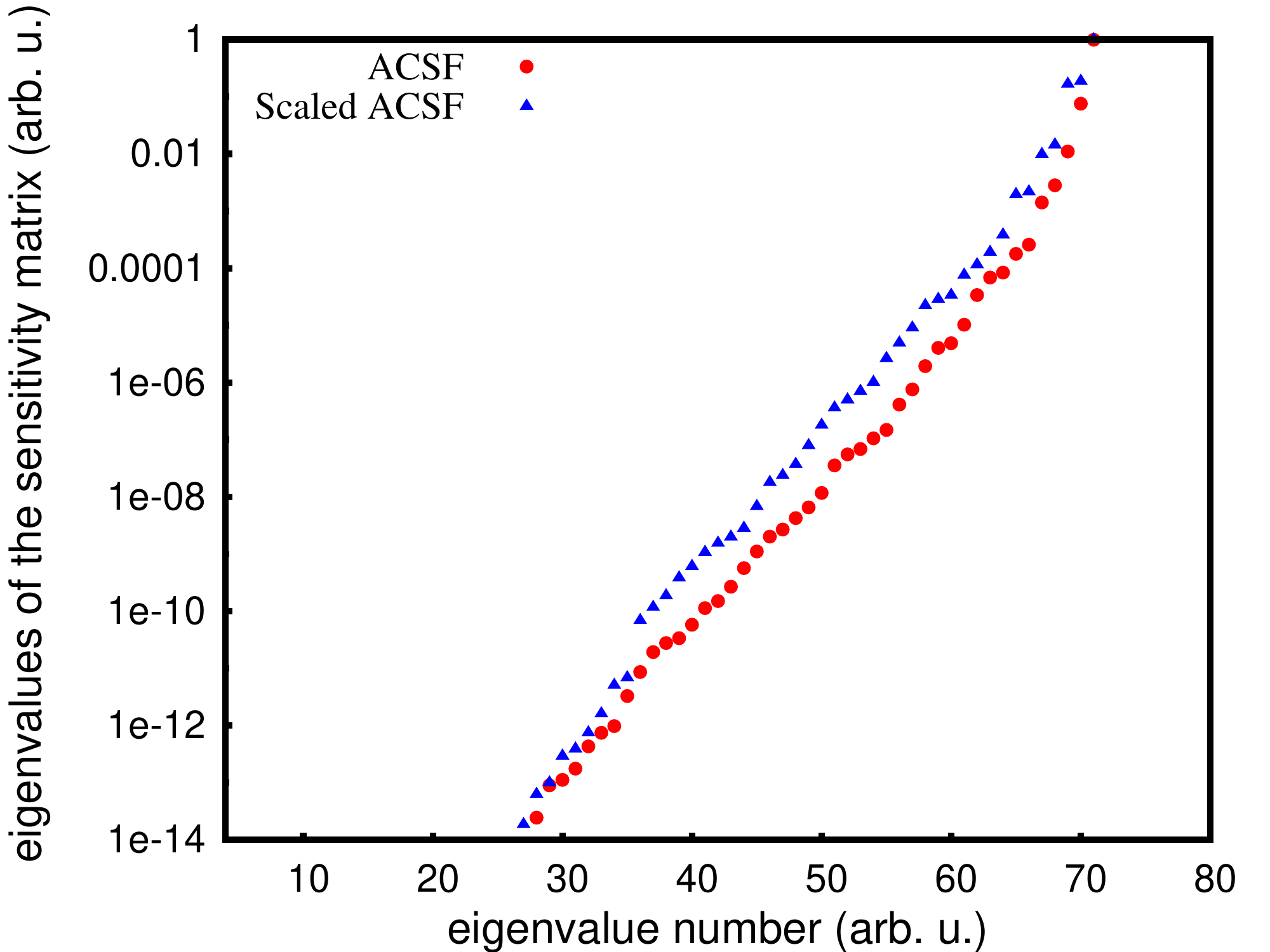}
         \caption{}
         \label{fig:1ascaledacsf}
     \end{subfigure}
     \begin{subfigure}[b]{0.48\textwidth}
         \centering
         \includegraphics[width=\textwidth]{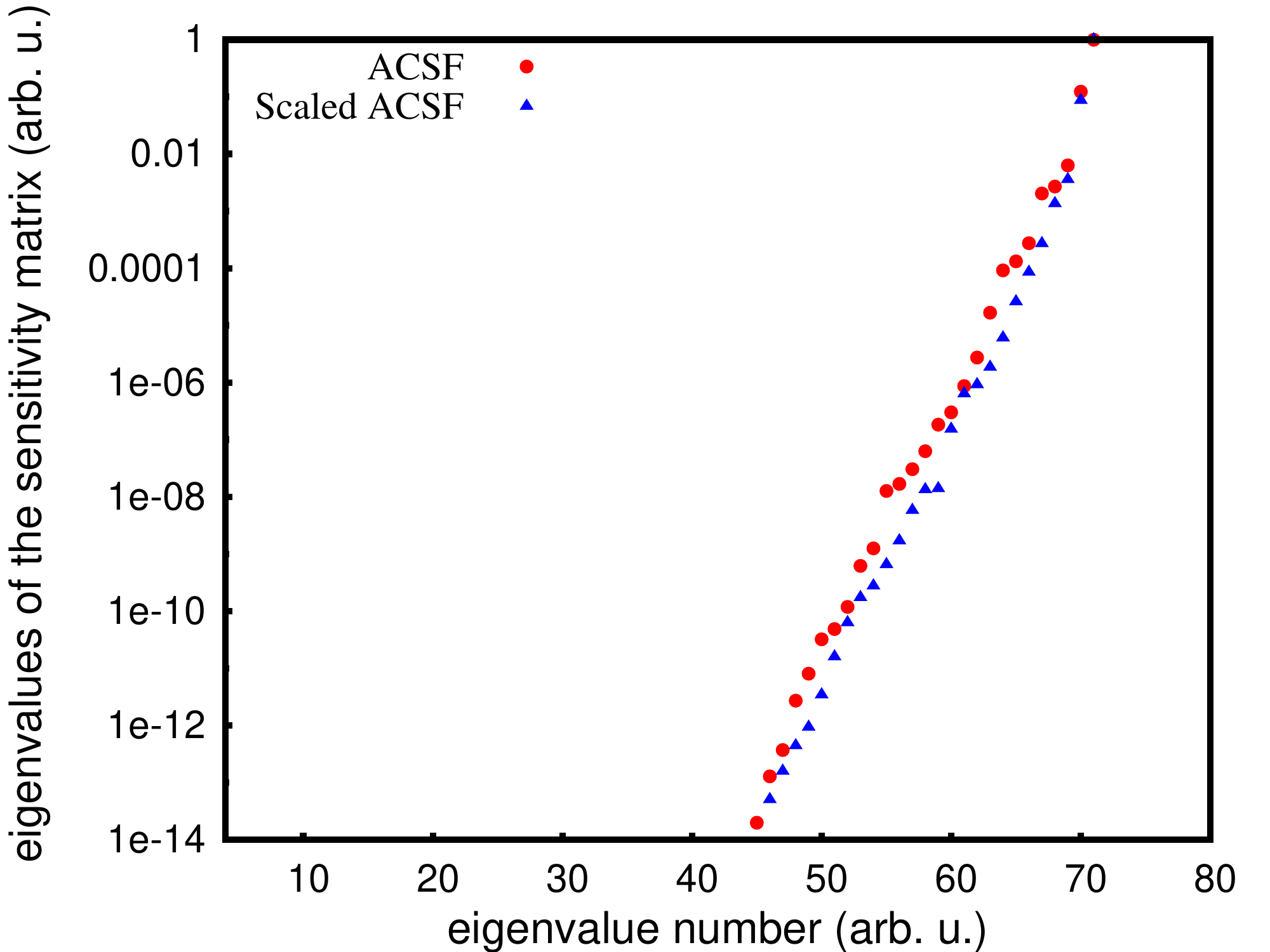}
         \caption{}
         \label{fig:1bscaledacsf}
     \end{subfigure}
    \caption{The eigenvalues of the sensitivity matrix for the ACSF vs. scaled ACSF in \textbf{a}: for the 
    reference atom of \ref{fig:conf} and in \textbf{b}: for the 
    reference atom in \ref{fig:conf22}.}
    \label{fig:scaledacsf}
\end{figure*}

However, these findings are strictly true only if fingerprint vectors of different environments are exactly the same and have to be treated with care in the context of machine learning for several reasons, if fingerprint vectors are only similar. While correlations between physical properties and fingerprints are certainly supporting the construction of a ML model, most ML algorithms are highly non-linear methods, which are able to distinguish fingerprint vectors even if they are overall very similar, as measured by the fingerprint difference, but are sufficiently different in at least one or a few components. For instance, this is the case for the ACSF fingerprint vectors of the reference atoms shown in Table~\ref{tab:compare0} \textbf{d}. In this case the radial symmetry functions with large $\eta$ parameters are rather sensitive to the local coordination and provide different numerical values for the exemplified one- and two-fold coordination of the reference atom. This is usually sufficient to distinguish these environments. Further, in ML applications fingerprint vectors are commonly scaled such that the values of each individual fingerprint component are normalized to a range between zero and one. We have not done this in the present work to avoid any bias in the comparison of the performance of different fingerprints. Further, any scaling, although common practice, depends on the fingerprint values in the available data set. 
We observed in Fig.~\ref{fig:scaledacsf} that scaling has some effect on ACSFs in terms of increasing the eigenvalues and therefore enhancing the sensitivity of the fingerprint overall, and similar effects are expected also for the other fingerprint types.

Finally, for instance in case of ML potentials, usually not only the total energy as a rather insensitive global property but also the atomic forces are used in the fitting process, which contain local atomic information about the potential energy surface. The inability to distinguish chemically different atomic environments thus results in large force errors, which can be used to improve the fingerprint set~\cite{behler2011atom}.

Irrespective of these aspects of ML applications, which reduce the effect of similar fingerprint vectors, it has been demonstrated in this work and elsewhere~\cite{pozdnyakov2020completeness}, that the detection of fingerprint vectors remaining exactly invariant upon structural changes is a major challenge and of utmost importance for many applications.

\section{Conclusions}

We have introduced stringent tests for the resolution power of atomic fingerprints describing the environment around a reference atoms.
First we introduced the  sensitivity matrix that can detect atomic displacement modes that leave the fingerprint invariant.
Based on a large data set of carbon structures we then 
investigated the correlation between fingerprint distances calculated 
with various fingerprints. For SOAP, ACSF, MBSF and FCHL, there exist 
atomic movements that leave the fingerprints invariant. This behavior can apparently only be found for 
some small molecules and it did not occur in our study of larger systems. 
For the symmetry function-related fingerprints, we found many movement modes that leave the fingerprint nearly invariant 
and we found many cases where environments that were classified as nearly identical were actually quite different.
In all the tests we saw an improvement when going from the ACSF and MBSF   
to the FCHL fingerprint. 
The OM fingerprint is the only fingerprint for which no atomic displacement was ever found that leaves 
the fingerprint invariant. It is also the fingerprint whose distance assignments corresponds best to basic chemical 
concepts. This comes from the fact that the OM fingerprint is obtained from a matrix diagonalization that 
is akin to the solution of the Schr\"{o}dinger equation and 
therefore naturally incorporates the full many-body character of the atomic environment.
However, the 
limited resolution of some atomic fingerprints for some environments 
is most critical for structural discrimination, while there is still a good correlation of global molecular fingerprints in case of the prediction of extensive 
properties such as total energies of systems that are composed of a large number of environments. 
Also applications like machine learning are less affected, as they are able to resolve even subtle differences in the fingerprints.

\section{acknowledgments} 
This research was performed within the NCCR MARVEL funded by the Swiss National Science Foundation. 
The calculations were done using the computational resources of the Swiss National Supercomputer (CSCS) 
under project s963 and on the Scicore computing center of the University of Basel. JB thanks the Deutsche Forschungsgemeinschaft for support (Be3264/13-1, project number 411538199).
We thank G\'{a}bor Cs\'{a}nyi for help in finding good parameters for the SOAP fingerprints, Michele Ceriotti for providing us with the problematic methane configurations and Jonas Finkler for the careful reading 
of the manuscript.

\bibliography{main}

\end{document}